\documentclass[letterpaper,twocolumn,10pt]{article}
\usepackage{usenix}

\usepackage{tikz}
\usepackage{amsmath}

\usepackage[nolist]{acronym}
\usepackage{xcolor}
\usepackage{diagbox}
\usepackage{multirow}
\usepackage{graphicx}
\usepackage{subcaption}
\usepackage{amsmath}
\usepackage{xcolor}
\usepackage{listings}
\usepackage[frozencache,cachedir=_minted]{minted}
\usepackage[most]{tcolorbox}
\usepackage{booktabs}
\usepackage{enumitem}
\usepackage[table]{xcolor}
\usepackage{array}
\usepackage{todonotes}

\begin{document}

\newcommand{\pouya}[1]{\textcolor{red}{#1}}

\newcommand{\POZZER}{\textsc{Pozzer}}
\newcommand{\FIRMWARENUMBER}{15}
\newcommand{\REALTARGETS}{two}
\newcommand{\MCUNUMBER}{two}

\newcommand{\meng}[1]{\textcolor{orange}{Meng: #1}}
\newcommand{\addison}[1]{\textcolor{blue}{Addison: #1}}
\newcommand{\pansilu}[1]{\textcolor{cyan}{Pansilu: #1}}
\newcommand{\ksenia}[1]{\textcolor{magenta}{Ksenia: #1}}
\newcommand{\ulysse}[1]{\textcolor{darkgreen}{Ulysse: #1}}
\newcommand{\ali}[1]{\textcolor{teal}{#1}}

\begin{acronym}
\acro{PSC}{power side channel}
\acro{MCU}{microcontroller}
\acro{SNR}{signal-to-noise-ratio}
\acro{ML}{machine learning}
\acro{PCB}{printed circuit board}
\acro{ISA}{instruction set architecture}
\acro{IP}{intellectual property}
\acro{EM}{electromagnetic}
\acro{NDA}{non disclosure agreement}
\acro{HSI}{high speed interface}
\acro{CW}{ChipWhisperer}
\acro{DWT}{debug and watchpoint trace}
\acro{CFG}{control flow graph}
\acro{COTS}{commercial off-the-shelf}
\acro{IoT}{internet of things}
\acro{GNSS}{global navigation satellite system}
\acro{CDF}{cumulative distribution function}
\acro{API}{application program interface}
\acro{FPGA}{field-programmable gate array}
\acro{EDG}{Execution Divergence Graph}
\end{acronym}

\date{}

\title{\Large \bf \POZZER: A Power Side Channel-guided Fuzzer for Black-Box Embedded Systems}

\author{
{\rm Pouya Narimani, Kseniia Rogova, Addison Crump, Martin Mohl, Meng Wang, Ulysse Planta,}\\
{\rm Pansilu Pitigalaarachchi and Ali Abbasi}\\ \\
CISPA Helmholtz Center for Information Security
} 

\maketitle

\begin{abstract}

Firmware fuzzing is an effective technique for discovering vulnerabilities in embedded systems. However, existing coverage-guided firmware fuzzers typically obtain feedback through firmware instrumentation, hardware debug interfaces, or firmware rehosting, which requires access to the firmware source code or binary image. Such requirements are often infeasible for off-the-shelf embedded devices, where firmware binaries are inaccessible, unrehostable, immodifiable, or undebuggable, necessitating fuzzing under black-box conditions.

In this paper, we present \POZZER{}, a power side-channel-guided fuzzer for black-box embedded systems. \POZZER{} uses power traces as feedback to identify previously unseen behavior via an incrementally constructed graph-based representation of observed executions, guiding the fuzzer toward unexplored execution paths. Its non-profiling design requires neither prior firmware knowledge nor a clone device, extracting meaningful feedback from a single power trace per execution while remaining robust to measurement noise.
We evaluate \POZZER{} on \FIRMWARENUMBER{} firmware targets across \MCUNUMBER{} platforms and two real-world commercial embedded devices. Across the resulting target-platform combinations, \POZZER{} outperforms a blind fuzzer under the same time budget in 26 out of 30 target-platform combinations. Furthermore, \POZZER{} discovers two previously unknown vulnerabilities in one of the commercial devices, both confirmed by the vendor, demonstrating its potential for identifying vulnerabilities in black-box embedded systems.
\end{abstract}

\section{Introduction}\label{section:introduction}

Embedded systems have become an integral part of modern society, powering applications ranging from consumer \ac{IoT} devices~\cite{ding2020iot} to industrial automation~\cite{xu2018survey}, automotive systems~\cite{rahim2021evolution}, and medical devices~\cite{huang2023internet}. As these systems process increasingly security-sensitive data and control safety-critical operations, vulnerabilities in their firmware can have severe security and safety consequences. Fuzzing is an effective technique for discovering software and hardware vulnerabilities. Coverage-guided fuzzing, in particular, is effective because it continuously uses execution feedback to guide input generation toward previously unexplored execution paths. Existing firmware fuzzers obtain such feedback through instrumentation~\cite{shift}, hardware debug interfaces~\cite{eisele2023fuzzing}, or rehosting~\cite{scharnowski2022fuzzware}. However, these approaches require some degree of access to the target firmware or its execution environment, which are often unavailable in practice.

Modern embedded systems commonly integrate third-party modules supplied by different vendors. Many of these modules, such as wireless communication chips~\cite{esp32}, \ac{GNSS} receivers~\cite{ubloxgnss}, secure elements~\cite{stsafe}, and sensor controllers~\cite{cansensor}, contain their own \ac{MCU} executing proprietary firmware. The firmware of these modules is often inaccessible to system integrators and end users, and debugging interfaces are often disabled to prevent firmware extraction~\cite{zheng2019firm,chen2018iotfuzzer}. Consequently, many real-world embedded systems can only be tested through their exposed communication interfaces, entrenching a strict \emph{black-box} setting in which the attacker has no access to the firmware source code or binary image. Under such conditions, existing coverage-guided fuzzers cannot obtain the execution feedback required to guide a fuzzer and therefore fall back to a blind fuzzer without execution feedback, which explores the input space significantly less efficiently.

Obtaining execution feedback, therefore, becomes the fundamental challenge for black-box embedded firmware fuzzing. Physical side-channels offer an alternative source of execution feedback, without requiring firmware instrumentation, access to the firmware binary or source code~\cite{hidinginplainsight}, making physical side-channels well-suited for strict black-box settings. In particular, \acp{PSC} leak information about executed instructions~\cite{glamovcanin2023instruction} and processed data~\cite{galli2025chameleon}, making them a promising feedback source for black-box embedded fuzzing. However, transforming raw \ac{PSC} traces into practical fuzzing feedback is challenging as measurement noise, temporal alignment, and execution-dependent signal variations make it difficult to distinguish execution-related discrepancies from measurement artifacts.
Therefore, practical \ac{PSC}-guided fuzzing requires a fast, online approach that extracts meaningful execution-related feedback from a single noisy power trace under black-box conditions. Although prior side-channel-guided fuzzers explore similar ideas~\cite{sperl2019side,vincentiduskfuzz,mcclintick2024side,barredo2025gaflerna}, they rely on the availability of ground-truth data to train models~\cite{sperl2019side}, repeating executions~\cite{vincentiduskfuzz}, or unrealistic threat models~\cite{sperl2019side}.

In this paper, we first empirically analyze \ac{PSC} traces from diverse embedded firmware targets to determine what execution-related information can be reliably extracted from raw \ac{PSC} traces in a black-box setting, and identify the challenges posed by noisy environments and high-throughput fuzzing. We analyze the relationship between firmware execution and the corresponding \ac{PSC} traces, characterizing execution-related features that consistently appear in the traces as well as artifacts introduced by the underlying hardware and firmware implementation. 
Furthermore, our analysis identifies several sources of variation in \ac{PSC} traces, including redundant code segments, localized measurement noise, and temporal shifts, which can obscure the distinction between previously observed and genuinely new execution behavior and thereby affect the stability of a \ac{PSC}-guided fuzzer. Guided by these observations, we design \POZZER{}, a \ac{PSC}-guided fuzzer for black-box settings.
\POZZER{} addresses the challenge of applying the general idea of graph-based feedback to \ac{PSC} traces collected from embedded systems to guide a black-box fuzzer. The key intuition behind \POZZER{} is that discrepancies between \ac{PSC} traces indicate potential discrepancies in firmware control flow. Building upon this intuition, \POZZER{} incrementally constructs a pseudo execution-flow graph by identifying discrepancy points among the observed traces. The constructed graph serves as a global reference for determining whether newly captured traces exhibit previously unseen behavior and deserve more attention. 
To address the challenges of \ac{PSC}-based feedback, \POZZER{} combines several \ac{PSC}-specific techniques for robust \ac{PSC} trace comparison, hardware noise handling, and efficient execution-flow graph construction.
The empirical findings from our analysis directly inform the design of \POZZER{}'s hardware and software-level noise-handling techniques, enabling reliable single power trace feedback without requiring repeated measurements, profiling, prior knowledge of the firmware, or firmware instrumentation.

To evaluate \POZZER{}, we conduct two sets of experiments. First, we fuzz \FIRMWARENUMBER{} firmware targets spanning \MCUNUMBER{} different hardware platforms. Although firmware source code is available for these targets, \POZZER{} operates on uninstrumented firmware, and it does not receive direct execution feedback during fuzzing; its guidance is derived solely from \ac{PSC} traces. We compare \POZZER{} against a blind fuzzer as the baseline and an instrumented coverage-guided fuzzer as an upper-bound fuzzer with direct execution feedback.
Under the same time budget, \POZZER{} outperforms the blind fuzzer in 26 out of 30 targets and, in some cases, achieves performance comparable to that of the instrumented coverage-guided fuzzer. Second, we conduct case studies on \REALTARGETS{} real-world \ac{COTS} devices under a black-box setting, where firmware instrumentation and binary extraction are unavailable. \POZZER{} identifies two previously unknown vulnerabilities in one of the commercial devices, both of which are confirmed by the vendors, whereas the blind fuzzer fails to discover them, demonstrating \POZZER{}'s applicability to real-world black-box embedded systems.

In summary, this paper makes the following contributions:
\begin{itemize}
\item
We present \POZZER{}, a \ac{PSC}-guided fuzzer for black-box embedded systems that outperforms a blind fuzzer without requiring firmware instrumentation, firmware binary extraction, or profiling.
\item
We conduct an empirical analysis of \ac{PSC} traces in the context of black-box embedded fuzzing, identifying hardware and software-level characteristics of embedded systems that affect the reliable extraction of execution-related feedback from \ac{PSC} traces.
\item
We develop and integrate a set of \ac{PSC}-specific techniques, including signature-based search, chunk-based comparison, first derivative comparison, and time-shift-aware neighbor search into \POZZER{}'s graph-based feedback pipeline, enabling meaningful feedback extraction from a single noisy \ac{PSC} trace per execution.

\item
We demonstrate both the effectiveness and practical applicability of \POZZER{}: across \FIRMWARENUMBER{} firmware targets spanning \MCUNUMBER{} hardware platforms, \POZZER{} improves average coverage by 12.06\% over a blind fuzzer under the same time budget, while case studies on \REALTARGETS{} real-world \ac{COTS} devices show that it can identify vulnerabilities under a black-box threat model without firmware instrumentation or binary access.

\end{itemize}

\section{Background}\label{section:background}
In this section, we first define \POZZER{}'s threat model, detailing the information available to \POZZER{} about the target and the constraints imposed by it. We then introduce the concepts and terminology underlying coverage-guided fuzzing and physical side-channels, providing the background necessary to understand the design of \POZZER{}.

\subsection{Threat model}\label{section:threatmodel}

\POZZER{} leverages \ac{PSC} leakage to fuzz embedded systems, operating at the intersection of hardware and software. Therefore, we define the threat model from both the hardware and software perspectives.

\textbf{Software.}
At the software level, we assume a strict black-box threat model. \POZZER{} has no access to the target firmware's source code or binary image. The attacker can interact with the target only through its exposed communication interfaces and observe its responses, if any. However, the attacker cannot obtain code-coverage feedback through source code or binary instrumentation. Thus, conventional instrumentation-based guidance is unavailable, making the setting equivalent to black-box blind fuzzing.

\textbf{Hardware.}
At the hardware level, we assume that the target's debug interfaces are inaccessible and no instruction traces or debugging information can be obtained from the device. However, the attacker has physical access to the device and can passively monitor \ac{PSC} traces by installing a shunt resistor on the target's power supply. We further assume that the attacker knows the exposed interface pins and can use them to send test inputs to the target during fuzzing.

\subsection{Power Side-Channels}

\acp{PSC} were initially exploited to recover cryptographic keys from smart cards by leveraging data-dependent variations in power consumption~\cite{kocher1999differential}. Such attacks exploit implementation-induced information leakage rather than weaknesses in the underlying cryptographic algorithms. Subsequently, \acp{PSC} have also been used to identify executed instructions, as different instructions exhibit distinguishable \ac{PSC} patterns~\cite{narimani2021side, park2018power}. Crucially, these distinct patterns mean that different execution paths and control-flow transitions induce measurable differences in a device's \ac{PSC} traces. This makes \acp{PSC} a highly promising modality for passively inferring program state. However, using \acp{PSC} to provide online execution feedback differs fundamentally from traditional side-channel analysis. Unlike cryptographic attacks, which typically target short execution windows and tolerate extensive offline processing, fuzzing requires extracting meaningful execution information from program execution in real time. This introduces new challenges, including handling measurement noise~\cite{narimani2024exploring}, clock synchronization~\cite{ferrufino2023fobos}, and efficient processing of long time-series traces~\cite{bursztein2023generalized}.

\subsection{Profiling vs. Non-Profiling Attacks}

Side-channel attacks are commonly classified as profiling or non-profiling, depending on the attacker's capabilities. Profiling attacks assume access to a clone device that is sufficiently similar to the target and controlled by the attacker. This enables the collection of ground-truth execution data to build a leakage model~\cite{standaert2009compare}, which is then applied to the target device. However, profiling a clone device is not feasible in black-box scenarios because it needs access to the source code or binary image of the firmware to obtain ground-truth data.

In contrast, a non-profiling attack assumes that the attacker has no access to a clone device. Instead, the attacker only possesses a set of side-channel measurements with their corresponding inputs and/or outputs~\cite{electronics12153279}, and performs the attack directly on these measurements~\cite {brier2004correlation}. Accordingly, \POZZER{} requires no profiling stage, training data, or leakage models from similar devices. Instead, it derives feedback solely from \ac{PSC} traces collected from the target during fuzzing, consistent with our black-box threat model.

\subsection{Coverage-guided fuzzing}

Coverage-guided fuzzing has been proven to be a powerful tool to discover vulnerabilities in software and hardware systems. It relies on code-coverage feedback to identify inputs that reach previously unseen basic blocks, add them to a \emph{corpus}, and iteratively mutate them to explore deeper program states. While highly effective, traditional coverage-guided fuzzers rely heavily on compile-time instrumentation or dynamic binary translation (e.g., QEMU~\cite{bellard2005qemu}) to retrieve this coverage feedback, both of which are intrusive and incompatible with our black-box threat model. In this work, we replace traditional software-based coverage feedback with a custom \ac{PSC}-based feedback mechanism. This enables \POZZER{} to perform coverage-guided fuzzing using only \ac{PSC} traces as passive side-channel leakage, remaining completely oblivious to the target's internal software constraints.

\section{Empirical analysis of \ac{PSC} traces as feedback}\label{section:systematicanalysis}
This section presents an empirical analysis of \ac{PSC} traces that motivates \POZZER{}'s design. As discussed in Section~\ref{section:background}, \ac{PSC} traces leak information about executed instructions and processed data~\cite{narimani2021side,ling2025fusiondisassembler}. However, their suitability as an online feedback source for fuzzing remains largely unexplored. Unlike traditional \ac{PSC} analysis, fuzzing requires extracting reliable execution-related information from a single \ac{PSC} trace, without prior knowledge of the target firmware or repeated measurements. 
Therefore, we investigate which execution-related characteristics are reliably observable in \ac{PSC} traces and which software- and hardware-level effects affect their interpretation.
Specifically, we ask:
\begin{itemize}
    \item
    Which firmware behaviors introduce variability or other artifacts into \ac{PSC} traces, and how do these effects influence the interpretation of execution-related information?
    \item
    Which hardware and measurement factors introduce temporal or voltage variations in \ac{PSC} measurements, and how do these factors affect the reliability of \ac{PSC} trace comparison?
\end{itemize}

To answer these questions, we design our empirical analysis as follows. We select and execute ten firmware programs on two hardware platforms, \emph{ST STM32F3} and \emph{Microchip SAM4S}. For each firmware, we manually construct ten test cases that exercise different execution paths. We execute each test case 1000 times while capturing the corresponding \ac{PSC} traces, resulting in 200,000 traces in total. 

To characterize firmware-induced effects in \ac{PSC} traces, we compare traces generated by different test cases that exercise different execution paths. This comparison allows us to identify execution dependent variations that can complicate feedback extraction. Motivated by observations reported in prior work~\cite{liu2016code}, we examine recurring similarities and differences across the traces. Our analysis reveals that similar code segments can appear at different temporal locations across \ac{PSC} traces, resulting in \emph{redundant code segments} and \emph{temporal shifts} that affect trace comparison. To characterize hardware and measurement-induced effects, we compare measurements for repeated executions of the same test case. Since repeated executions of the same test case are generally expected to follow the same firmware execution, differences between their traces should primarily arise from hardware and measurement effects. Comparing these repeated traces reveals \emph{localized measurement noise} and temporal or voltage variations, demonstrating how these effects can affect the reliability of \ac{PSC} trace comparison. 

The remainder of this section examines three key characteristics identified through the above empirical analysis: \emph{redundant code segments}, \emph{localized measurement noise}, and \emph{time shifts}.
These characteristics affect the reliability of \ac{PSC}-based feedback extraction and motivate the design choices presented in Section~\ref{section:design}.

\begin{figure}[!t]
    \centering
    \includegraphics[width=7.5cm]{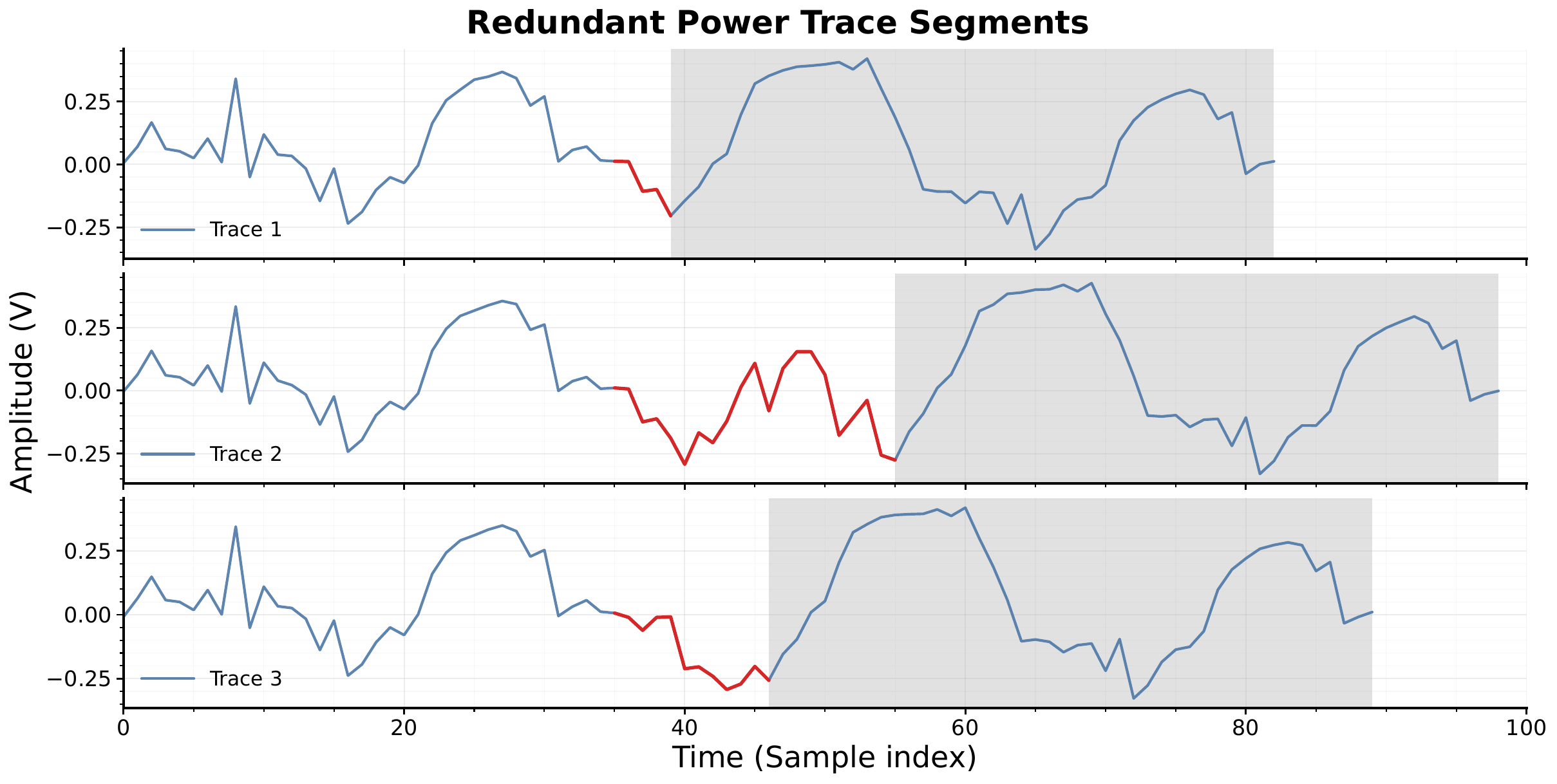}
    \caption{Three different power traces, all having the same code segment (inside the highlighted area) executed with different prior code execution paths (plotted in red).}
    \label{fig:redundanttraces}
\end{figure}

\textbf{Redundant Code Segments.}
Prior work has shown that identical code segments may occur at different temporal positions in \ac{PSC} traces~\cite{liu2016code}, a behavior we also observe. We refer to these recurring segments as redundant code segments because they represent previously observed code execution rather than new execution behavior. Consequently, two traces can differ substantially while still containing segments corresponding to the same code execution. Figure~\ref{fig:redundanttraces} illustrates an example in which multiple execution paths converge to an identical code segment. Although the preceding execution differs, the latter parts of the traces correspond to the same code execution. This behavior naturally arises when distinct firmware execution paths converge, causing identical code segments to appear at different temporal locations. As a result, direct time-domain comparison is insufficient for distinguishing previously observed execution behavior from genuinely new executions.

\begin{figure}[!t]
    \centering
    \includegraphics[width=7.5cm]{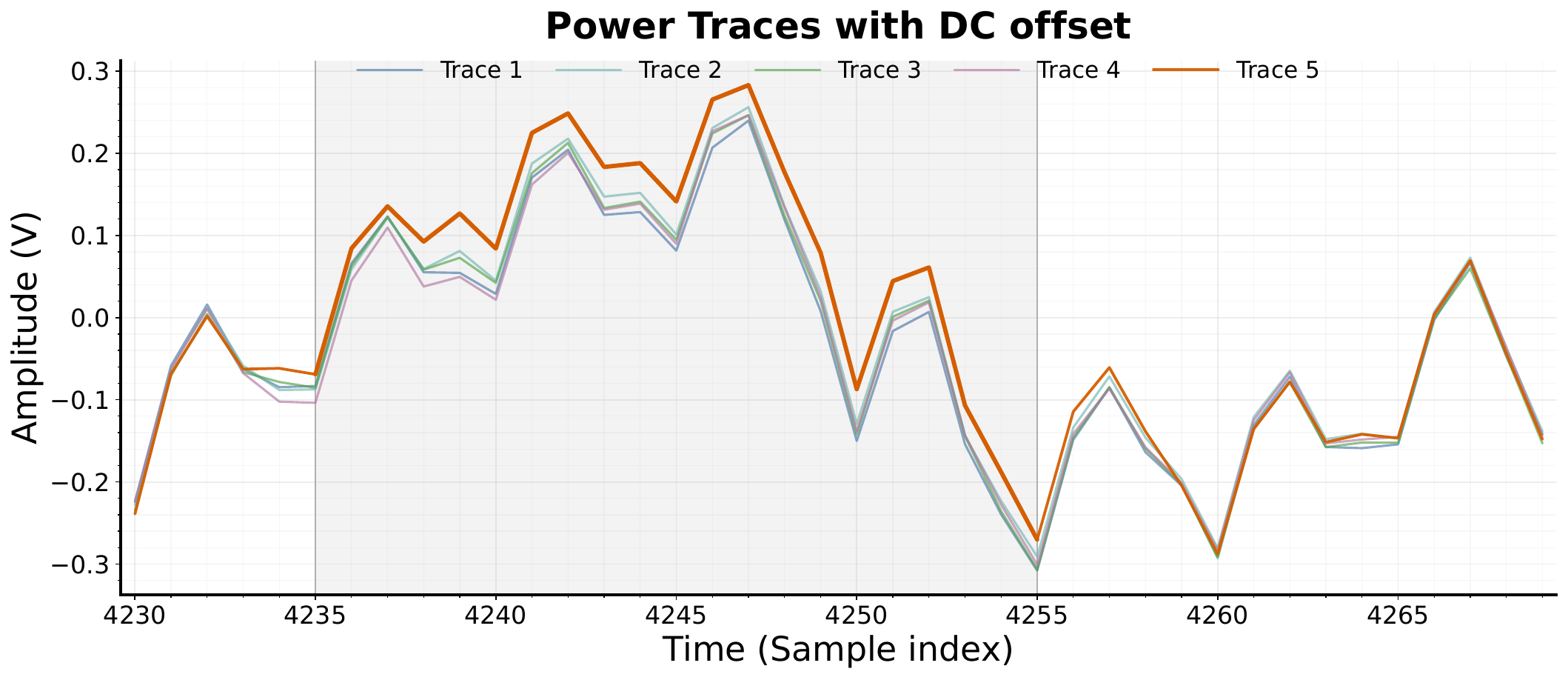}
    \caption{Five \ac{PSC} traces with identical code executions, with one execution
exhibiting a noticeable local amplitude shift.}
    \label{fig:dcshift}
\end{figure}

\textbf{Localized Measurement Noise.}
Repeated executions of the same test case reveal localized hardware-induced variations in the measured \ac{PSC} traces. Figure~\ref{fig:dcshift} shows an example in which one execution exhibits a local amplitude shift despite executing the same code. These localized distortions originate from the measurement process, such as quantization errors, rather than the firmware itself. Since they affect only a portion of the \ac{PSC} trace, they can obscure execution related features and make traces from identical executions appear different. Consequently, localized measurement noise can induce ambiguity when interpreting execution behavior directly from raw \ac{PSC} traces.

\textbf{Time Shifts.}
Another behavior consistently observed during our measurements is the presence of small time shifts between repeated executions of the same test case. Although the executed instructions remain identical, small timing variations cause corresponding features in the \ac{PSC} traces to become misaligned. As a result, traces generated from the same execution may no longer align sample by sample along the temporal axis, making it difficult to reliably compare execution behavior directly in the time domain.

\section{\ac{PSC}-based Graph Construction}\label{section:design}

In this section, we explain the design of \POZZER{} for using \ac{PSC} traces as real-time fuzzing feedback. \POZZER{} must derive reliable execution-related feedback from noisy \ac{PSC} measurements while operating under a strict black-box threat model. Our empirical analysis in Section~\ref{section:systematicanalysis} provides insights into several factors that affect \ac{PSC} traces. These observations constitute the design of \POZZER{} and its approach to extracting, comparing, and interpreting execution-related feedback.

In addition to these trace-specific challenges, \POZZER{} must satisfy two general requirements of online fuzzing. First, it must maintain high throughput. A fuzzing campaign executes many test cases, each producing a large \ac{PSC} trace that must be acquired and processed. Second, it requires a global reference against which each new observation can be compared to determine whether the corresponding test case exhibits previously unseen behavior. Therefore, \POZZER{} must derive execution-related feedback from \ac{PSC} traces while limiting the overhead of trace acquisition and processing in addition to accounting for measurement variability and noise. These requirements guide the design of \POZZER{} and lead to the following three design goals:

\begin{itemize}[nosep]
    \item
    \textbf{Efficient feedback extraction:}
    \POZZER{} should minimize the overhead of capturing and processing \ac{PSC} traces so the fuzzer can achieve high throughput within a given time budget.
    
    \item
    \textbf{Reliable single-trace feedback:}
    \POZZER{} should account for temporal and voltage variations in \ac{PSC} traces without relying on repeated measurements. In particular, it should extract stable execution-related information from a single trace for each test case.
    \item
    \textbf{Global execution-flow modeling:}
    \POZZER{} should maintain a global representation of execution flow behavior observed during fuzzing. This representation should allow each newly captured trace to be compared with previously observed behavior and support the identification of previously unseen execution-flow transitions.
\end{itemize}

In addition, non-profiling operation is a constraint imposed by our threat model rather than an independent design goal. Therefore, the feedback mechanism of \POZZER{} must construct and update the global execution-flow representation during fuzzing using the \ac{PSC} traces.

\subsection{Graph-based Feedback}

The concept of constructing a global reference for black-box fuzzers has been discussed in a prior paper~\cite{lin2026execution}. This paper introduces the concept of \ac{EDG} and uses it as a global reference for a black-box fuzzer. In this graph, nodes denote the execution traces, and edges denote transitions. For example, an edge \textsf{A}$\to$\textsf{B} denotes that an execution trace \textsf{A} followed by an execution trace \textsf{B} was observed. Consequently, the graph stores observed prefixes of execution traces as paths from an initial node. If a program execution produces an execution trace that is not a prefix of any existing path, the graph is updated by splitting nodes or adding new edges to represent the newly observed program behavior.

Although the \ac{EDG} provides the foundations for graph construction in a noise-free environment, using it on \ac{PSC} traces requires significant design improvements and the introduction of new techniques to address the challenges that arise when using \ac{PSC} traces. These challenges include introducing a new similarity metric for \ac{PSC} traces, designing techniques to mitigate hardware noise as discussed in Section~\ref{section:systematicanalysis}, and controlling the graph size. Following, we first introduce the similarity metric that we use to detect discrepancies between \ac{PSC} traces, then we explain the techniques used by \POZZER{} to mitigate hardware-induced noise in the next section.

\textbf{Similarity metric.}
\ac{PSC} traces are floating-point time series that are heavily affected by noise. Therefore, exact equivalence is unsuitable for detecting discrepancies due to measurement variations. Hence, \POZZER{} requires a similarity metric to detect discrepancies between two \ac{PSC} traces and consequently detect novelty. In this paper, we use \textit{L$_1$} distance metric to compare the points in \ac{PSC} traces. If the distance between two points is higher than a threshold, then the corresponding point is a discrepancy point. The \textit{L$_1$} distance is inexpensive to compute, making it a fast and efficient metric for the repeated trace comparisons required by \POZZER{}'s online feedback.

\subsection{Hardware Noise Handling}
\label{subsection:hw-noise-handle}

In this section, we explain the techniques used by \POZZER{} to mitigate hardware-induced noise of \ac{PSC} traces. Our empirical analysis in Section~\ref{section:systematicanalysis} identified two types of hardware noise, localized amplitude shifts (Figure~\ref{fig:dcshift}) and small time shifts. These variations can cause the discrepancy detection procedure to incorrectly identify measurement artifacts as execution differences. \POZZER{} therefore incorporates several essential noise-handling techniques at different stages of the feedback workflow to improve robustness while preserving execution-related information.

\begin{figure}[!t]
    \centering

    \begin{subfigure}{0.22\textwidth}
        \centering
        \includegraphics[width=\linewidth]{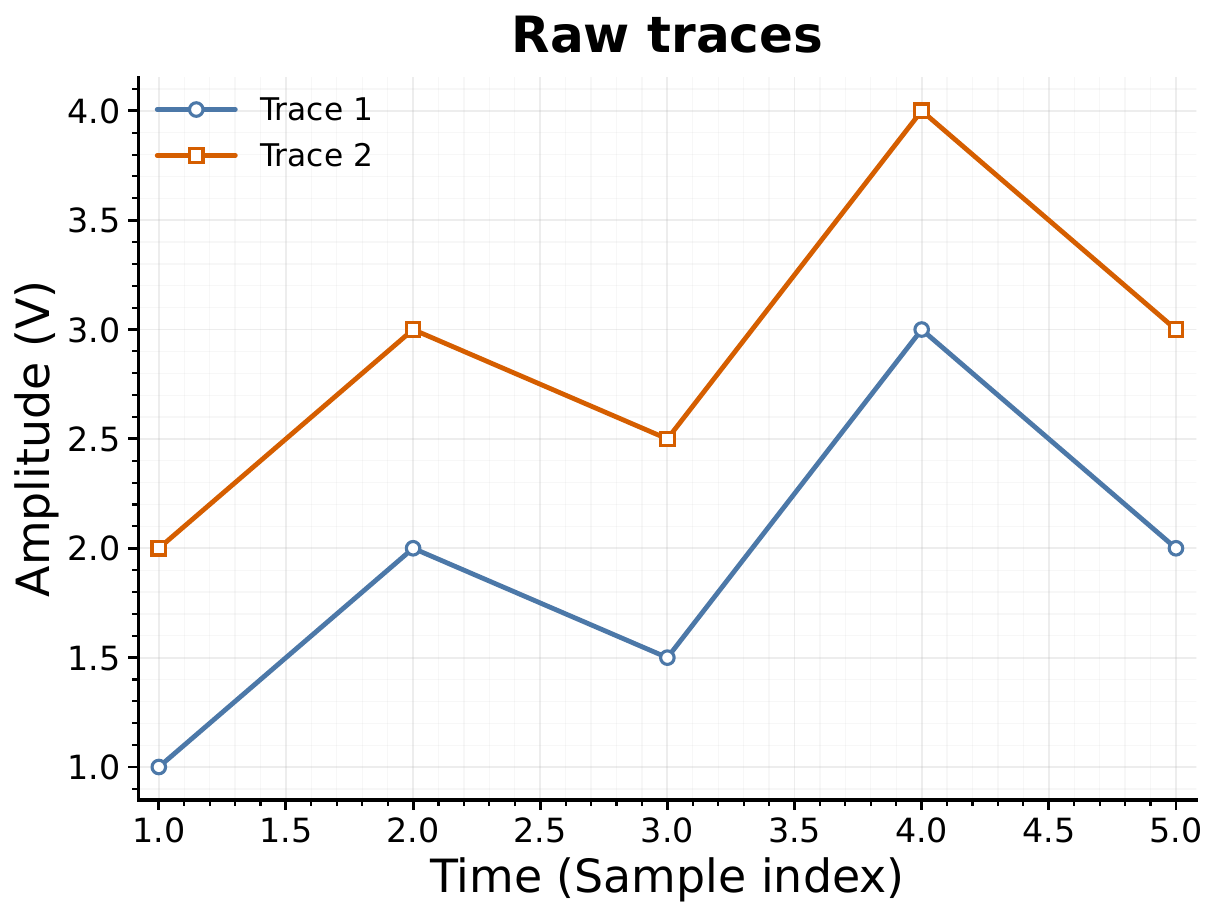}
        \caption{Raw \ac{PSC} traces}
        \label{fig:rawtraces}
    \end{subfigure}
    \hspace{0.02\textwidth}
    \begin{subfigure}{0.22\textwidth}
        \centering
        \includegraphics[width=\linewidth]{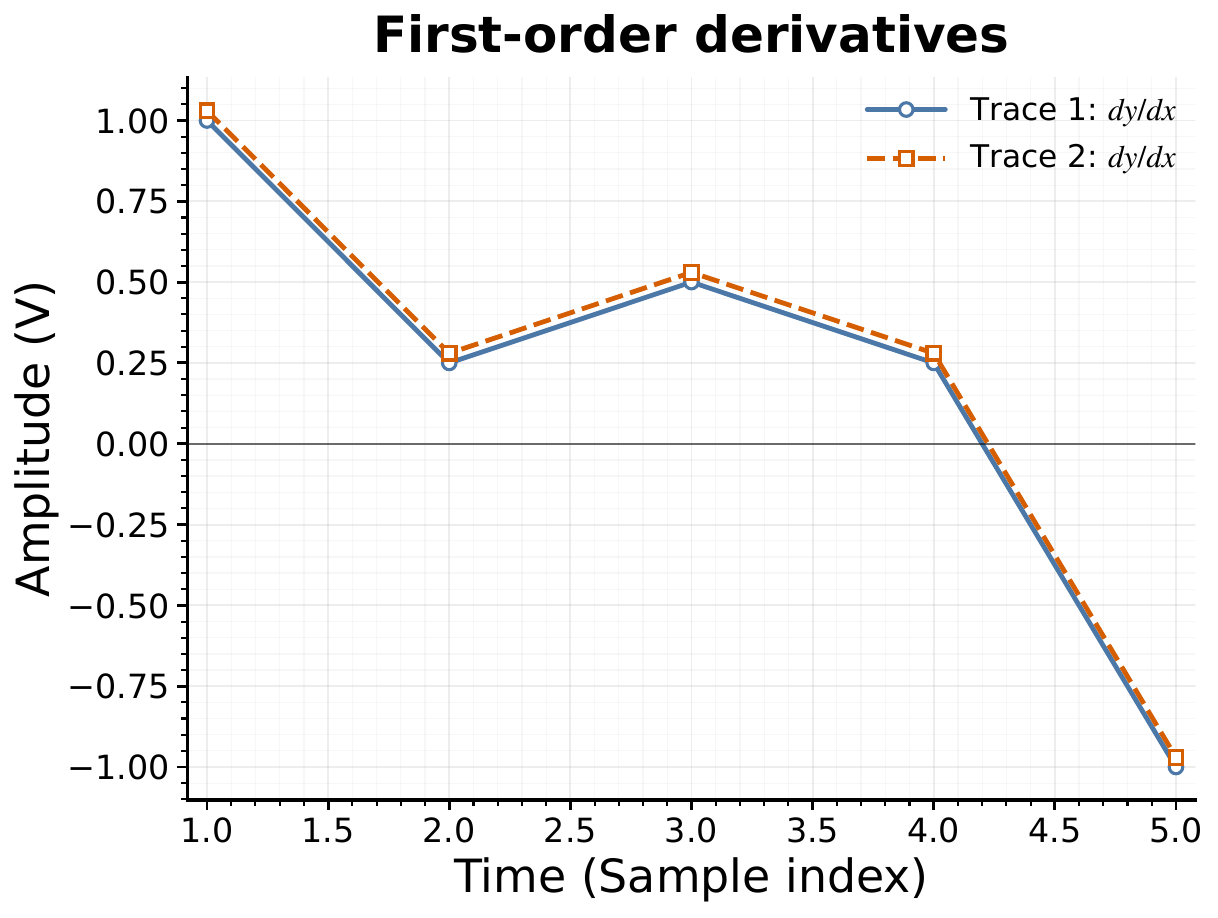}
        \caption{First-order derivatives}
        \label{fig:afterderivatives}
    \end{subfigure}

    \caption{\ac{PSC} traces before and after first-order differentiation. Differentiation suppresses the local amplitude shift, allowing point-wise comparison to correctly detect similarity.}
    \label{fig:firstorderderivatives}
\end{figure}

\textbf{Chunk-based Discrepancy Detection.} 
Point-wise comparison is sensitive to localized amplitude variations because a disturbance affecting only a small portion of a trace can produce a sequence of apparent discrepancies. \POZZER{} therefore compares fixed-size chunks rather than individual samples. Each chunk represents a sequence of consecutive samples corresponding to a portion of the execution. Given two chunks $X$ and $Y$, \POZZER{} computes their normalized \textit{L$_1$} distance:

\begin{equation*}
D_{L_1}(X,Y)=\frac{1}{|CS|}\sum_{i=0}^{|CS|-1}|x_i-y_i|,
\end{equation*}

where $|CS|$ is the number of samples in a chunk. A chunk is considered discrepant when its distance exceeds a predefined threshold. Comparing chunks makes discrepancy detection less sensitive to localized variations and enables the graph construction algorithm to operate at a lower temporal resolution, as it compares trace segments corresponding to instruction sequences rather than individual instruction profiles. Recall that Figure~\ref{fig:dcshift} already features a localized amplitude shift in \ac{PSC} traces. A point-wise comparison would incorrectly identify this region as a discrepancy region and, consequently, as an interesting execution behavior. In contrast, \POZZER{}'s chunk-based discrepancy detection is considerably more robust to such disturbances.

\textbf{First-Order Derivatives.}
After identifying the discrepant chunk, \POZZER{} performs a point-wise comparison within that chunk to determine the exact discrepancy point. At this stage, point-wise comparison is more sensitive to local amplitude shifts because it operates at the sample level. Therefore, \POZZER{} performs this comparison in the first-order derivative domain rather than directly on the raw traces. Differentiation suppresses constant and slowly varying amplitude offsets while preserving changes in the trace signal. Figure~\ref{fig:firstorderderivatives} illustrates this effect, and Appendix~\ref{section:appendix:derivatives} provides a mathematical analysis of the approach.

\begin{figure}[!t]
    \centering
    \includegraphics[width=7.5cm]{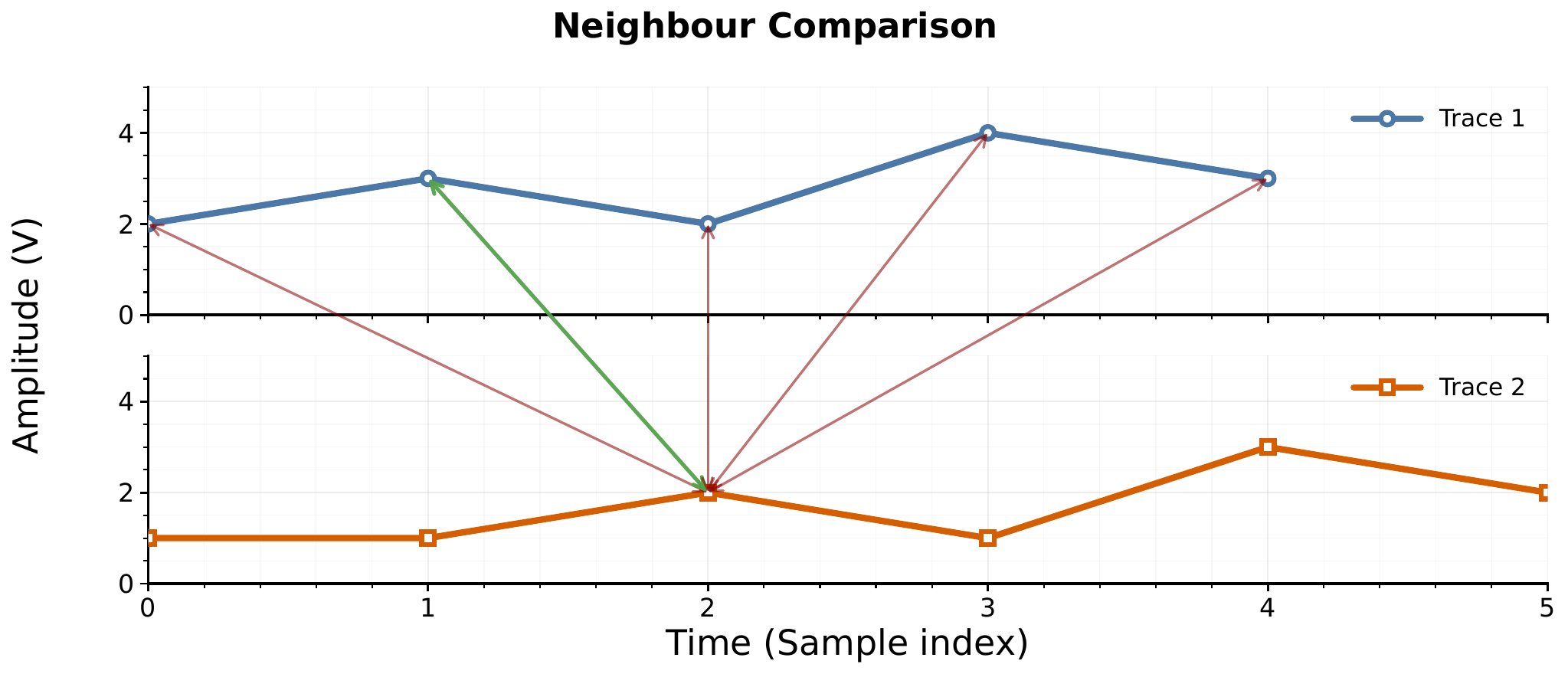}
    \caption{Robustness of neighbor comparison to time shifts. Each point is compared with neighboring points within a radius of two samples, enabling correct matching despite small time shifts.}
    \label{fig:neighborshifts}
\end{figure}

\textbf{Time Shifts.} \looseness=-1
Small time shifts can also cause false discrepancy points because corresponding features may be displaced by one or more samples between traces. Since point-wise comparison ultimately determines the precise discrepancy location, such misalignment can introduce spurious graph nodes that do not represent new execution behavior. \POZZER{} therefore extends point-wise comparison with a neighbor search. For each sample in one trace, it compares the sample with the aligned sample and neighboring samples within a predefined window in the other trace. This allows corresponding features to match despite small temporal misalignments and reduces false discrepancy point detections. Figure~\ref{fig:neighborshifts} illustrates the approach using a neighborhood of two samples on either side.

\section{Implementation}\label{section:implementation}

\begin{figure}[!t]
    \centering
    \includegraphics[width=7.5cm]{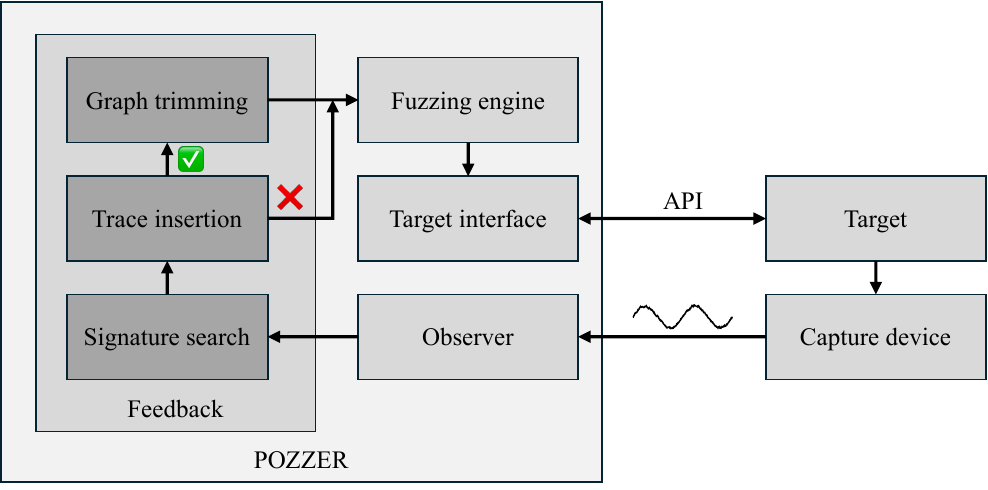}
    \caption{An overview of \POZZER{}'s architecture}
    \label{fig:overview}
\end{figure}

In this section, we first explain the high-level implementation of \POZZER{}, then we explain each stage of the feedback module, including graph construction, trace insertion, and graph trimming in a fuzzing loop.

\subsection{Overview}
\label{subsection:design-overview}
In this subsection, we provide an overview of \POZZER{} and its online feedback workflow. Figure~\ref{fig:overview} shows the main components of \POZZER{} and their interactions. At each iteration, the fuzzing engine selects and mutates an input from the corpus to generate a new test case. The target interface sends the test case to the target through the target API, while the capture device collects the resulting \ac{PSC} trace. The observer receives the trace and passes it to the feedback module, which compares the new trace with the execution flow observed so far and determines whether the test case exhibits previously unseen execution behavior.

The feedback module maintains a graph-based representation of the observed execution behavior. It first uses a signature search to locate the corresponding portions of the new trace in the existing graph. If the trace contains previously unseen behavior, the trace insertion procedure incorporates the new execution into the graph, after which graph trimming removes redundant portions of the graph. The feedback module then reports whether new execution behavior was observed. If so, the fuzzing engine adds the corresponding test case to the corpus and continues with the next iteration; otherwise, it proceeds without adding the test case.

The following subsections describe these mechanisms in detail. Subsection~\ref{subsection:graph-cons} presents \POZZER{}'s graph construction procedure and explains how it addresses challenges arising from firmware execution behavior and the interpretation of \ac{PSC} traces.

\subsection{Graph Construction}\label{subsection:graph-cons}
As described in Section~\ref{subsection:design-overview}, \POZZER{} represents observed execution behavior as a graph and uses this representation as the global reference for online feedback. This section describes how \POZZER{} constructs and maintains this graph from \ac{PSC} traces. The construction process consists of three stages: \emph{signature-based search}, \emph{trace insertion}, and \emph{graph trimming}. Signature-based search addresses the non-sequential appearance of execution segments in \ac{PSC} traces, trace insertion incorporates previously unseen execution behavior into the graph, and graph trimming removes redundant representations introduced during insertion.

\subsubsection{\textbf{Signature-based Search}}

A newly captured \ac{PSC} trace cannot be compared only with the local descendants of its current position in the graph. As discussed in Section~\ref{section:systematicanalysis}, firmware execution paths can converge, causing the same code segment to appear at different temporal locations in different traces. Consequently, a new trace may deviate from the current graph path and later rejoin an already observed path. Restricting the search to the current branch would interpret such a transition as new behavior and unnecessarily expand the graph.

\POZZER{} addresses this issue using a signature-based search that searches for matching signatures across the entire graph. Since all \ac{PSC} traces share the same entry point, \POZZER{} first compares a new trace with the graph from the root until the first discrepancy point. Rather than restricting subsequent comparison to the descendants of that discrepancy point, it searches a global pool of signatures collected from the entire graph. A signature is a fixed-length segment extracted from the beginning of each graph branch and represents the entry point of that branch. Searching across signatures throughout the entire graph enables \POZZER{} to identify execution-flow transitions that cannot be detected by considering only the local descendants of the current discrepancy point, as program execution does not necessarily follow a single sequential path. 

\begin{figure}[!t]
    \centering
    \begin{subfigure}{0.22\textwidth}
        \centering
        \includegraphics[width=\linewidth]{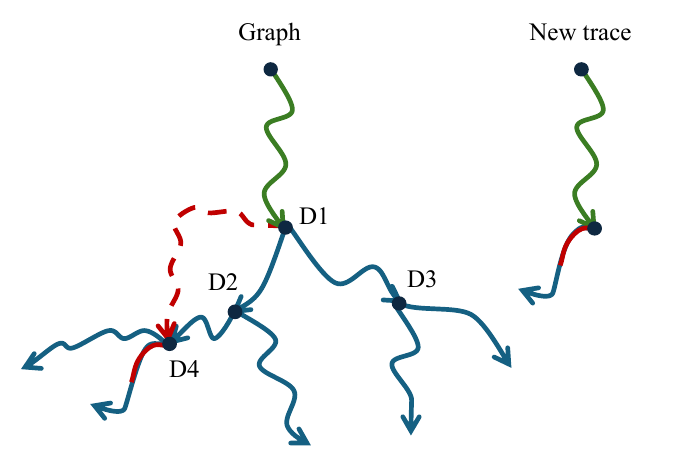}
        \caption{Jump detection}
        \label{fig:jump}
    \end{subfigure}
    \hspace{0.03\textwidth}
    \begin{subfigure}{0.185\textwidth}
        \centering
        \includegraphics[width=\linewidth]{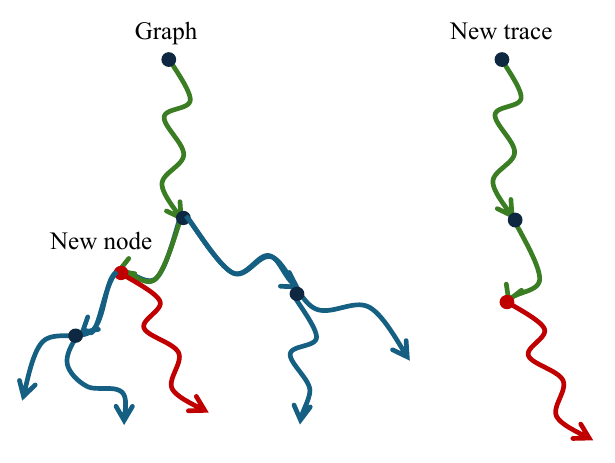}
        \caption{New discrepancy point}
        \label{fig:newdivergingpoint}
    \end{subfigure}
    
    \begin{subfigure}{0.20\textwidth}
        \centering
        \includegraphics[width=\linewidth]{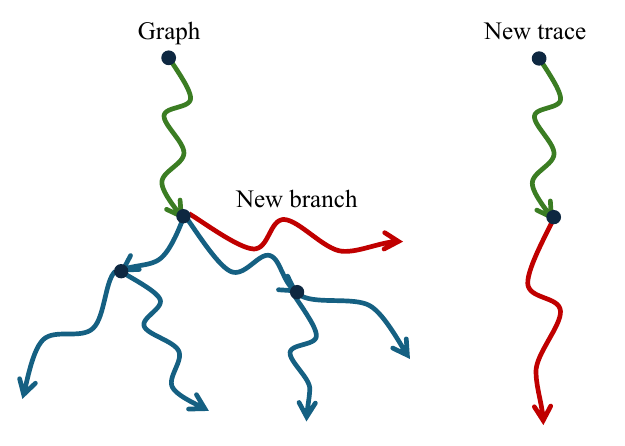}
        \caption{New branch}
        \label{fig:newbranch}
    \end{subfigure}
    \hspace{0.03\textwidth}
    \begin{subfigure}{0.21\textwidth}
        \centering
        \includegraphics[width=\linewidth]{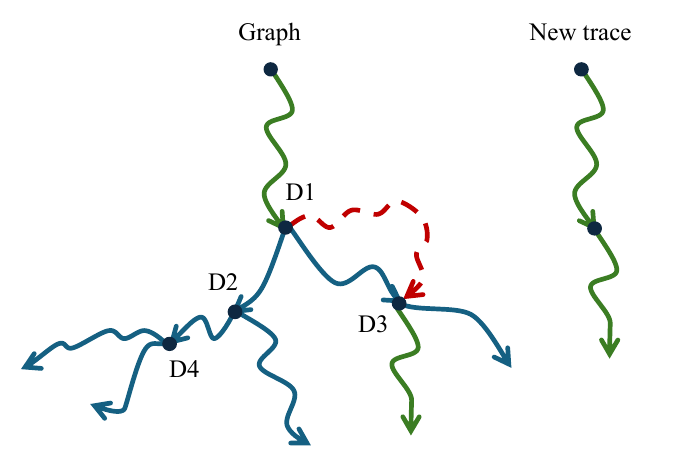}
        \caption{New path}
        \label{fig:newpath}
    \end{subfigure}

    \caption{Four trace-insertion cases handled by \POZZER{} and the corresponding updates to the execution-flow graph.}
    \label{fig:pozzeralgorithm}
\end{figure}

This global search also improves the efficiency of trace comparison. Signatures are substantially smaller than the full \ac{PSC} traces stored in the graph, allowing candidate matches to be identified without comparing the full new trace against every branch in the graph. Once a matching signature is found, \POZZER{} resumes trace comparison from the corresponding graph location.
Figure~\ref{fig:jump} illustrates this process. If a new trace transitions from discrepancy point \textit{D1} directly to a segment represented by \textit{D4}, bypassing \textit{D2}, signature matching identifies the transition and allows the trace to rejoin the existing execution path. Without this search, the transition would be integrated as a new branch from \textit{D1}, resulting in unnecessary graph expansion.

\subsubsection{\textbf{Trace Insertion}}

After capturing a new \ac{PSC} trace and comparing it against the constructed graph, \POZZER{} determines whether the trace exhibits previously unseen behavior; if so, it incorporates that behavior into the graph. This step is necessary to maintain the global execution reference used by subsequent fuzzing iterations. The decision falls into one of the following cases:

\textbf{New discrepancy point.}
If the new \ac{PSC} trace matches the signature of an existing node but exhibits a discrepancy at a location not represented by an edge, \POZZER{} creates the corresponding nodes and edge. Figure~\ref{fig:newdivergingpoint} shows this case. Since the trace reveals a previously unseen execution branch (red), the associated test case is considered interesting and is added to the corpus.

\textbf{New branch.}
If from a discrepancy point onward, the new \ac{PSC} trace does not match any existing signature, \POZZER{} creates a new branch from that discrepancy point. Since all traces share the same entry point, they necessarily match the graph for at least an initial segment. This case introduces previously unseen execution behavior without introducing another point of discrepancy. Figure~\ref{fig:newbranch} illustrates this case, where a new branch (red) is added to an already existing discrepancy point. The corresponding test case is also marked as interesting and added to the corpus.

\textbf{New Path.}
A new \ac{PSC} trace may also consist entirely of segments already represented in the graph while connecting them in a previously unseen order. Such a trace does not introduce a new segment, but it represents a previously unobserved execution path. \POZZER{} therefore adds the new edge to the graph and retains the corresponding test input. Figure~\ref{fig:newpath} shows an example in which the trace connects discrepancy points \textit{D1} and \textit{D3} through a path not previously represented in the graph.

\textbf{Full match.}
If the entire trace is represented by a prefix on an existing path in the graph, the execution behavior has already been observed. \POZZER{} therefore discards the corresponding test input and does not modify the graph.

\subsubsection{\textbf{Graph Trimming}}

Trace insertion can introduce redundant graph structure when the same execution segment appears at different temporal locations in \ac{PSC} traces. As discussed in Section~\ref{section:systematicanalysis}, such redundant segments arise naturally when different firmware execution paths converge. If the converged segment starts at an already existing discrepancy point, trace insertion identifies it as a new path, as described in the previous subsection. However, if the convergence point is not an already existing discrepancy point, then the corresponding \ac{PSC} trace is added to the graph, leading to unnecessary graph expansion. Therefore, \POZZER{} applies graph trimming after inserting a new trace to identify redundant execution segments and merge them. This reduces the graph size while preserving the paths. Trimming is performed only when the graph is modified, avoiding additional processing for traces that do not introduce new behavior.

\section{Evaluation}\label{section:evaluation}

In this section, we evaluate \POZZER{} against a blind fuzzer and compare its performance with a coverage-guided fuzzer that provides a reference point for performance under instrumentation. We first describe the experimental setup and evaluation metrics. 
We then use a simple example program to showcase the internal operation of \POZZER{} and provide a detailed analysis of its discrepancy detection algorithm using a representative execution flow. Next, we evaluate \POZZER{} on \FIRMWARENUMBER{} firmware targets using the \ac{CW} platform to compare its performance with the blind fuzzer, followed by a comparison with the related work. Finally, we evaluate \POZZER{} on \REALTARGETS{} real-world black-box embedded systems whose firmware is not publicly available and investigate its ability to discover vulnerabilities. 
Throughout these experiments, we address the following research questions:

\begin{enumerate}[nosep,label={\bfseries R\arabic*:}]
    \item
    Can \POZZER{} outperform a blind fuzzer under the same time budget, and is the performance difference statistically significant?
    
    \item
    How does the overhead introduced by \POZZER{}'s feedback affect its fuzzing throughput compared with the blind fuzzer, and what is the resulting corpus size of \POZZER{}?

    \item
    Can \POZZER{} fuzz real-world black-box embedded systems and discover vulnerabilities?
\end{enumerate}

\subsection{Experimental setup}

In this work, we develop a prototype of \POZZER{} on top of the LibAFL framework~\cite{libafl}. The \ac{PSC}-trace acquisition, processing, graph construction, and feedback logic are implemented within \POZZER{}'s prototype. \POZZER{} requires a calibration stage to determine a set of setup-specific parameters. The calibration stage does not require any ground-truth data from the target; hence, it does not violate our black-box threat model. The details of this stage are provided in Appendix~\ref{section:appendix:calibration}.

We use two experimental setups to evaluate \POZZER{}. First, we use the widely adopted \ac{CW}~\cite{chipwhisperer} platform to validate the effectiveness of \POZZER{}'s approach. Specifically, we use four \ac{CW} Husky platforms with CW313 target boards. As target \acp{MCU}, we use an \emph{ST} ARM Cortex-M4 \emph{STM32F303} and a \emph{Microchip} ARM Cortex-M4 \emph{SAM4S}. We evaluate \FIRMWARENUMBER{} firmware targets spanning a diverse set of embedded firmware components, including protocol and data parsers, binary file parsers, media decoders, and command interpreters. Our target set also includes benchmark programs commonly used by firmware rehosting frameworks~\cite{feng2020p2im,scharnowski2022fuzzware,mera2021dice}, enabling a direct comparison with prior embedded firmware fuzzing work, despite \POZZER{} operating under a more restrictive black-box threat model. The fuzzer runs on an x86 host machine with 32~GB of memory and communicates with the \ac{CW} platforms over USB. Further details of the experimental setup are provided in Appendix~\ref{section:appendix:evalmachine}.

Second, to demonstrate \POZZER{}'s flexibility and independence from the \ac{CW} platform, we evaluate \POZZER{} on two real-world black-box targets whose firmware source code and binaries are not publicly available. The first target is a \emph{U-Blox ZED-F9P} GNSS receiver module. The second is a security chip that cannot be explicitly identified due to its sensitive nature and a restrictive \ac{NDA}; throughout this paper, we refer to it as the \emph{Anonymous Target}. 
Unlike \ac{CW} target boards, the real-world targets neither include an on-board shunt resistor for \ac{PSC} measurements nor provide an internal trigger signal. Therefore, we install an external shunt resistor in the power supply path and measure the resulting \ac{PSC} traces across the resistor using a Tektronix MSO66B oscilloscope equipped with a \ac{HSI}. To generate the trigger signal, we run \POZZER{} on a Raspberry Pi 5 and use its GPIO pins. The Raspberry Pi serves as the host, sending test cases to the target while receiving the captured \ac{PSC} traces from the oscilloscope and processing the feedback.

\subsection{Evaluation metric}
We evaluate \POZZER{} using three primary metrics: \emph{code coverage}, \emph{fuzzing throughput}, and \emph{corpus size}. These metrics capture complementary aspects of fuzzing effectiveness and efficiency.

\textbf{Code Coverage.}
Since the source code of the firmware targets in the \ac{CW} setup is available, we use the source code line coverage as the primary measure of fuzzing effectiveness. We use line coverage following established fuzzing evaluation practices~\cite{moesok} to enable consistent comparisons across targets running on \acp{MCU} from different vendors. We collect coverage using a unified replay procedure for all experiments. Specifically, after each fuzzing campaign, we replay the resulting corpus on a gcov~\cite{gcov}-instrumented version of the target and record the lines the corpus exercises. Because we use the same replay procedure and source code for all campaigns, the resulting coverage measurements are directly comparable across targets and fuzzers. 

\textbf{Fuzzing Throughput.}
We measure fuzzing throughput as the number of test cases executed per unit of time. This metric captures the execution overhead introduced by \POZZER{}'s \acs{PSC}-based feedback mechanism and is used to assess its efficiency relative to the blind fuzzer. Because execution rates can vary across hardware platforms and firmware targets, we report throughput separately for each target and platform. 

\textbf{Corpus Size.}
We also measure the number of test cases retained in the final corpus of each fuzzing campaign. Corpus size indicates how selectively the feedback mechanism identifies inputs as interesting and complements coverage when evaluating \POZZER{}'s feedback effectiveness. 

\textbf{Fuzzing Configuration.} 
For all experiments, we use the same mutation strategies and initial seeds across the compared fuzzers. We conduct five independent 12-hour campaigns for each target and report the average results across these repeated runs to account for the inherent randomness of fuzzing~\cite{moesok}. All fuzzers are given the same 12-hour time budget, enabling a time-based comparison despite differences in fuzzing throughput.

For the comparisons, we use a blind fuzzer as the baseline because it operates under the same black-box threat model and attacker capabilities as \POZZER{}. The blind fuzzer uses the same fuzzing configuration as \POZZER{}, with only the observer and feedback modules disabled. Consequently, every generated test case is retained as interesting. We also use a reference coverage-guided fuzzer executed on a conventional computer as a reference point for performance achievable with direct execution feedback. This fuzzer uses GCC coverage instrumentation and therefore has access to information unavailable to \POZZER{} under the black-box threat model. Since this fuzzing cannot be performed directly on \ac{CW} target hardware, we project its results onto the same 12-hour time axis based on the number of test cases executed by the blind fuzzer. This comparison favors the traditional coverage-guided fuzzer by ignoring instrumentation overhead and therefore acts as a conservative baseline for \POZZER{}'s performance.

\begin{figure}[!t]
    \centering
    \includegraphics[width=7cm]{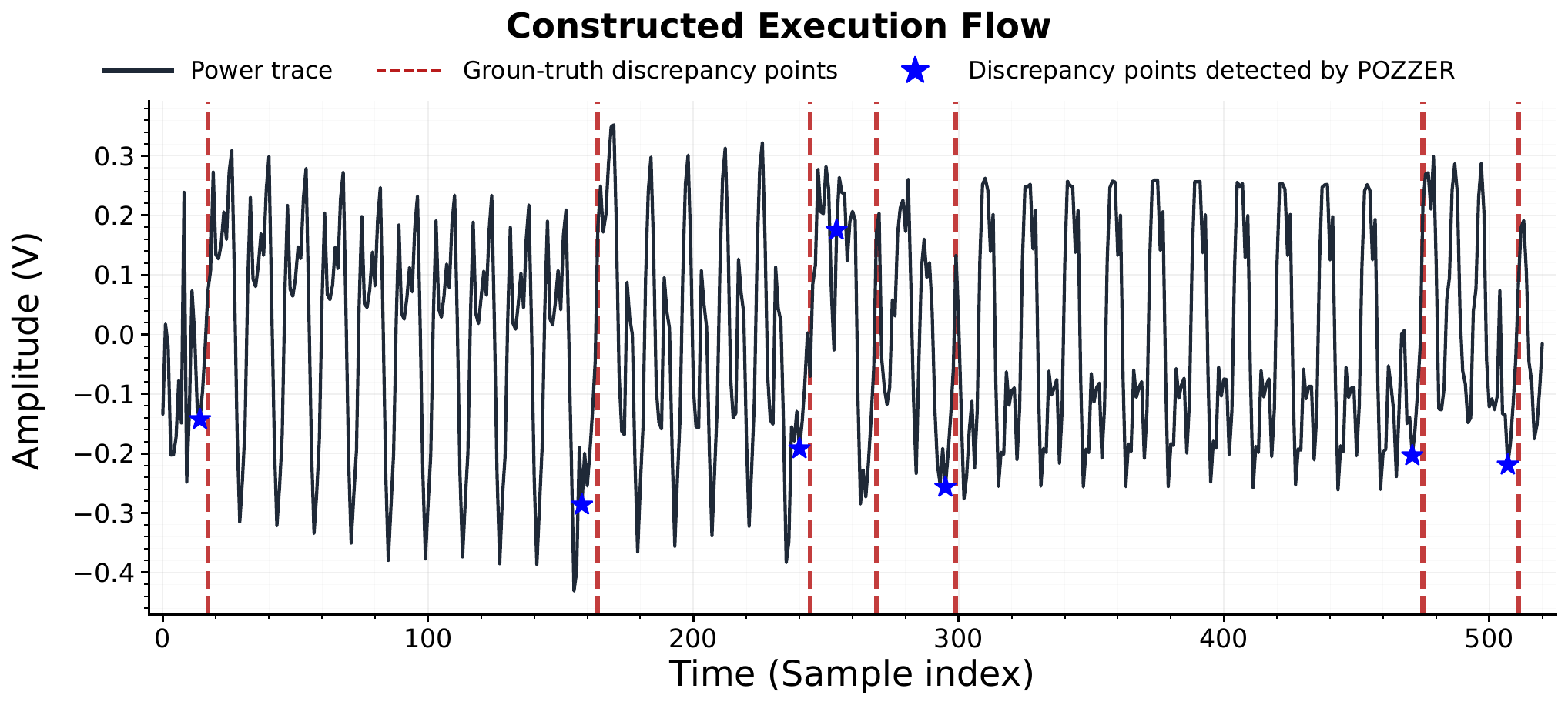}
    \caption{Comparison of ground-truth execution discrepancies and the discrepancy points detected by \POZZER{} for the nested-\emph{if} target.}
    \label{fig:casestudy}
\end{figure}

\subsection{Example of graph construction }
To analyze \POZZER{}'s graph construction algorithm in a controlled environment, we implement a contrived target firmware with a nested \texttt{if} structure (see Appendix~\ref{section:appendix:nestedif}). The program checks the input against a fixed character sequence, with each correctly matched character reaching the next \texttt{if} condition. Although simple, this structure provides a basic execution flow that reveals how \POZZER{} constructs execution-flow information from \ac{PSC} traces. We use the ARM \ac{DWT} unit on the \textit{STM32F3} \ac{MCU} to obtain cycle-accurate instruction traces as ground truth.

We select eight test cases that progressively satisfy the nested if conditions by providing increasingly longer correct input prefixes. We then execute these test cases, capture the corresponding \ac{PSC} traces, and record the instruction traces using the \ac{DWT}.
\POZZER{} incrementally constructs the execution-flow graph. The first test case does not satisfy the first \texttt{if}. Starting with the second trace, which reaches and satisfies the first \texttt{if}, it identifies a discrepancy in the \ac{PSC} trace that corresponds to the branch point in the instruction trace. When a trace extending to the second \texttt{if} is observed, \POZZER{} matches the previously identified discrepancy and detects the new discrepancy corresponding to the second \texttt{if}. It repeats this process for progressively longer executions, incrementally constructing the execution-flow graph and identifying the discrepancy associated with each branch.
Figure~\ref{fig:casestudy} shows the resulting \ac{PSC} trace and the eight discrepancy points identified by \POZZER{}. The red dashed lines indicate the ground-truth branch locations obtained from the \ac{DWT} traces, while the blue stars indicate the discrepancy points detected by \POZZER{}. Their close correspondence illustrates how \POZZER{} recovers the execution-flow structure from \ac{PSC} traces and identifies newly observed execution behavior.

\begin{figure*}[!t]
    \centering
    \begin{subfigure}{0.23\textwidth}
        \centering
        \includegraphics[width=\linewidth]{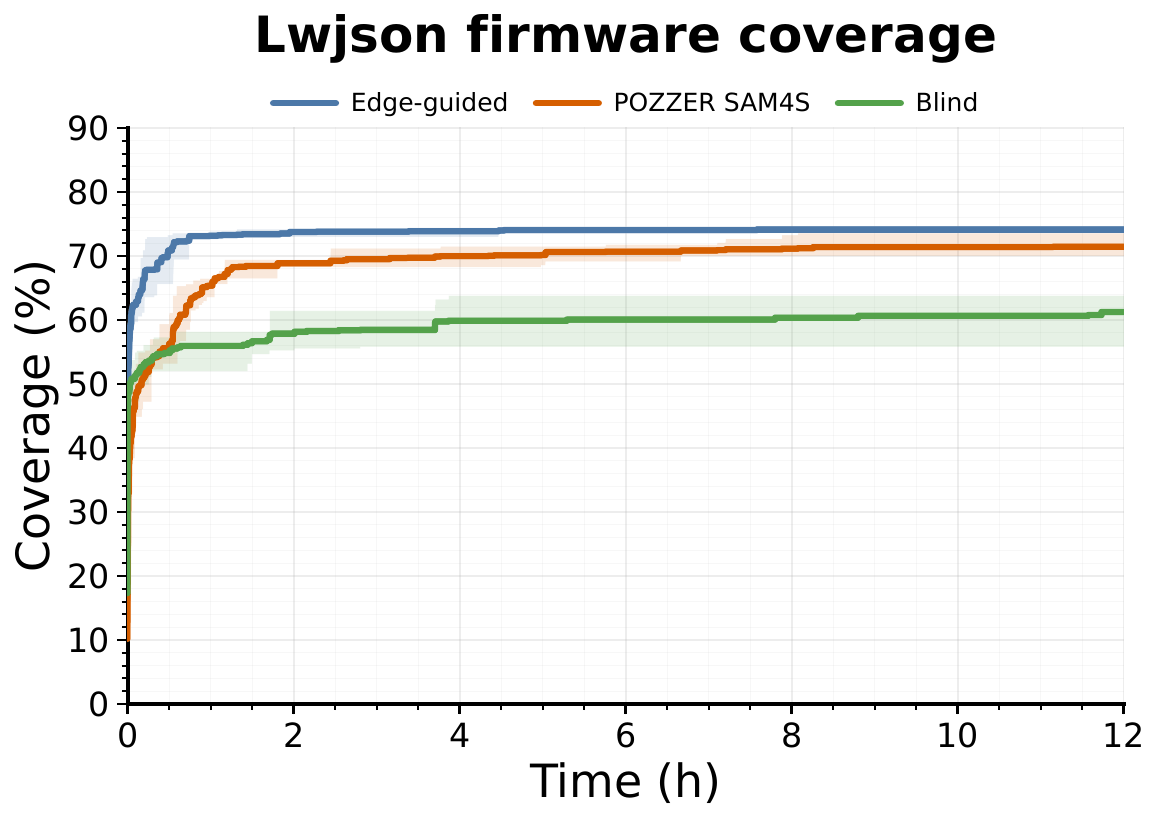}
    \end{subfigure}
    \begin{subfigure}{0.23\textwidth}
        \centering
        \includegraphics[width=\linewidth]{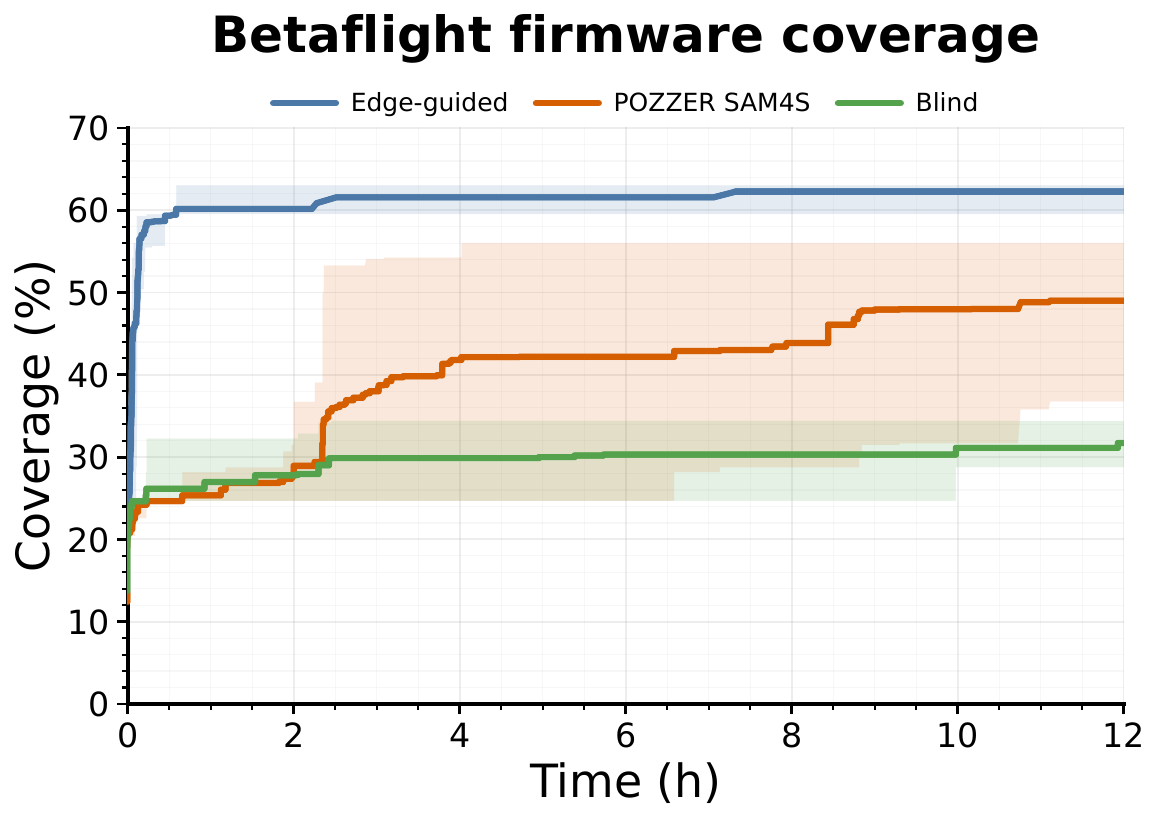}
    \end{subfigure}
    \begin{subfigure}{0.23\textwidth}
        \centering
        \includegraphics[width=\linewidth]{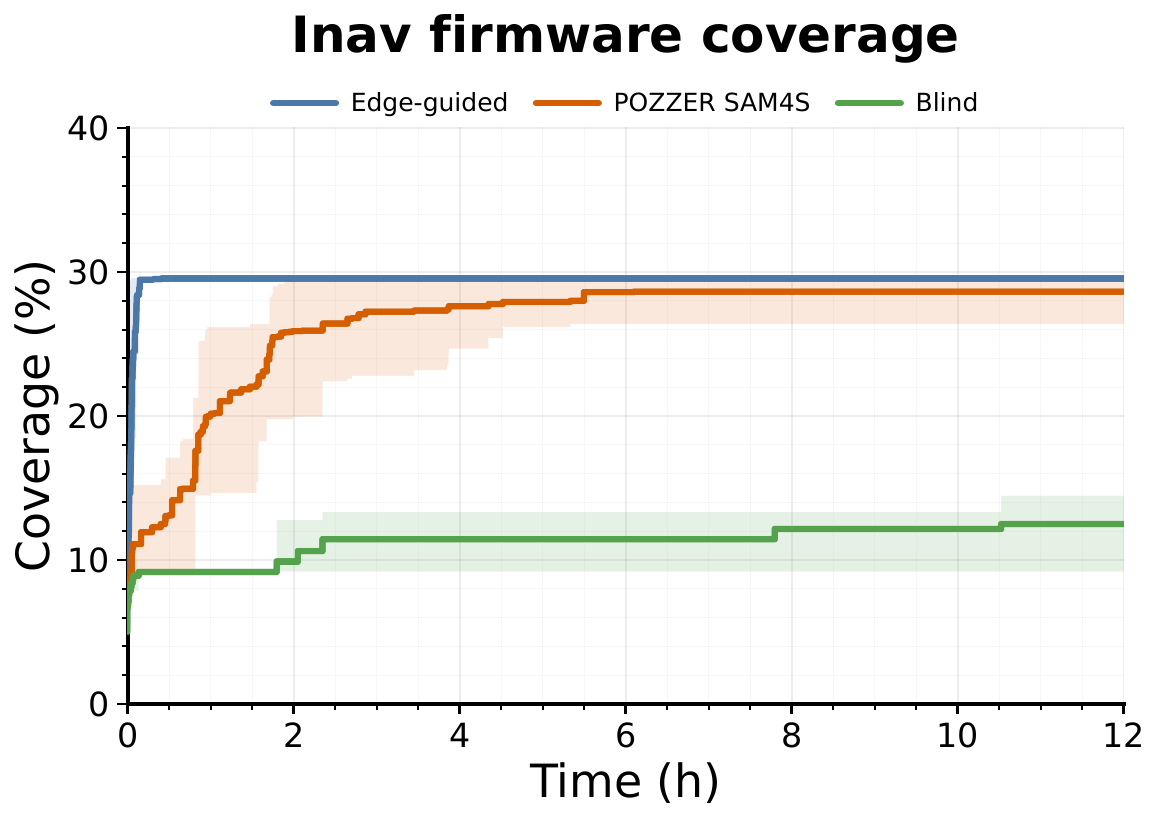}
    \end{subfigure}
    \begin{subfigure}{0.23\textwidth}
        \centering
        \includegraphics[width=\linewidth]{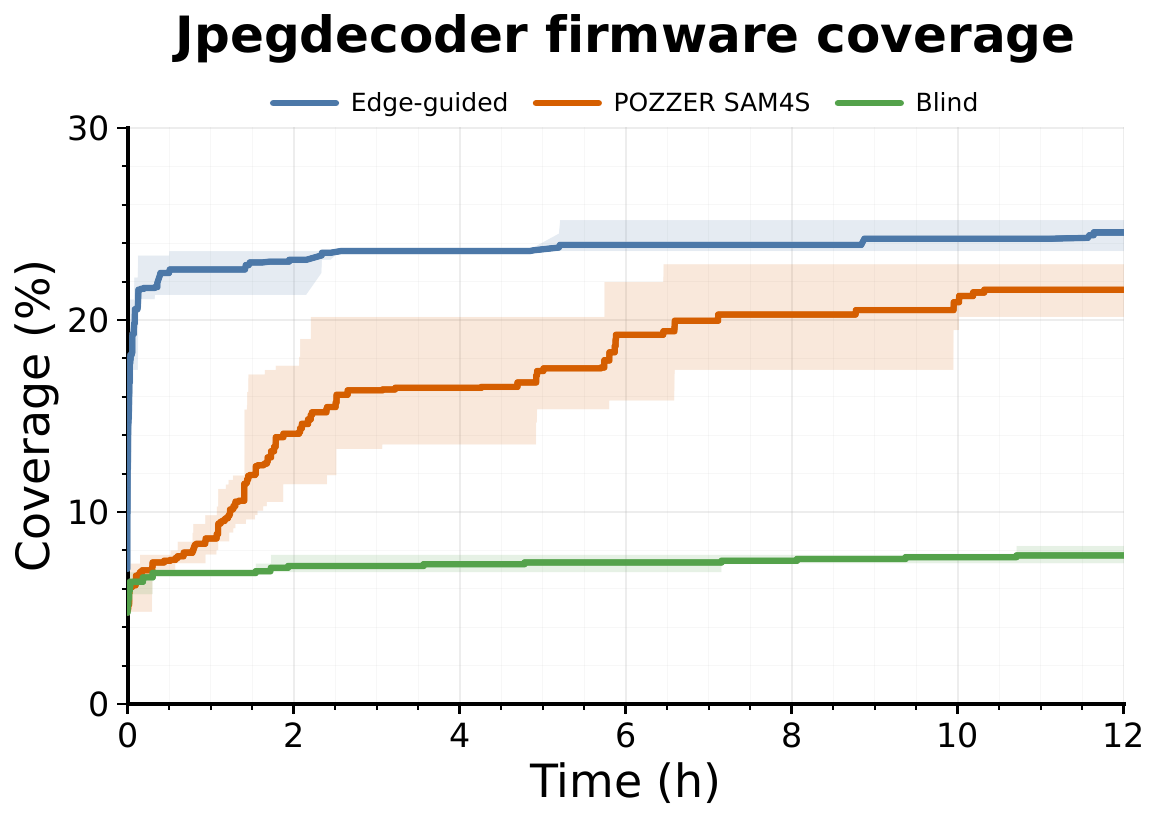}
    \end{subfigure}
    
    \begin{subfigure}{0.23\textwidth}
        \centering
        \includegraphics[width=\linewidth]{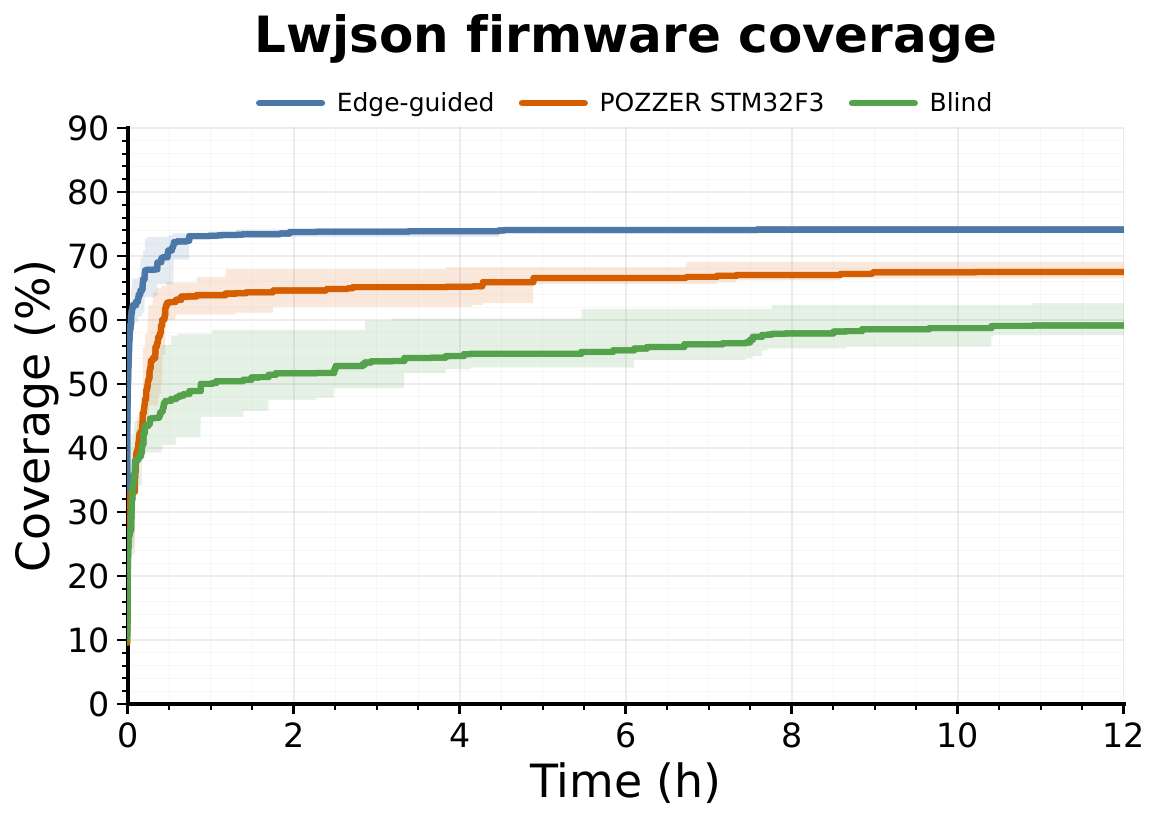}
    \end{subfigure}
    \begin{subfigure}{0.23\textwidth}
        \centering
        \includegraphics[width=\linewidth]{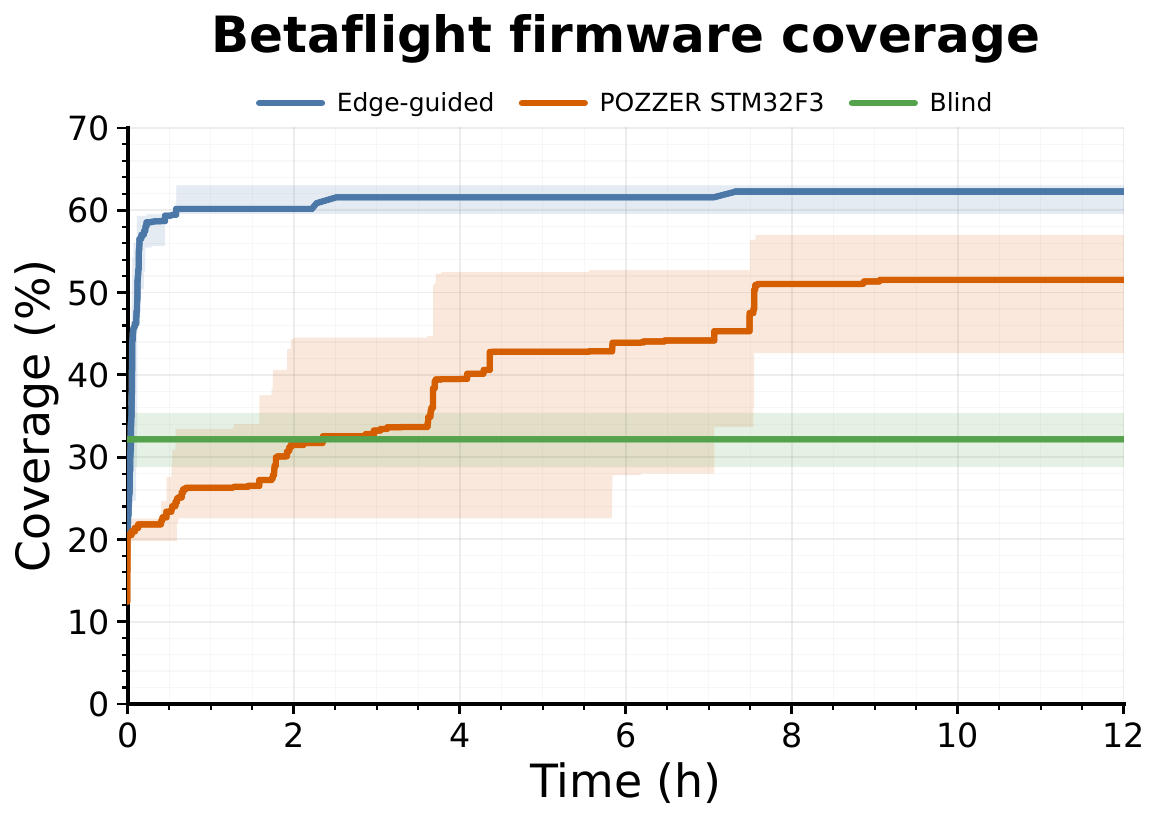}
    \end{subfigure}
    \begin{subfigure}{0.23\textwidth}
        \centering
        \includegraphics[width=\linewidth]{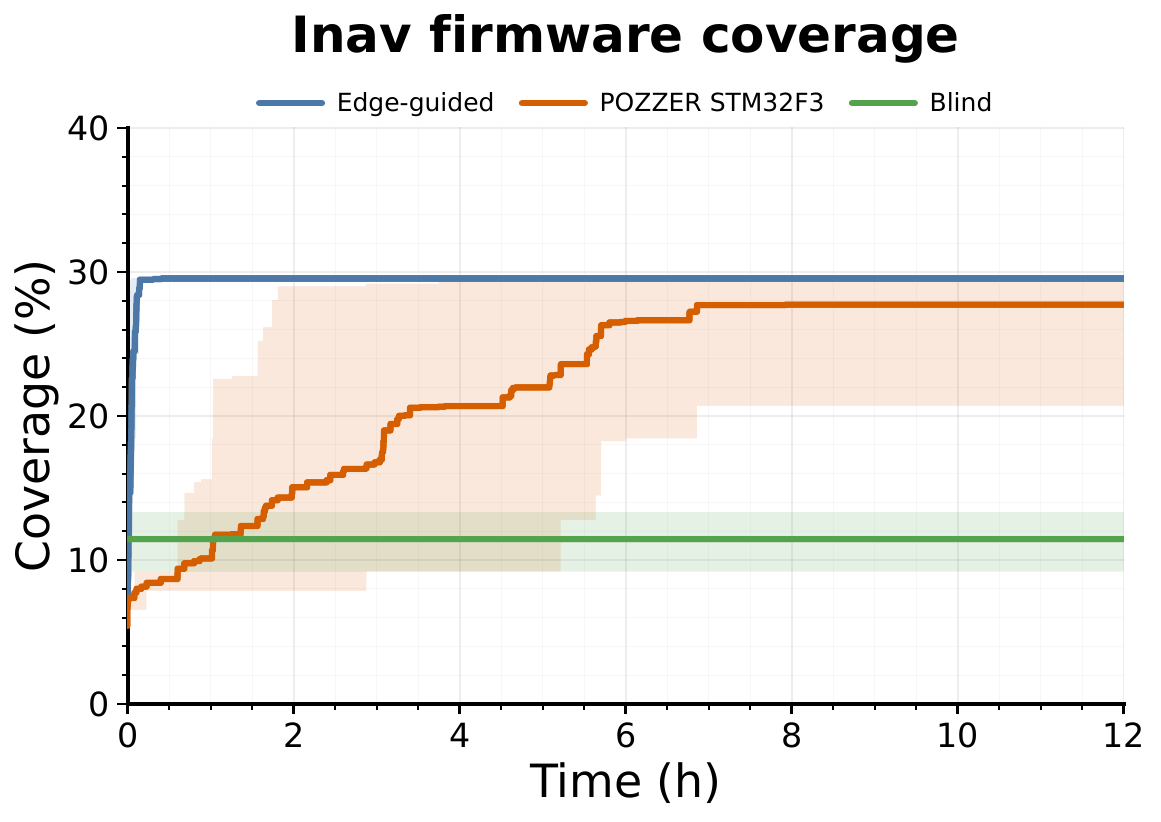}
    \end{subfigure}
    \begin{subfigure}{0.23\textwidth}
        \centering
        \includegraphics[width=\linewidth]{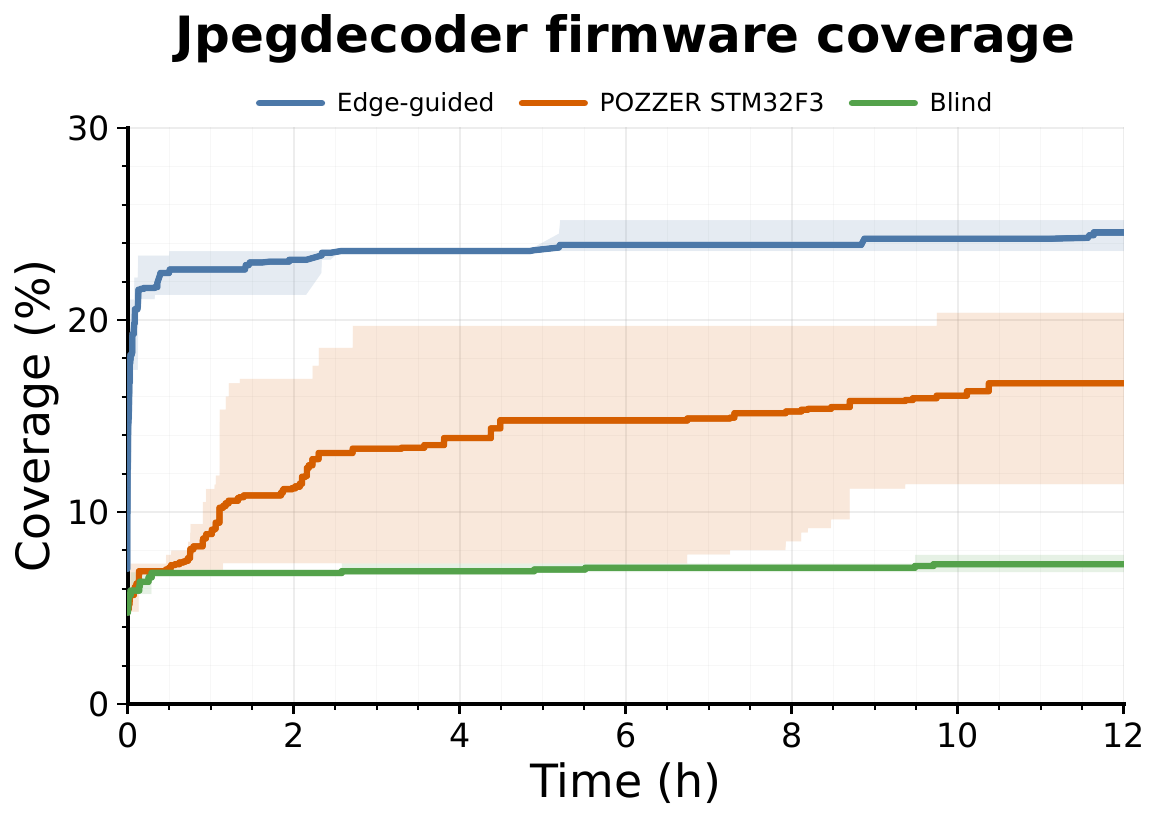}
    \end{subfigure}
    \caption{Coverage plots for evaluated targets. The first row shows targets running on \emph{SAM4S} and the second row shows targets running on \emph{STM32F3}.}
    \label{fig:coverageplots}
\end{figure*}

To illustrate the usefulness of this feedback, we run the blind fuzzer, \POZZER{}, and the greybox-guided fuzzer for five independent 12-hour campaigns on two target \acp{MCU}, \emph{STM32F3} and \emph{SAM4S}, and compare how accurately they recover the eight discrepancy points. The execution-flow graph reconstructed from the blind fuzzer's corpus contains, on average, three of the eight discrepancy points. 
\POZZER{} identifies all eight discrepancy points in two of the five campaigns on \emph{STM32F3} and three of the five campaigns on \emph{SAM4S}; in the remaining campaigns on both platforms, it misses one discrepancy point. The greybox-guided fuzzer identifies all eight discrepancy points in every campaign. 
While this simple example is not intended to support broad performance conclusions, the results illustrate that \POZZER{}'s feedback can surpass the blind baseline and approach the coverage achieved by an instrumented feedback-driven fuzzer.

\subsection{Results on \ac{CW}}\label{section:evaluation:cwresults}

\begin{table*}[!th]
    \centering
    \caption{Coverage (Cov.) achieved by \POZZER{}, the blind and the guided fuzzers on STM32F3 and SAM4S platforms. Also shown are the execution speed (executions per second) of the blind fuzzer and \POZZER{}, the corpus size (Corp. size) of \POZZER{}, and the Mann-Whitney $U$-test p-values and $A_{12}$ effect sizes comparing \POZZER{} and the blind-fuzzer corpora.}
        \resizebox{\textwidth}{!}{
        \begin{tabular}{ccc|ccc|cc|cc|ccc|cc|c}
        \toprule
        \multirow[c]{3}{*}[-1.4ex]{Target} & \multicolumn{7}{c|}{SAM4S} &
        \multicolumn{7}{c|}{STM32F3} & \multirow[c]{2}{*}[-0.6ex]{\shortstack[c]{Greybox\\Guided}} \\
        \cmidrule{2-8}
        \cmidrule{9-15}
        & \multicolumn{2}{c|}{Blind} & \multicolumn{3}{c|}{\POZZER{}} & \multicolumn{1}{c}{\multirow[c]{2}{*}[-0.6ex]{p-value}} & \multicolumn{1}{c|}{\multirow[c]{2}{*}[-0.6ex]{$A_{12}$}} & \multicolumn{2}{c|}{Blind} & \multicolumn{3}{c|}{\POZZER{}} & \multicolumn{1}{c}{\multirow[c]{2}{*}[-0.6ex]{p-value}} & \multicolumn{1}{c|}{\multirow[c]{2}{*}[-0.6ex]{$A_{12}$}} & \\
        \cmidrule{2-6}
        \cmidrule{9-13}
        \cmidrule{16-16}
        & Cov. & Speed & Cov. & Speed & Corp. size & & & Cov. & Speed & Cov. & Speed & Corp. size & & & Cov. \\
        \midrule
        lwjson~\cite{githubGitHubMaJerlelwjson} & 60.71\% & 70.20 & \textbf{71.39}\% & 31.98 & 2760.60 & \textbf{0.0119} & 1.0 & 59.12\% & 69.54 & \textbf{67.43}\% & 32.35 & 1660.20 & \textbf{0.0116} & 1.0 & 74.04\% \\
        \midrule
        cAT~\cite{githubGitHubMarcinbor85cAT} & 18.18\% & 69.41 & \textbf{20.98}\% & 23.02 & 270.20 & \textbf{0.0361} & 0.92& 17.95\% & 65.69 & \textbf{27.30}\% & 22.10 & 1150.60 & \textbf{0.0079} & 1.0 & 43.48\% \\
        \midrule
        minmea~\cite{githubGitHubKosmaminmea} & 28.22\% & 70.83 & \textbf{62.68}\% & 32.19 & 3906.00 & \textbf{0.0094} & 1.0 & 29.30\% & 67.48 & \textbf{58.22}\% & 31.57 & 1313.80 & \textbf{0.0111} & 1.0 & 71.46\%  \\
        \midrule
        lwgps~\cite{githubGitHubMaJerlelwgps} & 24.91\% & 70.00 & \textbf{47.98}\% & 30.82 & 507.20 & \textbf{0.0211} & 0.96 & 23.33\% & 67.56 & \textbf{30.00}\% & 28.45 & 152.80 & 0.2358 & 0.74 & 84.74\% \\
        \midrule
        jsonparser~\cite{githubGitHubRafagafetinyjson} & 79.72\% & 69.82 & \textbf{90.09}\% & 31.37 & 3265.00 & \textbf{0.0106} & 1.0 & 78.44\% & 65.94 & \textbf{87.16}\% & 31.23 & 1369.20 & \textbf{0.0116} & 1.0 & 91.74\% \\
        \midrule
        nanomodbus~\cite{githubGitHubDebevvnanoMODBUS} & 25.43\% & 68.35 & \textbf{26.11}\% & 28.40 & 2584.00 & 0.2873 & 0.72 & \textbf{26.08}\% & 70.89 & 24.85\% & 26.10 & 3644.60 & \textbf{0.0231} & 0.06 & 30.95\% \\
        \midrule
        regex~\cite{githubGitHubKokketinyregexc} & 84.04\% & 69.03 & \textbf{87.17}\% & 25.00 & 4412.60 & \textbf{0.0116} & 1.0 & \textbf{83.84}\% & 66.13 & 81.82\% & 23.88 & 2104.40 & 0.1732 & 0.22 & 87.88\% \\
        \midrule
        libelf~\cite{githubGitHub0introlibelf} & 7.19\% & 66.86 & \textbf{33.35}\% & 30.68 & 1771.00 & \textbf{0.0085} & 1.0 & \textbf{7.07}\% & 69.19 & 6.83\% & 29.21 & 1624.80 & 0.2330 & 0.32 & 35.29\% \\
        \midrule
        jpegdecoder~\cite{githubGitHubCmumfordTJpgDec} & 7.78\% & 64.11 & \textbf{21.56}\% & 26.86 & 3306.80 & \textbf{0.0109} & 1.0 & 7.32\% & 67.66 & \textbf{16.70}\% & 24.05 & 2614.80 & \textbf{0.0116} & 1.0 & 24.53\% \\
        \midrule
        microshell~\cite{githubGitHubNrushernr_micro_shell} & 82.88\% & 65.82 & \textbf{88.00}\% & 24.94 & 4046.80 & 0.0593 & 0.88 & \textbf{84.24}\% & 66.60 & 77.76\% & 24.42 & 2247.20 & \textbf{0.0119} & 0.0 & 95.60\% \\
        \midrule
        betaflight~\cite{githubGitHubBetaflightbetaflight} & 31.69\% & 69.90 & \textbf{48.97}\% & 24.65 & 2480.40 & \textbf{0.0119} & 1.0 & 32.19\% & 69.18 & \textbf{51.50}\% & 24.14 & 1238.40 & \textbf{0.0119} & 1.0 & 62.21\% \\
        \midrule
        inav~\cite{githubGitHubINavFlightinav} & 12.52\% & 67.49 & \textbf{28.61}\% & 27.87 & 2112.60 & \textbf{0.0116} & 1.0 & 11.47\% & 67.45 & \textbf{27.71}\% & 23.24 & 2297.40 & \textbf{0.0114} & 1.0 & 29.51\% \\
        \midrule
        drone~\cite{githubGitHubRiS3Labp2imreal_firmware} & 9.14\% & 65.79 & \textbf{56.12}\% & 24.91 & 2383.00 & \textbf{0.0101} & 1.0 & 8.79\% & 71.10 & \textbf{35.69}\% & 24.40 & 709.40 & \textbf{0.0109} & 1.0 & 60.34\% \\
        \midrule
        cnc~\cite{githubGitHubRiS3Labp2imreal_firmware} & 50.76\% & 68.95 & \textbf{60.48}\% & 25.43 & 4359.40 & \textbf{0.0079} & 1.0 & 50.97\% & 70.12 & \textbf{54.82}\% & 26.57 & 2378.00 & \textbf{0.0159} & 0.98 & 79.40\% \\
        \midrule
        stepper~\cite{githubGitHubRiS3LabDICEDMAEmulation} & 36.38\% & 67.75 & \textbf{43.28}\% & 15.95 & 2353.00 & \textbf{0.0106} & 1.0 & 36.38\% & 68.17 & \textbf{43.20}\% & 13.17 & 1466.20 & \textbf{0.0109} & 1.0 & 57.10\% \\
        \bottomrule
        \end{tabular}
        }
    \label{table:cwresults}
\end{table*}

To validate that \POZZER{} performs well on real targets, we conduct an extensive evaluation of \POZZER{} on \FIRMWARENUMBER{} firmware targets across \MCUNUMBER{} hardware platforms. Table~\ref{table:cwresults} reports the average coverage achieved by each fuzzer over five campaigns for each target and \ac{MCU}. Coverage is computed only over the target source-code files and normalized by the number of lines considered instrumentable by \texttt{gcov}, rather than by the total number of lines across all firmware files. The \emph{Cov.} columns under \emph{Blind} report the average coverage achieved by the blind fuzzer on each target and platform, while the \emph{Cov.} columns under \emph{\POZZER{}} report the corresponding coverage achieved by \POZZER{}. Since the greybox-guided fuzzer runs on an x86 machine, we report a common reference coverage with instrumentation-based feedback for both platforms in the \emph{Greybox guided} column. Overall, \POZZER{} outperforms the blind fuzzer on 26 out of 30 target-platform combinations. Figure~\ref{fig:coverageplots} presents representative coverage plots; the remaining plots are provided in Appendix~\ref{section:appendix:plots} due to space constraints.

\begin{table*}[!th]
    \centering
    \vspace{0mm}
    \caption{Comparison of edge coverage and fuzzing speed (executions per second) between FuzzEMup and \POZZER{}.}
    \scriptsize
        \begin{tabular}{ccc|cc|cc|cc}
        \toprule
        \multirow{2}{*}{Fuzzer} & \multicolumn{2}{c}{GPS} & \multicolumn{2}{c}{Soldering} & \multicolumn{2}{c}{CNC} & \multicolumn{2}{c}{Stepper} \\
        \cmidrule{2-9}
        & Edge Cov. & Speed & Edge Cov. & Speed & Edge Cov. & Speed & Edge Cov. & Speed \\
        \midrule
        FuzzEMup~\cite{moradihaghighi2026fuzz} & 909 & 0.04 & 1086 & 0.36 & 1164 & 0.03 & 1341 & 0.09 \\
        \midrule
        \POZZER{} (FuzzEMup duration) &
        \textbf{1380} & \textbf{23.96} &
        1076 & \textbf{4.40} &
        \textbf{1175} & \textbf{23.44} &
        \textbf{1346} & \textbf{15.16} \\
        \midrule
        POZZER{} (12-hours) &
        \textbf{1509} & \textbf{23.26} &
        \textbf{1110} & \textbf{4.20} &
        \textbf{1176} & \textbf{23.00} &
        \textbf{1431} & \textbf{13.90} \\
        \bottomrule
        \end{tabular}
    \label{table:comparison}
\end{table*}

\textbf{Statistical Analysis.}
To further assess \POZZER{}'s performance, we statistically compare its coverage with that of the blind fuzzer across repeated campaigns. Since fuzzing is inherently randomized, a single campaign is insufficient to characterize a fuzzer's performance. Unlike prior work~\cite{moradihaghighi2026fuzz}, which reports results from a single campaign, we repeat each experiment five times to account for the effect of randomness. Following the statistical evaluation methodology recommended by Schloegel et. al.~\cite{moesok}, we use the Mann-Whitney \emph{U}-test~\cite{sachs2012applied} to assess statistical significance and the Vargha-Delaney $A_{12}$ effect size~\cite{vargha2000critique} to quantify the magnitude of the observed differences between \POZZER{} and the blind fuzzer. We use a significance level of \textit{0.05}. The $A_{12}$ effect size represents the probability that a randomly selected run of \POZZER{} achieves higher coverage than a randomly selected run of the blind fuzzer. An $A_{12}$ value of \textit{0.5} indicates no difference between the two fuzzers, values greater than \textit{0.5} favor \POZZER{}, and values less than \textit{0.5} favor the blind fuzzer.

Table~\ref{table:cwresults} reports both the p-values obtained from the Mann-Whitney \emph{U}-test and the corresponding $A_{12}$ effect sizes for each target-platform combination. Reporting both metrics is important because statistical significance alone does not indicate the magnitude of the observed improvement. As shown in the table, 25 out of 30 target-platform combinations achieve a p-value below \textit{0.05}, indicating that the difference in coverage between \POZZER{} and the blind fuzzer is statistically significant. Furthermore, 26 out of 30 targets achieve an $A_{12}$ value larger than \textit{0.5}, indicating that a randomly selected run of \POZZER{} is more likely to achieve higher coverage than a randomly selected run of the blind fuzzer.

\begin{tcolorbox}[
    colback=gray!10,
    colframe=black,
    boxrule=0.5pt,
    left=6pt,
    right=6pt,
    top=6pt,
    bottom=6pt
]
\textbf{R1:} \POZZER{} outperforms the blind fuzzer under the same time budget on 26 of the 30 target-platform combinations, with the improvement being statistically significant on 25 of them.
\end{tcolorbox}

\textbf{Speed.}
A guided fuzzer generally incurs overhead compared with a blind fuzzer because it must process feedback and determine whether each test case is interesting. For a \ac{PSC}-guided fuzzer, this overhead is particularly significant because \ac{PSC} traces are large time-series signals whose processing can substantially reduce fuzzing throughput. To limit this cost, \POZZER{} uses the fast \emph{L1} distance for discrepancy detection. Its graph construction and trimming stage further removes redundancy from the constructed graph, thereby controlling the growth of the corpus and graph and reducing the cost of subsequent feedback decisions. While previous work~\cite{moradihaghighi2026fuzz} does not report speed or corpus size, the \emph{speed} column in Table~\ref{table:cwresults} reports the average fuzzing throughput and corpus size for each target over the five 12-hour campaigns. The \emph{Corp. size} column also reports the corpus size as a percentage of the total number of executions performed by \POZZER{}. The results show that \POZZER{}'s throughput varies across targets, reflecting differences in target execution times. Nevertheless, the resulting corpus remains reasonably small in all campaigns.

\begin{tcolorbox}[
    colback=gray!10,
    colframe=black,
    boxrule=0.5pt,
    left=6pt,
    right=6pt,
    top=6pt,
    bottom=6pt
]
\textbf{R2:} Despite the overhead introduced by its feedback mechanism, \POZZER{} achieves an average throughput of 26.30 executions per second and an average corpus size of 2216.35 test cases.
\end{tcolorbox}

\subsection{Comparison with related work}
Several works have explored side channel-guided fuzzing for embedded systems~\cite{vincentiduskfuzz,moradihaghighi2026fuzz}. However, they either provide insufficient information about performance or contain inconsistencies in their threat models or evaluations. According to established fuzzing practices~\cite{moesok,klees2018evaluating,bohme2020fuzzing}, a fair comparison of fuzzers should satisfy the following criteria:

\begin{enumerate}[nosep]
    \item
    Allocating an equal time budget to all fuzzers
    \item
    Running the fuzzers for sufficiently long periods, ranging from several hours to a full day
    \item
    Repeating fuzzing campaigns to account for randomness
\end{enumerate}

However, the related works do not satisfy these criteria. First, they evaluate fuzzers based on an equal input budget. Because capturing and processing \ac{PSC} traces introduces additional overhead, this comparison ignores the difference in throughput between blind and \ac{PSC}-guided fuzzers. Second, the comparisons use a limited number of test cases, typically around 1,000 inputs. However, fuzzing is inherently randomized and requires longer execution times to explore the program space effectively. Finally, the existing studies report results from only a single campaign, whereas a standard evaluation requires repeated campaigns to account for randomness~\cite{moesok}.

In our evaluation, we address all of these limitations. We compare \POZZER{} with FuzzEMup~\cite{moradihaghighi2026fuzz}.
To ensure a fair comparison with FuzzEMup, we contacted its authors and obtained their artifacts, including the seeds, corpus for each target, coverage collection script, and target source code. We then ported the targets to our \ac{CW} setup and fuzzed them with \POZZER{} using the same seeds. Finally, we replayed the corpus generated by \POZZER{} using FuzzEMup's coverage collection script and collected the corresponding edge coverage. Table~\ref{table:comparison} reports the execution speed and edge coverage of both fuzzers for each target. Because FuzzEMup's execution speed is not reported directly in its paper, we obtained this information from its authors. In addition, FuzzEMup does not conduct standard 12-hour campaigns; hence, comparing its results directly with those from a 12-hour \POZZER{} campaign would be unfair. Therefore, we report the results in two different rows. The \emph{\POZZER{} (FuzzEMup duration)} row reports the results when \POZZER{} is allocated the same time budget as FuzzEMup, whereas the \emph{\POZZER{} (12-hour)} row reports the results from a full 12-hour campaign.

As shown in Table~\ref{table:comparison}, under an equal time budget, \POZZER{} achieves a higher execution rate and covers more edges than FuzzEMup on three out of four targets. With a standard 12-hour campaign, \POZZER{} outperforms FuzzEMup on all four targets. These experiments also highlight the importance of high-quality seeds. By measuring the unique edges contributed by each test case, we observe that most of the unique edge coverage is achieved by the initial high-quality seeds provided by FuzzEMup. However, such high-quality seeds might not be available in real-world black-box scenarios.

\subsection{Real-world case study}

After evaluating \POZZER{} extensively on the \ac{CW} setup, we evaluate it on two real-world black-box embedded systems for which neither the firmware source code nor the binary image is publicly available. We fuzz a \emph{U-Blox ZED-F9P} \ac{GNSS} receiver module and an Anonymous Target. On the \ac{GNSS} receiver, \POZZER{} achieves a throughput of 7.19 executions per second despite the substantially larger \ac{PSC} traces compared with the \ac{CW} setup. After three 4-hour fuzzing campaigns, \POZZER{} discovers two distinct crashes on the ZED-F9P. Both crashes were confirmed by the vendor, and one CVE\footnote{CVE id redacted for double blind review.} assigned so far. Due to the sensitive nature of the Anonymous Target and the restrictions imposed by our \ac{NDA}, we can not disclose the results for this target.

\textbf{Bug description.}
The ZED-F9P module communicates through u-blox's proprietary \emph{UBX} protocol, a modular binary protocol with a low-overhead checksum. UBX defines more than 100 message types organized into message classes for configuring the module and querying navigation information. We use the UBX interface as the input channel for fuzzing. 
According to the UBX protocol specification, floating-point fields are encoded according to the IEEE 754 standard. Our analysis shows that the module crashes when one of these fields contains an IEEE 754 encoding of \emph{NaN}, a non-finite value that the firmware does not handle correctly. When the module receives a crafted input containing such a value, it crashes and reboots, indicating insufficient validation or handling of special floating-point values.

\begin{tcolorbox}[
    colback=gray!10,
    colframe=black,
    boxrule=0.5pt,
    left=6pt,
    right=6pt,
    top=6pt,
    bottom=6pt
]
\textbf{R3:} \POZZER{} operates independently of the \ac{CW} platform, can fuzz real-world black-box embedded systems, and can discover vulnerabilities in such targets.
\end{tcolorbox}

\section{Discussion}\label{section:discussion}

In this work, we present \POZZER{}, the first \ac{PSC}-guided black-box fuzzer to outperform a blind fuzzer under the same time budget. Our evaluation highlights three aspects that warrant further discussion: seed selection, performance variation of \POZZER{} across hardware platforms, and scalability challenges arising from long executions and large graphs.

\textbf{Seed selection.}
Seed selection can strongly influence fuzzing performance, particularly for targets with structured input formats. In Section~\ref{section:evaluation:cwresults}, we initialize all campaigns with a random seed, except for \emph{lwgps} and \emph{minmea}. These targets require inputs that conform to a strict grammar and begin with specific magic values. \POZZER{}, the blind fuzzer, and the traditional coverage-guided fuzzer generate inputs based on existing seeds, so both benefit equally from seeding with magic values.
For the remaining targets, \POZZER{} starts from random seeds yet outperforms the blind fuzzer, suggesting that its feedback can guide input generation toward valid input formats and previously unexplored behavior.

This result contrasts with the evaluation of FuzzEMup~\cite{moradihaghighi2026fuzz}, which uses meaningful seeds for every target. As reported in Table~\ref{table:comparison}, when we fuzz the FuzzEMup targets with \POZZER{} using the same seeds, \POZZER{} achieves higher edge coverage than FuzzEMup. Table~\ref{table:comparison} also shows that \POZZER{} achieves higher execution throughput. This higher throughput provides a plausible explanation for its improved coverage: by executing more test cases within the same time budget, \POZZER{} has more opportunities to explore the input space and discover inputs that exercise additional behavior. These results highlight the importance of feedback-processing efficiency and execution speed in \ac{PSC}-guided fuzzing.

\textbf{Performance variation.}
A second discussion point is the performance variation of \POZZER{} across hardware platforms. Table~\ref{table:cwresults} shows that \POZZER{}'s performance varies across platforms. In particular, it achieves lower average coverage on \emph{STM32F3} compared to \emph{SAM4S}. To investigate this variation, we compare the compiled binary images for both platforms for the same targets. Although the binary images implement semantically similar firmware behavior, they differ in their instruction sequences and execution characteristics because the target \acp{MCU} uses different vendor-specific toolchains and implementations. These differences can affect the power characteristics of the observed executions, thereby influencing the quality of the \ac{PSC}-based feedback and \POZZER{}'s resulting performance. Identifying the precise cause of this platform-dependent behavior would therefore require detailed knowledge of the target \acp{MCU}'s hardware implementations and power characteristics.

\textbf{Scalability.}
A third issue concerns scalability with execution length and graph size. Longer executions produce larger \ac{PSC} traces and graphs, reducing \POZZER{}'s throughput. For the Anonymous Target, the execution length is 20~ms. At a sampling frequency of 25~MS/s, each test case therefore produces approximately 500,000 samples. Processing such a large \ac{PSC} trace for every test case and constructing a graph from it further reduces \POZZER{}'s throughput. Moreover, when an oscilloscope is used, the trigger signal must also be transmitted to and processed by \POZZER{}'s observer, introducing additional time overhead. 
These limitations motivate future work on implementing \POZZER{}'s processing pipeline on a high-performance \ac{FPGA}. Such an implementation could capture and process each \ac{PSC} trace, compare it with the execution-flow graph stored on the \ac{FPGA}, and notify the host-side fuzzer when a trace exhibits previously unseen behavior. The host would then add the corresponding test case to the corpus. This design would eliminate the need to transfer every \ac{PSC} trace to the host and allow the \ac{FPGA} to maintain the graph and perform the feedback computation locally.

\section{Related work}\label{section:relatedworks}

This section reviews the work most closely related to \POZZER{}. We first discuss existing embedded firmware fuzzing techniques under different threat models. We then review black-box fuzzing approaches that rely on software-generated feedback before discussing prior work on side-channel-guided fuzzing.

Embedded firmware fuzzers differ primarily in the assumptions they make about the target system and the execution feedback available during fuzzing. When firmware source code is available, instrumentation-based approaches execute the instrumented firmware directly on the target device to obtain code coverage~\cite{shift}. Alternatively, hardware debugging interfaces can be used to extract execution feedback without modifying the firmware~\cite{eisele2023fuzzing}. Firmware rehosting provides another widely adopted solution by executing the firmware inside an emulator. Unlike instrumentation-based methods, rehosting does not require source code. For example, Fuzzware~\cite{scharnowski2022fuzzware} performs coverage-guided fuzzing directly on firmware binaries through emulation. Although these approaches differ in their assumptions, they all require some level of access to the firmware, either through source code, firmware binaries, or hardware debugging interfaces.

To relax these assumptions, recent work has explored fuzzing techniques that operate under black-box settings. Rather than relying on code coverage, these approaches leverage software-generated information exposed by the target, such as grammar specifications~\cite{zamudio2025fandango}, protocol descriptions, system logs~\cite{aafer2021android}, or protocol snippets~\cite{feng2021snipuzz}, to improve input generation. While effective in their respective scenarios, they still assume that execution-related information is available through the software interface. Consequently, they cannot exploit undocumented commands, hidden execution paths, or implementation-specific behaviors that are not reflected in the exposed interface.

Several studies have investigated physical side channels as a source of fuzzing feedback. Sperl et al.~\cite{sperl2019side} first proposed employing \ac{PSC} together with a supervised machine-learning classifier to identify firmware branches. However, their approach relies on a profiling stage and assumes a white-box threat model. McClintick et al.~\cite{mcclintick2024side} proposed an \ac{EM} side-channel-guided fuzzer based on trace clustering and, to the best of our knowledge, presented the first timing-based comparison against blind fuzzing. Nevertheless, their evaluation is limited to a manually constructed parser and requires high sampling resolutions. DuskFuzz~\cite{vincentiduskfuzz} integrates \ac{EM} side-channel feedback into LibAFL~\cite{libafl} by exploiting a microarchitectural leakage originating from the instruction prefetch buffer. It uses this leakage to infer execution-flow discrepancies and guide the fuzzing process. However, the approach requires five executions for each test case, significantly reducing fuzzing throughput, and does not provide a timing-based comparison with blind fuzzing. Fuzz'EMup~\cite{moradihaghighi2026fuzz} applies a frequency-domain feature extractor on \ac{EM} side-channels and implements a divergence-based feedback mechanism for an \ac{EM}-guided fuzzer. They evaluate their work on four firmware benchmarks and achieve better performance than a blind fuzzer, but they do not provide information about their fuzzer's speed. Furthermore, they do not give the same time budget to the fuzzers and stop the fuzzers after at most 1000 executions. This makes their evaluation vague. Other side-channel-guided fuzzers have also been proposed~\cite{barredo2025gaflerna,su2024fuzz}; however, they either assume stronger attacker models or do not demonstrate practical improvements over blind fuzzing under comparable time budgets.

Unlike previous work, \POZZER{} is designed specifically for a strict black-box threat model. It requires neither firmware instrumentation, firmware binaries, nor profiling data, extracts meaningful feedback from a single power trace, and, to the best of our knowledge, is the first \ac{PSC}-guided fuzzer to consistently outperform blind fuzzing under the same time budget.

\section{Conclusion}\label{section:conclusion}

In this paper, we presented \POZZER{}, a \ac{PSC}-guided fuzzer for black-box embedded systems, and demonstrated that it achieves superior fuzzing performance compared with a baseline blind fuzzer under the same time budget. \POZZER{} uses \ac{PSC} traces to construct a pseudo execution-flow graph of the target firmware and integrates hardware and software-level noise-handling techniques into its feedback mechanism. We extensively evaluated \POZZER{} on the \ac{CW} platform using \MCUNUMBER{} different hardware platforms and \FIRMWARENUMBER{} firmware targets. Our results show that, across more than 3,600 hours of fuzzing, \POZZER{} achieves, on average, 12.06\% better performance than the baseline blind fuzzer. We further demonstrate that \POZZER{} is not limited to the \ac{CW} platform by connecting it to an oscilloscope and evaluating it on \REALTARGETS{} real-world black-box embedded systems. On one of these targets, \POZZER{} identified two crashes, which were both confirmed by the vendor.

\clearpage




\appendix

\section*{Ethical Considerations}

We organize our ethical analysis around the stakeholders affected by the research, the potential benefits and harms to each, the measures taken to reduce those harms, and our decision to conduct and publish the work.

\textbf{Stakeholders and benefits.}
The principal stakeholders are users and operators of the evaluated software, project maintainers and contributors, the security-research community, and the research team. 
Users benefit when previously unknown defects are identified and corrected before independent discovery or exploitation. 
Maintainers receive limited input and technical evidence to support diagnosis and remediation, although processing reports requires engineering effort and may create reputational concerns. 
The research community benefits from a reproducible method for testing black-box embedded devices. 
At the same time, the resulting system is dual-use: it could help an adversary discover defects in software for which the source code or binary image is unavailable. We believe that the security improvements in hard-to-test devices resulting from publishing our tool outweigh the risks of potential adversarial use. We choose not to publish the harnesses for our real-world case studies.

\textbf{Research conduct.}
Our experiments were performed in a controlled lab environment. For the Chipwhisperer targets, we used publicly available projects and compiled them for the corresponding target platform. For the real-world case studies, we had two targets. The experiments for both targets were conducted in a controlled environment in our lab.
For the anonymous target, we have shared the draft of our paper with the corresponding vendor, and they have confirmed that the current version is okay for publication.

\textbf{Vulnerability disclosure.}
For the Chipwhisperer experiment, we did not find any unknown vulnerability. For the real-world case study, we have found two vulnerabilities on the UBlox \ac{GNSS} receiver, and we disclosed the findings to the vendor. They confirmed their reproducibility, and we have assigned one CVE so far.
Reports included a minimal reproducer, affected versions, observed impact, and available root cause information, as described in the paper. 
We coordinated subsequent disclosure with the vendor. 
For the Anonymous Target, reports remain private. Therefore, we omit project identities and other information that would materially facilitate exploitation. 

\section*{Open Science}

We will release our implementation's source code and the results of the paper after its acceptance.

\clearpage



\bibliographystyle{plainurl}
\bibliography{ref}

@article{vincentiduskfuzz,
  title={DuskFuzz: Encoding Side-Channel Information to Improve Blackbox Fuzzing},
  author={Vincenti, Ulysse and Hiscock, Thomas and Hely, David}
}

@inproceedings{narimani2021side,
  title={Side-channel based disassembler for AVR micro-controllers using convolutional neural networks},
  author={Narimani, Pouya and Akhaee, Mohammad Ali and Habibi, Seyed Amin},
  booktitle={2021 18th International ISC Conference on Information Security and Cryptology (ISCISC)},
  pages={75--80},
  year={2021},
  organization={IEEE}
}

@inproceedings{ling2025fusiondisassembler,
  title={FusionDisassembler: A Cross-Device Approach for Effective Instruction Disassembly in Side-Channel Attacks},
  author={Ling, Chen and Zhang, Jinyuan and Hai, Ouchang and Liu, Hangcheng and Han, Xingshuo},
  booktitle={2025 IEEE 31th International Conference on Parallel and Distributed Systems (ICPADS)},
  pages={1--8},
  year={2025},
  organization={IEEE}
}

@inproceedings{libafl,
 author       = {Andrea Fioraldi and Dominik Maier and Dongjia Zhang and Davide Balzarotti},
 title        = {{LibAFL: A Framework to Build Modular and Reusable Fuzzers}},
 booktitle    = {Proceedings of the 29th ACM conference on Computer and communications security (CCS)},
 series       = {CCS '22},
 year         = {2022},
 month        = {November},
 location     = {Los Angeles, U.S.A.},
 publisher    = {ACM},
}

@InProceedings{chipwhisperer,
author="O'Flynn, Colin
and Chen, Zhizhang (David)",
editor="Prouff, Emmanuel",
title="ChipWhisperer: An Open-Source Platform for Hardware Embedded Security Research",
booktitle="Constructive Side-Channel Analysis and Secure Design",
year="2014",
publisher="Springer International Publishing",
address="Cham",
pages="243--260",
}

@inproceedings {shift,
author = {Alejandro Mera and Changming Liu and Ruimin Sun and Engin Kirda and Long Lu},
title = {{SHiFT}: Semi-hosted Fuzz Testing for Embedded Applications},
booktitle = {33rd USENIX Security Symposium (USENIX Security 24)},
year = {2024},
isbn = {978-1-939133-44-1},
address = {Philadelphia, PA},
pages = {5323--5340},
url = {https://www.usenix.org/conference/usenixsecurity24/presentation/mera},
publisher = {USENIX Association},
month = aug
}

@inproceedings{moesok,
  title={Sok: Prudent evaluation practices for fuzzing},
  author={Schloegel, Moritz and Bars, Nils and Schiller, Nico and Bernhard, Lukas and Scharnowski, Tobias and Crump, Addison and Ale-Ebrahim, Arash and Bissantz, Nicolai and Muench, Marius and Holz, Thorsten},
  booktitle={2024 IEEE Symposium on Security and Privacy (SP)},
  pages={1974--1993},
  year={2024},
  organization={IEEE}
}

@inproceedings{kocher1999differential,
  title={Differential power analysis},
  author={Kocher, Paul and Jaffe, Joshua and Jun, Benjamin},
  booktitle={Annual international cryptology conference},
  pages={388--397},
  year={1999},
  organization={Springer}
}

@inproceedings{scharnowski2022fuzzware,
  title={Fuzzware: Using precise $\{$MMIO$\}$ modeling for effective firmware fuzzing},
  author={Scharnowski, Tobias and Bars, Nils and Schloegel, Moritz and Gustafson, Eric and Muench, Marius and Vigna, Giovanni and Kruegel, Christopher and Holz, Thorsten and Abbasi, Ali},
  booktitle={31st USENIX Security Symposium (USENIX Security 22)},
  pages={1239--1256},
  year={2022}
}

@inproceedings{brier2004correlation,
  title={Correlation power analysis with a leakage model},
  author={Brier, Eric and Clavier, Christophe and Olivier, Francis},
  booktitle={Cryptographic Hardware and Embedded Systems-CHES 2004: 6th International Workshop Cambridge, MA, USA, August 11-13, 2004. Proceedings 6},
  pages={16--29},
  year={2004},
  organization={Springer}
}

@inproceedings{park2018power,
  title={Power-based side-channel instruction-level disassembler},
  author={Park, Jungmin and Xu, Xiaolin and Jin, Yier and Forte, Domenic and Tehranipoor, Mark},
  booktitle={Proceedings of the 55th annual design automation conference},
  pages={1--6},
  year={2018}
}

@inproceedings{standaert2009compare,
  title={How to compare profiled side-channel attacks?},
  author={Standaert, Fran{\c{c}}ois-Xavier and Koeune, Fran{\c{c}}ois and Schindler, Werner},
  booktitle={International Conference on Applied Cryptography and Network Security},
  pages={485--498},
  year={2009},
  organization={Springer}
}

@Article{electronics12153279,
AUTHOR = {Pu, Kangran and Dang, Hua and Kong, Fancong and Zhang, Jingqi and Wang, Weijiang},
TITLE = {A Quantitative Analysis of Non-Profiled Side-Channel Attacks Based on Attention Mechanism},
JOURNAL = {Electronics},
VOLUME = {12},
YEAR = {2023},
NUMBER = {15},
ARTICLE-NUMBER = {3279},
}

@misc{gcov,
  author={{Free Software Foundation}},
  title={{Gcov (Using the GNU Compiler Collection (GCC))}},
  year={2021},
  howpublished = {\url{https://gcc.gnu.org/onlinedocs/gcc/Gcov.html}},
}

@book{sachs2012applied,
  title={Applied statistics: a handbook of techniques},
  author={Sachs, Lothar},
  year={2012},
  publisher={Springer Science \& Business Media}
}

@article{vargha2000critique,
  title={A critique and improvement of the CL common language effect size statistics of McGraw and Wong},
  author={Vargha, Andr{\'a}s and Delaney, Harold D},
  journal={Journal of Educational and Behavioral Statistics},
  volume={25},
  number={2},
  pages={101--132},
  year={2000},
  publisher={Sage Publications Sage CA: Los Angeles, CA}
}

@inproceedings{bellard2005qemu,
  title={QEMU, a fast and portable dynamic translator.},
  author={Bellard, Fabrice and others},
  booktitle={Usenix ATC, Freenix Track},
  pages={41--46},
  year={2005}
}

@inproceedings{eisele2023fuzzing,
  title={Fuzzing embedded systems using debug interfaces},
  author={Eisele, Max and Ebert, Daniel and Huth, Christopher and Zeller, Andreas},
  booktitle={Proceedings of the 32nd ACM SIGSOFT International Symposium on Software Testing and Analysis},
  pages={1031--1042},
  year={2023}
}

@inproceedings {hidinginplainsight,
author = {Yi Han and Matthew Chan and Zahra Aref and Nils Ole Tippenhauer and Saman Zonouz},
title = {Hiding in Plain Sight? On the Efficacy of Power Side {Channel-Based} Control Flow Monitoring},
booktitle = {31st USENIX Security Symposium (USENIX Security 22)},
year = {2022},
isbn = {978-1-939133-31-1},
address = {Boston, MA},
pages = {661--678},
url = {https://www.usenix.org/conference/usenixsecurity22/presentation/han},
publisher = {USENIX Association},
month = aug
}

@article{galli2025chameleon,
  title={Chameleon: A dataset for segmenting and attacking obfuscated power traces in side-channel analysis},
  author={Galli, Davide and Chiari, Giuseppe and Zoni, Davide and others},
  journal={IACR Transactions on Cryptographic Hardware and Embedded Systems},
  volume={2025},
  number={3},
  pages={389--412},
  year={2025}
}

@article{glamovcanin2023instruction,
  title={Instruction-level power side-channel leakage evaluation of soft-core CPUs on shared FPGAs},
  author={Glamo{\v{c}}anin, Ognjen and Shrivastava, Shashwat and Yao, Jinwei and Ardo, Nour and Payer, Mathias and Stojilovi{\'c}, Mirjana},
  journal={Journal of Hardware and Systems Security},
  volume={7},
  number={2},
  pages={72--99},
  year={2023},
  publisher={Springer}
}

@article{ding2020iot,
  title={IoT connectivity technologies and applications: A survey},
  author={Ding, Jie and Nemati, Mahyar and Ranaweera, Chathurika and Choi, Jinho},
  journal={IEEE Access},
  volume={8},
  pages={67646--67673},
  year={2020},
  publisher={IEEE}
}

@article{xu2018survey,
  title={A survey on industrial Internet of Things: A cyber-physical systems perspective},
  author={Xu, Hansong and Yu, Wei and Griffith, David and Golmie, Nada},
  journal={Ieee access},
  volume={6},
  pages={78238--78259},
  year={2018},
  publisher={IEEE}
}

@article{huang2023internet,
  title={Internet of medical things: A systematic review},
  author={Huang, Chenxi and Wang, Jian and Wang, Shuihua and Zhang, Yudong},
  journal={Neurocomputing},
  volume={557},
  pages={126719},
  year={2023},
  publisher={Elsevier}
}

@article{rahim2021evolution,
  title={Evolution of IoT-enabled connectivity and applications in automotive industry: A review},
  author={Rahim, Md Abdur and Rahman, Md Arafatur and Rahman, Md Mustafizur and Asyhari, A Taufiq and Bhuiyan, Md Zakirul Alam and Ramasamy, Devarajan},
  journal={Vehicular communications},
  volume={27},
  pages={100285},
  year={2021},
  publisher={Elsevier}
}

@inproceedings{feng2021snipuzz,
  title={Snipuzz: Black-box fuzzing of iot firmware via message snippet inference},
  author={Feng, Xiaotao and Sun, Ruoxi and Zhu, Xiaogang and Xue, Minhui and Wen, Sheng and Liu, Dongxi and Nepal, Surya and Xiang, Yang},
  booktitle={Proceedings of the 2021 ACM SIGSAC conference on computer and communications security},
  pages={337--350},
  year={2021}
}

@inproceedings{aafer2021android,
  title={Android $\{$SmartTVs$\}$ vulnerability discovery via $\{$log-guided$\}$ fuzzing},
  author={Aafer, Yousra and You, Wei and Sun, Yi and Shi, Yu and Zhang, Xiangyu and Yin, Heng},
  booktitle={30th USENIX Security Symposium (USENIX Security 21)},
  pages={2759--2776},
  year={2021}
}

@online{esp32,
	author = {{ESPRESSIF}},
	title = {{ESP32} wireless module},
	howpublished = {\url{https://www.espressif.com/en/products/modules}},
}

@online{ubloxgnss,
	author = {{U-Blox}},
	title = {{U-B}lox {GNSS} {R}eceiver},
	howpublished = {\url{https://www.u-blox.com/en/positioning-chips-and-modules}},
}

@online{stsafe,
	author = {{STM}icroelectronics},
	title = {{STSAFE}},
	howpublished = {\url{https://www.st.com/en/secure-mcus/stsafe-a100.html}},
}

@online{cansensor,
	author = {{RENESAS}},
	title = {{CAN} {S}ensor {N}etwork {D}evelopment {K}it},
	howpublished = {\url{https://www.renesas.com/en/design-resources/boards-kits/tw007-indcanpocz}},
}

@inproceedings{feng2020p2im,
  title={$\{$P2IM$\}$: Scalable and hardware-independent firmware testing via automatic peripheral interface modeling},
  author={Feng, Bo and Mera, Alejandro and Lu, Long},
  booktitle={29th USENIX Security Symposium (USENIX Security 20)},
  pages={1237--1254},
  year={2020}
}

@inproceedings{sperl2019side,
  title={Side-channel aware fuzzing},
  author={Sperl, Philip and B{\"o}ttinger, Konstantin},
  booktitle={European Symposium on Research in Computer Security},
  pages={259--278},
  year={2019},
  organization={Springer}
}

@article{mcclintick2024side,
  title={Side-Channel Assisted Real-Time Fuzzing for Embedded Systems (SCARFES): FY23 Cyber Security Line-Supported Program},
  author={Mcclintick, Kyle W and Binyamin, Erez and John, Brandon V and Ingols, Kyle W and Chetwynd, Brendon R and Shields, Emily K},
  year={2024}
}

@inproceedings{barredo2025gaflerna,
  title={GAFLERNA Ahoy! Integrating EM Side-Channel Analysis into Traditional Fuzzing Workflows},
  author={Barredo, Jorge and Petke, Justyna and Clark, David and Blackwell, Daniel and Eceiza, Maialen and Flores, Jose Luis and Iturbe, Mikel},
  booktitle={Proceedings of the 33rd ACM International Conference on the Foundations of Software Engineering},
  pages={550--554},
  year={2025}
}

@inproceedings{su2024fuzz,
  title={Fuzz Wars: The Voltage Awakens--Voltage-Guided Blackbox Fuzzing on FPGAs},
  author={Su, Kai and Giraud, Mark and Borcherding, Anne and Krautter, Jonas and Nenninger, Philipp and Tahoori, Mehdi},
  booktitle={2024 IEEE 42nd VLSI Test Symposium (VTS)},
  pages={1--7},
  year={2024},
  organization={IEEE}
}

@article{zamudio2025fandango,
  title={FANDANGO: evolving language-based testing},
  author={Zamudio Amaya, Jos{\'e} Antonio and Smytzek, Marius and Zeller, Andreas},
  journal={Proceedings of the ACM on Software Engineering},
  volume={2},
  number={ISSTA},
  pages={894--916},
  year={2025},
  publisher={ACM New York, NY, USA}
}

@article{lin2026execution,
  title={Execution Divergence Graphs: Effective Discovery of Control-Flows from Execution Traces as Fuzzing Feedback},
  author={Lin, Yu-De and Tippenhauer, Nils Ole},
  journal={arXiv preprint arXiv:2607.03396},
  year={2026}
}

@article{papagiannopoulos2023side,
  title={The side-channel metrics cheat sheet},
  author={Papagiannopoulos, Kostas and Glamo{\v{c}}anin, Ognjen and Azouaoui, Melissa and Ros, Dorian and Regazzoni, Francesco and Stojilovi{\'c}, Mirjana},
  journal={ACM Computing Surveys},
  volume={55},
  number={10},
  pages={1--38},
  year={2023},
  publisher={ACM New York, NY}
}

@article{narimani2024exploring,
  title={Exploring Power Side-Channel Challenges in Embedded Systems Security},
  author={Narimani, Pouya and Wang, Meng and Planta, Ulysse and Abbasi, Ali},
  journal={arXiv preprint arXiv:2410.11563},
  year={2024}
}

@article{bursztein2023generalized,
  title={Generalized power attacks against crypto hardware using long-range deep learning},
  author={Bursztein, Elie and Invernizzi, Luca and Kr{\'a}l, Karel and Moghimi, Daniel and Picod, Jean-Michel and Zhang, Marina},
  journal={arXiv preprint arXiv:2306.07249},
  year={2023}
}

@inproceedings{ferrufino2023fobos,
  title={FOBOS 3: An open-source platform for side-channel analysis and benchmarking},
  author={Ferrufino, Eduardo and Beckwith, Luke and Abdulgadir, Abubakr and Kaps, Jens-Peter},
  booktitle={Proceedings of the 2023 Workshop on Attacks and Solutions in Hardware Security},
  pages={5--14},
  year={2023}
}

@inproceedings{liu2016code,
  title={On code execution tracking via power side-channel},
  author={Liu, Yannan and Wei, Lingxiao and Zhou, Zhe and Zhang, Kehuan and Xu, Wenyuan and Xu, Qiang},
  booktitle={Proceedings of the 2016 ACM SIGSAC conference on computer and communications security},
  pages={1019--1031},
  year={2016}
}

@inproceedings {zheng2019firm,
author = {Yaowen Zheng and Ali Davanian and Heng Yin and Chengyu Song and Hongsong Zhu and Limin Sun},
title = {{FIRM-AFL}: {High-Throughput} Greybox Fuzzing of {IoT} Firmware via Augmented Process Emulation},
booktitle = {28th USENIX Security Symposium (USENIX Security 19)},
year = {2019},
isbn = {978-1-939133-06-9},
address = {Santa Clara, CA},
pages = {1099--1114},
url = {https://www.usenix.org/conference/usenixsecurity19/presentation/zheng},
publisher = {USENIX Association},
month = aug
}

@inproceedings{chen2018iotfuzzer,
  title={IoTFuzzer: Discovering memory corruptions in IoT through app-based fuzzing.},
  author={Chen, Jiongyi and Diao, Wenrui and Zhao, Qingchuan and Zuo, Chaoshun and Lin, Zhiqiang and Wang, XiaoFeng and Lau, Wing Cheong and Sun, Menghan and Yang, Ronghai and Zhang, Kehuan},
  booktitle={NDSS},
  pages={1--15},
  year={2018}
}

@inproceedings{moradihaghighi2026fuzz,
  title={Fuzz'EMup: Leveraging EM Side-Channel Emanation to Guide Black-Box Embedded Firmware Fuzzing},
  author={Moradihaghighi, Fatemeh and Zhan, Zihao and Guo, Yanan and Zhao, Ziming and Chowdhury, Mashrur and Zhang, Zhenkai},
  booktitle={2026 IEEE International Symposium on Hardware Oriented Security and Trust (HOST)},
  pages={174--185},
  year={2026},
  organization={IEEE}
}

@article{bohme2020fuzzing,
  title={Fuzzing: Challenges and reflections},
  author={B{\"o}hme, Marcel and Cadar, Cristian and Roychoudhury, Abhik},
  journal={IEEE Software},
  volume={38},
  number={3},
  pages={79--86},
  year={2020},
  publisher={IEEE}
}

@inproceedings{klees2018evaluating,
  title={Evaluating fuzz testing},
  author={Klees, George and Ruef, Andrew and Cooper, Benji and Wei, Shiyi and Hicks, Michael},
  booktitle={Proceedings of the 2018 ACM SIGSAC conference on computer and communications security},
  pages={2123--2138},
  year={2018}
}

@inproceedings{mera2021dice,
  title={DICE: Automatic emulation of DMA input channels for dynamic firmware analysis},
  author={Mera, Alejandro and Feng, Bo and Lu, Long and Kirda, Engin},
  booktitle={2021 IEEE Symposium on Security and Privacy (SP)},
  pages={1938--1954},
  year={2021},
  organization={IEEE}
}

@misc{githubGitHubMaJerlelwjson,
	author = {},
	title = {{G}it{H}ub - {M}a{J}erle/lwjson: {L}ightweight {J}{S}{O}{N} parser for embedded systems --- github.com},
	howpublished = {\url{https://github.com/{M}a{J}erle/lwjson}},
	year = {2026},
}

@misc{githubGitHubMarcinbor85cAT,
	author = {},
	title = {{G}it{H}ub - marcinbor85/c{A}{T}: {P}lain {C} library for parsing {A}{T} commands for use in host devices. --- github.com},
	howpublished = {\url{https://github.com/marcinbor85/cAT}},
	year = {2023},
}

@misc{githubGitHubKosmaminmea,
	author = {},
	title = {{G}it{H}ub - kosma/minmea: a lightweight {G}{P}{S} {N}{M}{E}{A} 0183 parser library in pure {C} --- github.com},
	howpublished = {\url{https://github.com/kosma/minmea}},
	year = {2026},
}

@misc{githubGitHubMaJerlelwgps,
	author = {},
	title = {{G}it{H}ub - {M}a{J}erle/lwgps: {L}ightweight {G}{P}{S} {N}{M}{E}{A} parser for embedded systems --- github.com},
	howpublished = {\url{https://github.com/{M}a{J}erle/lwgps}},
	year = {2026},
}

@misc{githubGitHubRafagafetinyjson,
	author = {},
	title = {{G}it{H}ub - rafagafe/tiny-json: {T}he tiny-json is a versatile and easy to use json parser in {C} suitable for embedded systems. {I}t is fast, robust and portable. --- github.com},
	howpublished = {\url{https://github.com/rafagafe/tiny-json}},
	year = {2021},
}

@misc{githubGitHub0introlibelf,
	author = {},
	title = {{G}it{H}ub - 0intro/libelf: {L}ibelf is a simple library to read {E}{L}{F} files. --- github.com},
	howpublished = {\url{https://github.com/0intro/libelf}},
	year = {2026},
}

@misc{githubGitHubCmumfordTJpgDec,
	author = {},
	title = {{G}it{H}ub - cmumford/{T}{J}pg{D}ec: {T}iny {J}{P}{E}{G} {D}ecompressor --- github.com},
	howpublished = {\url{https://github.com/cmumford/{T}{J}pg{D}ec}},
	year = {2021},
}

@misc{githubGitHubNrushernr_micro_shell,
	author = {},
	title = {{G}it{H}ub - {N}rusher/nr\_micro\_shell: shell for {M}{C}{U}. --- github.com},
	howpublished = {\url{https://github.com/{N}rusher/nr\_micro\_shell/tree/master}},
	year = {2025},
}

@misc{githubGitHubBetaflightbetaflight,
	author = {},
	title = {{G}it{H}ub - betaflight/betaflight: {O}pen {S}ource {F}light {C}ontroller {F}irmware --- github.com},
	howpublished = {\url{https://github.com/betaflight/betaflight}},
	year = {2026},
}

@misc{githubGitHubINavFlightinav,
	author = {},
	title = {{G}it{H}ub - i{N}av{F}light/inav: {I}{N}{A}{V}: {N}avigation-enabled flight control software --- github.com},
	howpublished = {\url{https://github.com/i{N}av{F}light/inav.git}},
	year = {2026},
}

@misc{githubGitHubKokketinyregexc,
	author = {},
	title = {{G}it{H}ub - kokke/tiny-regex-c: {S}mall portable regex in {C} --- github.com},
	howpublished = {\url{https://github.com/kokke/tiny-regex-c}},
	year = {2024},
}

@misc{githubGitHubRiS3Labp2imreal_firmware,
	author = {},
	title = {{G}it{H}ub - {R}i{S}3-{L}ab/p2im-real\_firmware at d4c7456574ce2c2ed038e6f14fea8e3142b3c1f7 --- github.com},
	howpublished = {\url{https://github.com/{R}i{S}3-{L}ab/p2im-real\_firmware/tree/d4c7456574ce2c2ed038e6f14fea8e3142b3c1f7}},
	year = {},
	note = {[Accessed 25-08-2026]},
}

@misc{githubGitHubRiS3LabDICEDMAEmulation,
	author = {},
	title = {{G}it{H}ub - {R}i{S}3-{L}ab/{D}{I}{C}{E}-{D}{M}{A}-{E}mulation: {D}{I}{C}{E}: {A}utomatic {E}mulation of {D}{M}{A} {I}nput {C}hannels for {D}ynamic {F}irmware {A}nalysis --- github.com},
	howpublished = {\url{https://github.com/{R}i{S}3-{L}ab/{D}{I}{C}{E}-{D}{M}{A}-{E}mulation/tree/master}},
	year = {},
	note = {[Accessed 25-08-2026]},
}

@misc{githubGitHubDebevvnanoMODBUS,
	author = {},
	title = {{G}it{H}ub - debevv/nano{M}{O}{D}{B}{U}{S}: {A} compact {M}{O}{D}{B}{U}{S} {R}{T}{U}/{T}{C}{P} {C} library for embedded/microcontrollers --- github.com},
	howpublished = {\url{https://github.com/debevv/nano{M}{O}{D}{B}{U}{S}/tree/master}},
	year = {},
	note = {[Accessed 25-08-2026]},
}

\section{First order derivatives}\label{section:appendix:derivatives}
In this appendix, we mathematically show the effect of first-order derivatives on low-frequency amplitude offsets. Consider a measured signal composed of the desired signal $s[n]$ and an amplitude offset $o[n]$:

\begin{equation*}
x[n] = s[n] + o[n].
\end{equation*}

The first-order discrete derivative is defined as:

\begin{equation*}
\Delta x[n] = x[n] - x[n-1].
\end{equation*}

Substituting $x[n]$ gives:

\begin{align*}
\Delta x[n]
&= (s[n] + o[n]) - (s[n-1] + o[n-1]) \\
&= \Delta s[n] + \Delta o[n].
\end{align*}

For a constant (DC) offset,

\begin{equation*}
o[n] = C,
\end{equation*}

where $C$ is a constant. Therefore,

\begin{equation*}
\Delta o[n] = C - C = 0
\end{equation*}

Hence,

\begin{equation*}
\Delta x[n] = \Delta s[n]
\end{equation*}

showing that the DC offset is eliminated by the first-order derivative.

More generally, suppose the offset varies slowly with time,

\begin{equation*}
o[n] \approx o[n-1].
\end{equation*}

Then,

\begin{equation*}
\Delta o[n] = o[n]-o[n-1] \approx 0,
\end{equation*}

which implies

\begin{equation*}
\Delta x[n]
= \Delta s[n] + \Delta o[n]
\approx \Delta s[n].
\end{equation*}

Therefore, the influence of low-frequency amplitude variations is significantly reduced after differentiation. Only rapidly varying components contribute appreciably to the derivative.


\section{Calibration}\label{section:appendix:calibration}

As discussed earlier, \POZZER{} operates under a black-box threat model and therefore does not rely on any prior knowledge of the target firmware or profiling data. Nevertheless, a set of setup-specific parameters must be configured to account for differences in the measurement setup, such as the target \ac{MCU}, \ac{PCB}, and the acquisition device. These parameters include \emph{chunk size}, \emph{chunk threshold}, \emph{signature size}, \emph{signature threshold}, \emph{time-shift window size}, \emph{point-wise discrepancy threshold}, and \emph{minimum node size}.

Among all parameters, the \emph{minimum node size} plays a particularly important role in graph construction. It specifies the minimum temporal distance allowed between two consecutive discrepancy points in the constructed graph. Although the neighbor-search mechanism compensates for time shifts during the point-wise comparison stage, small time misalignments can still introduce spurious discrepancy points during chunk-based comparison. Enforcing a minimum node size prevents such closely spaced discrepancy points from being inserted into the graph, thereby reducing false positives and avoiding redundant graph nodes. The remaining of the size-related parameters are determined directly from the temporal resolution of the acquired \ac{PSC} traces and therefore depend only on the sampling rate of the measurement device and the clock frequency of the target \ac{MCU}.

On the other hand, the threshold parameters depend on the electrical characteristics of the target and the measurement setup. To estimate these parameters, \POZZER{} performs a one-time calibration stage before each fuzzing campaign. During calibration, \POZZER{} automatically generates a set of test cases, repeatedly executes each test case on the target, and captures the corresponding \ac{PSC} traces. Since each test case is repeated under the same conditions, we attribute the variation among the resulting traces to measurement noise. For calibration, \POZZER{} generates $M=100$ random test cases and captures the corresponding \ac{PSC} trace $N=100$ times for each test case. Let $X_m=\{x_{m,n}\}_{n=1}^{N}$ denote the set of repeated \ac{PSC} traces corresponding to the $m$-th test case. Following~\cite{papagiannopoulos2023side}, we assume the measurement noise follows a Gaussian distribution for each sample. The sample covariance matrix of the repeated measurements for test case $m$ is therefore estimated as

\begin{equation*}
\Sigma_m=\frac{1}{N-1}\sum_{n=1}^{N}(x_{m,n}-\bar{x}_m)(x_{m,n}-\bar{x}_m)^T,
\end{equation*}

where the mean of \ac{PSC} trace for the $m$-th test case is

\begin{equation*}
\bar{x}_m=\frac{1}{N}\sum_{n=1}^{N}x_{m,n}
\end{equation*}

Repeating this procedure for all calibration test cases yields $M$ covariance matrices. Since threshold estimation only requires the variance of each sample, we retain only the diagonal entries of $\Sigma_m$. For each test case, we then select the maximum variance as $\sigma_m^2=\max(\operatorname{diag}(\Sigma_m))$, which provides a conservative estimate of the worst-case measurement noise. Finally, the average worst-case noise variance over the entire calibration set is computed as

\begin{equation*}
\bar{\sigma}^2=\frac{1}{M}\sum_{m=1}^{M}\sigma_m^2.
\end{equation*}

The discrepancy threshold is then selected such that the probability of the measurement noise remaining below the threshold is upper-bounded to 0.9999. Under the zero-mean Gaussian assumption, this corresponds approximately to four standard deviations,

\begin{equation*}
P\left(x_{m,n,i}<\mathrm{th}\right)
=
\Phi\left(\frac{\mathrm{th}}{\bar{\sigma}}\right)
=
0.9999
\quad\Longrightarrow\quad
\boxed{\mathrm{th}\approx4\bar{\sigma}}
\end{equation*}

where $\Phi$ denotes the standard normal \ac{CDF} and $\bar \sigma$ is the average worst-case noise standard deviation over the entire calibration set.

The calibration stage, therefore, produces the setup-specific parameter set used during fuzzing. Once calibration is complete, all parameter values remain fixed throughout the campaign. \POZZER{} then uses this fixed configuration to process each captured \ac{PSC} trace and determine whether it represents previously unseen execution behavior.


\section{Evaluation machine}\label{section:appendix:evalmachine}

In this appendix, we describe the machine used for the fuzzing experiments. The machine is equipped with an \texttt{Intel Core Ultra 5 225} processor with 10 CPU cores and a maximum dynamic frequency of 4.4~GHz. It has 32~GB of memory and runs an \texttt{Ubuntu 24.04.4 LTS}.


\section{A representative example: Nested-\emph{if} structure}\label{section:appendix:nestedif}

\begin{listing}[h]
\begin{minted}
[
frame=lines,
framesep=2mm,
baselinestretch=1.0,
fontsize=\footnotesize,
linenos,
numbersep=1pt
]
{c}
if (input[0] == 'P')
  if (input[1] == 'O')
    if (input[2] == 'P')
      if (input[3] == 'F')
        if (input[4] == 'U')
          if (input[5] == 'Z')
            if (input[6] == 'Z')
              if (input[7] == '!')
                crash();
\end{minted}
\caption{Nested-if code structure}
\label{listing:ifelse}
\end{listing}


\section{Coverage plots}\label{section:appendix:plots}
\begin{figure*}[!t]
    \centering
    \begin{subfigure}{0.23\textwidth}
        \centering
        \includegraphics[width=\linewidth]{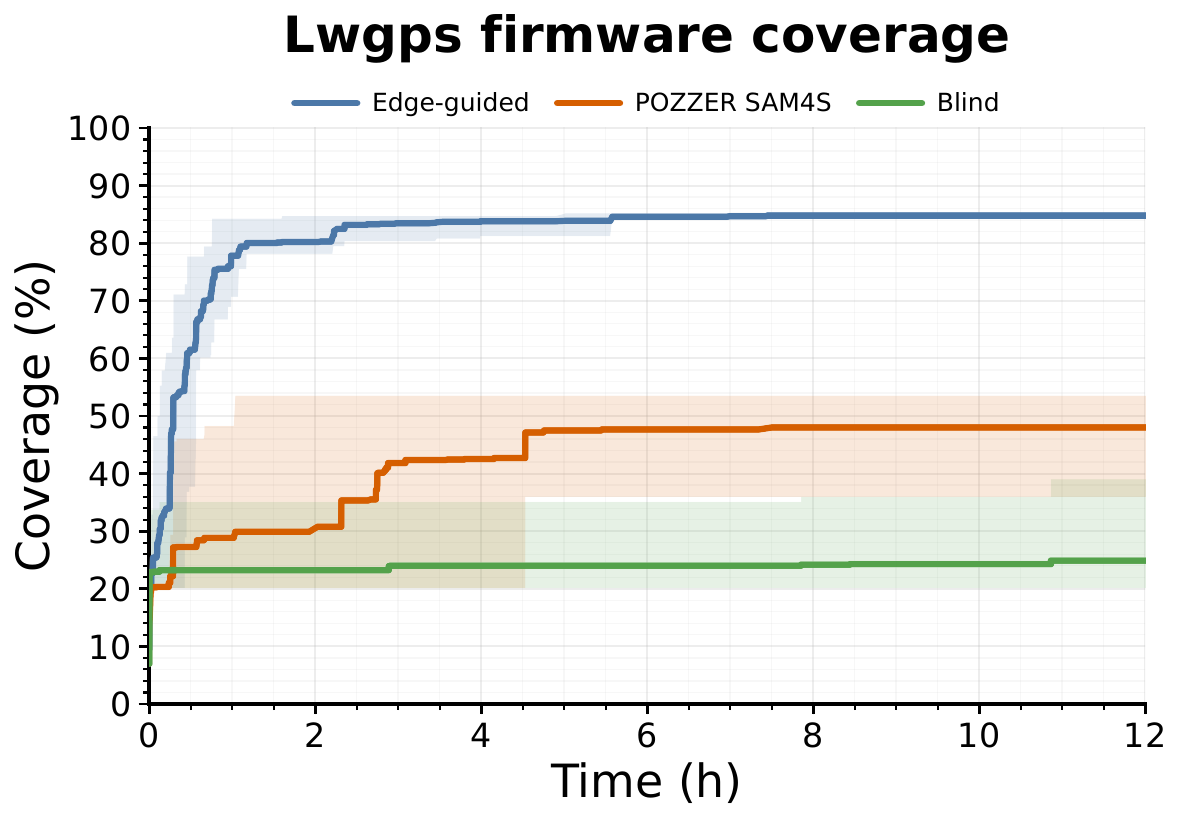}
    \end{subfigure}
    \begin{subfigure}{0.23\textwidth}
        \centering
        \includegraphics[width=\linewidth]{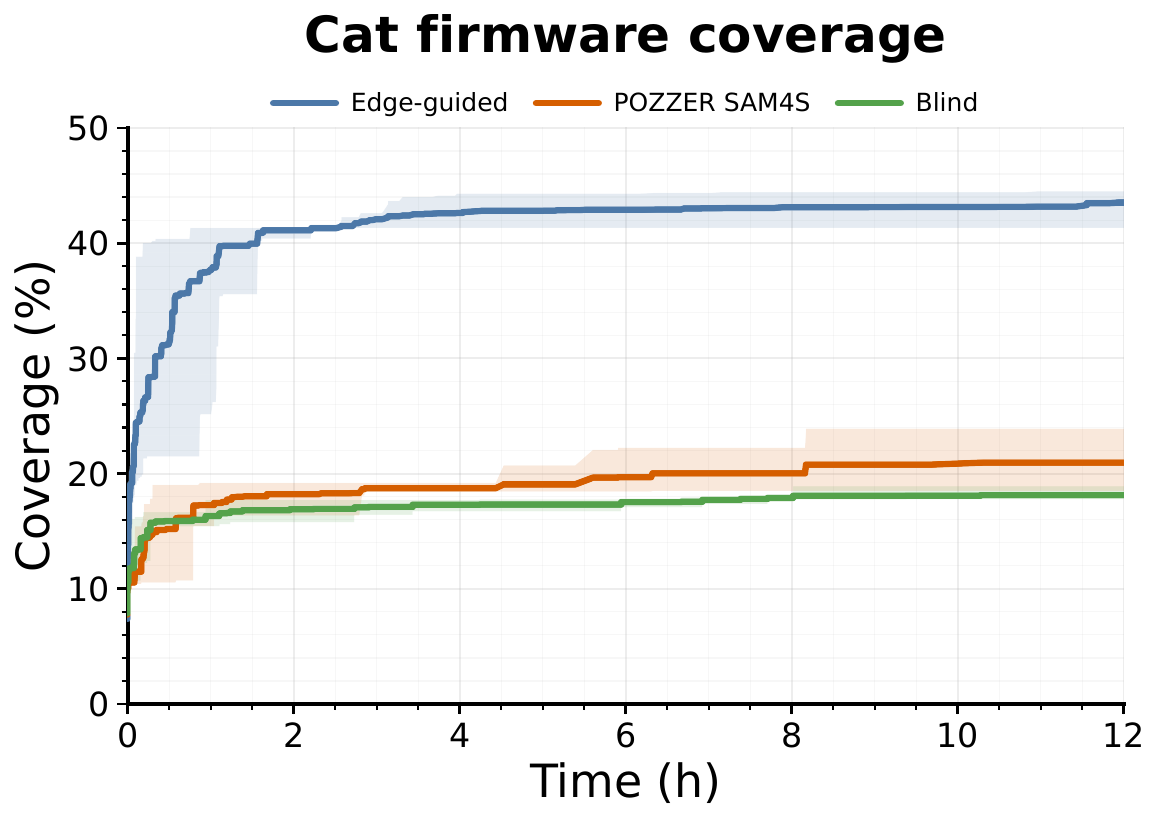}
    \end{subfigure}
    \begin{subfigure}{0.23\textwidth}
        \centering
        \includegraphics[width=\linewidth]{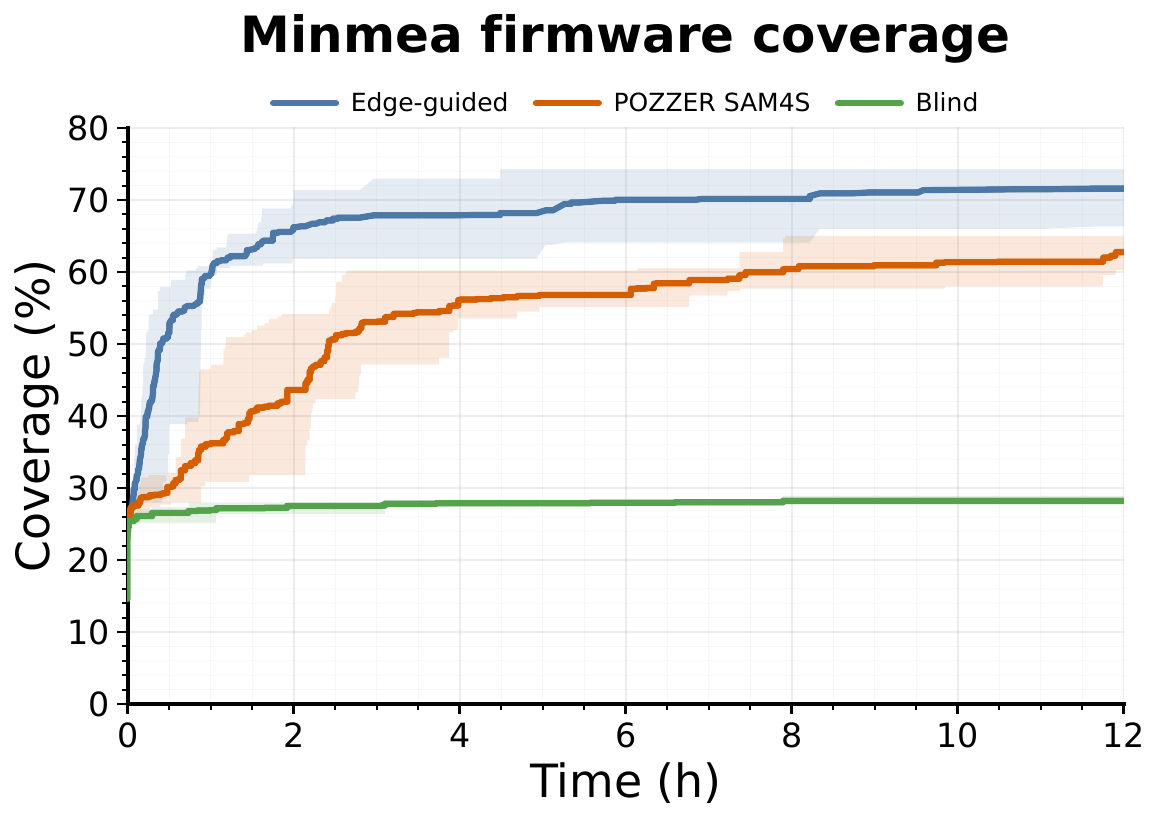}
    \end{subfigure}
    \begin{subfigure}{0.23\textwidth}
        \centering
        \includegraphics[width=\linewidth]{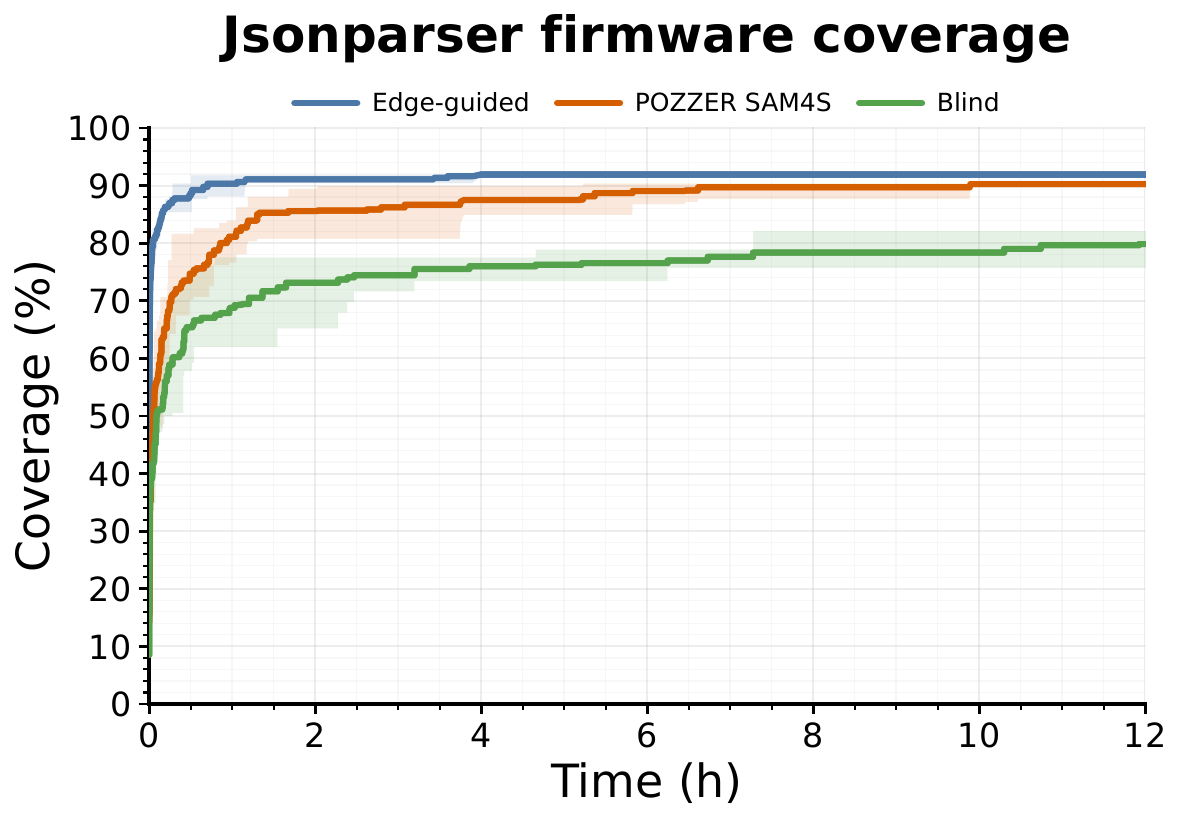}
    \end{subfigure}
    
    \begin{subfigure}{0.23\textwidth}
        \centering
        \includegraphics[width=\linewidth]{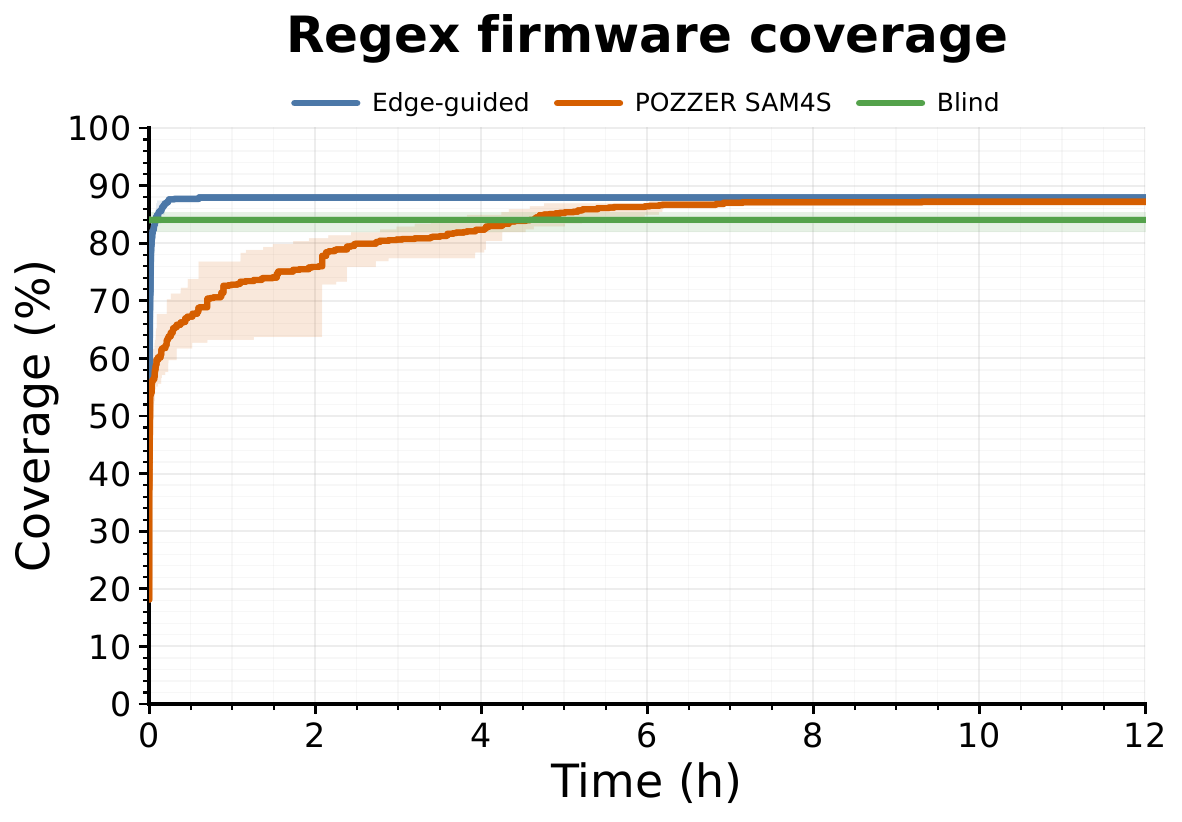}
    \end{subfigure}
    \begin{subfigure}{0.23\textwidth}
        \centering
        \includegraphics[width=\linewidth]{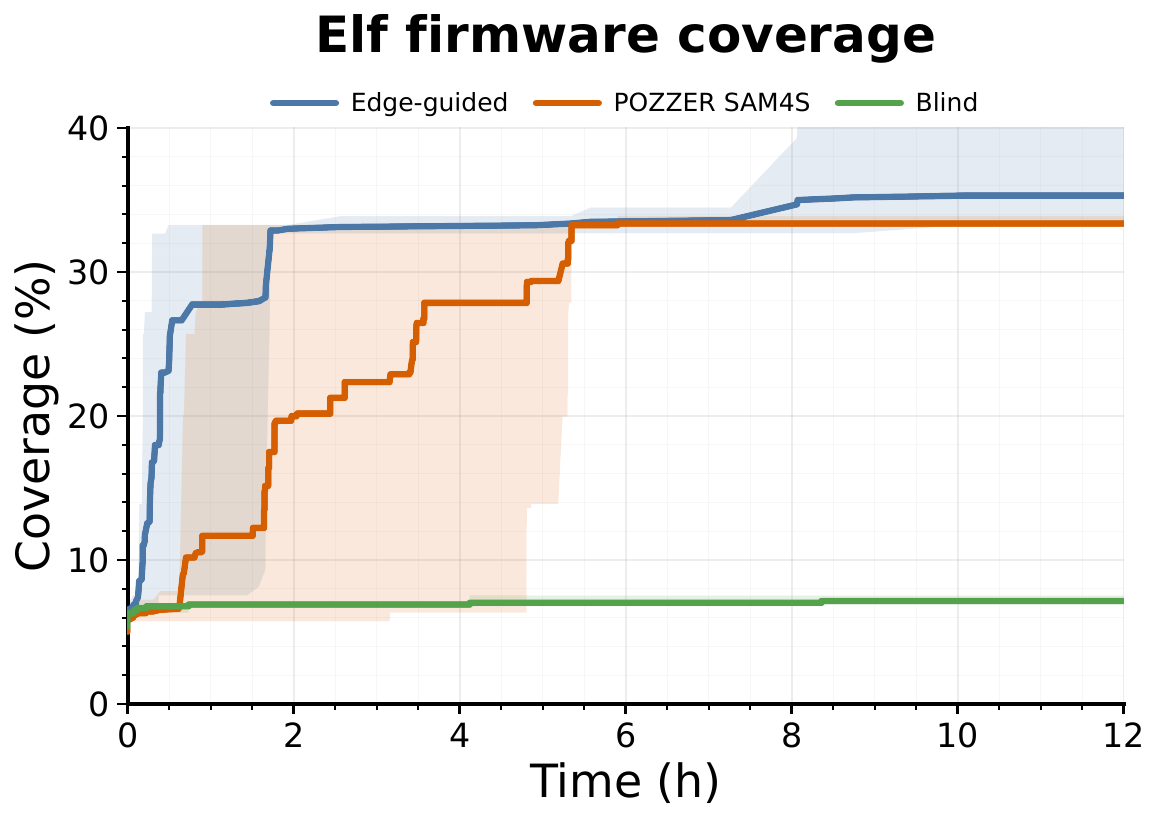}
    \end{subfigure}
    \begin{subfigure}{0.23\textwidth}
        \centering
        \includegraphics[width=\linewidth]{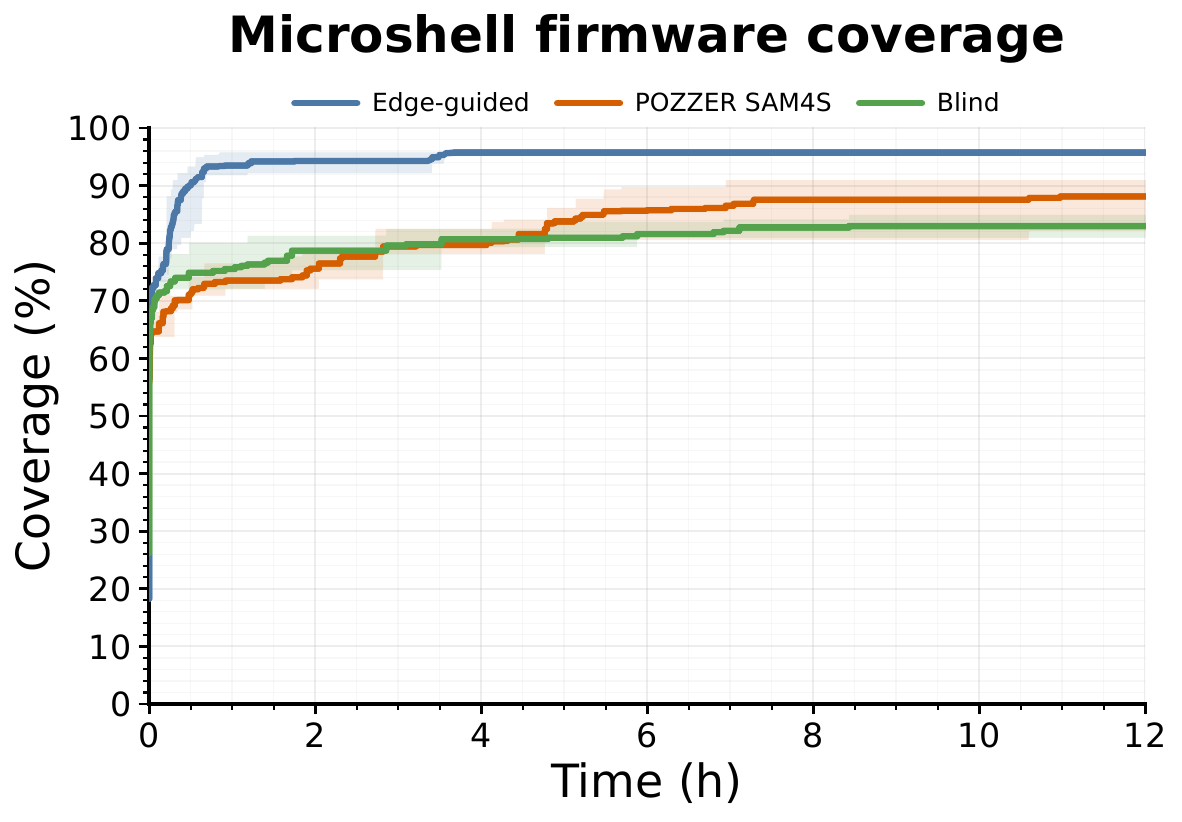}
    \end{subfigure}
    \begin{subfigure}{0.23\textwidth}
        \centering
        \includegraphics[width=\linewidth]{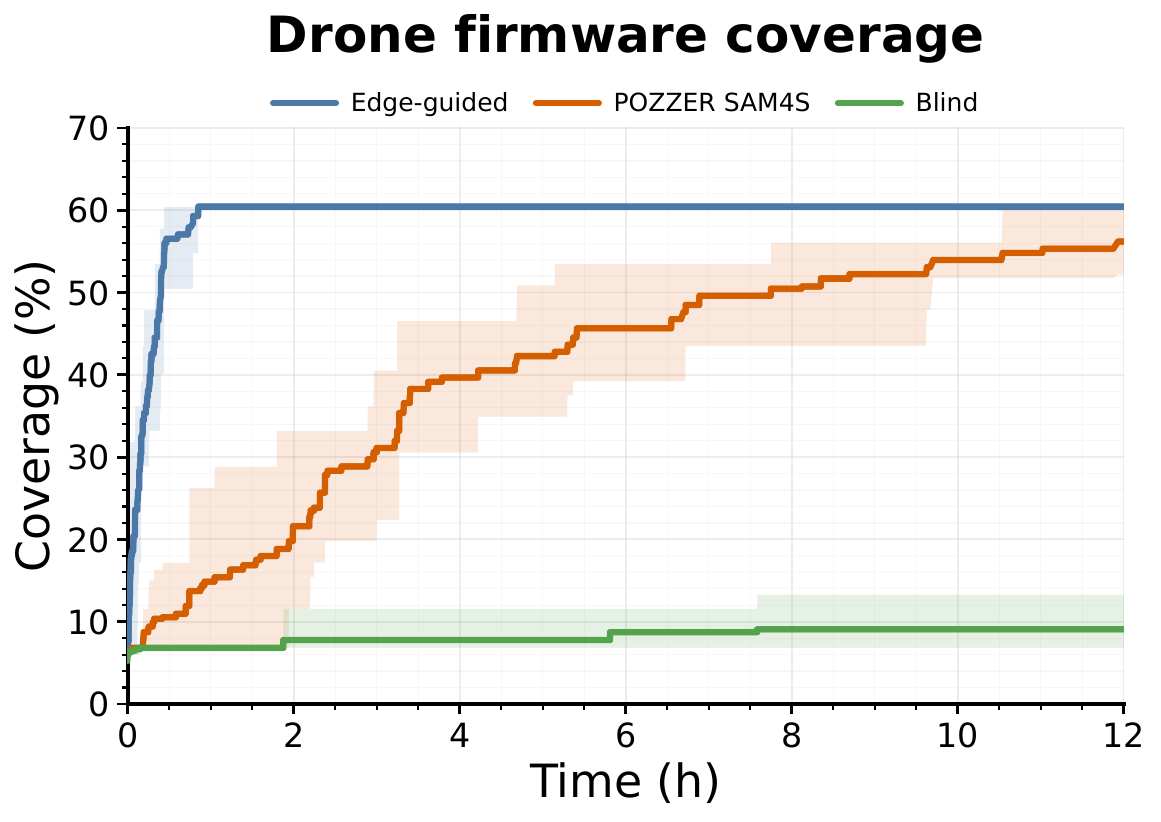}
    \end{subfigure}

    \begin{subfigure}{0.23\textwidth}
        \centering
        \includegraphics[width=\linewidth]{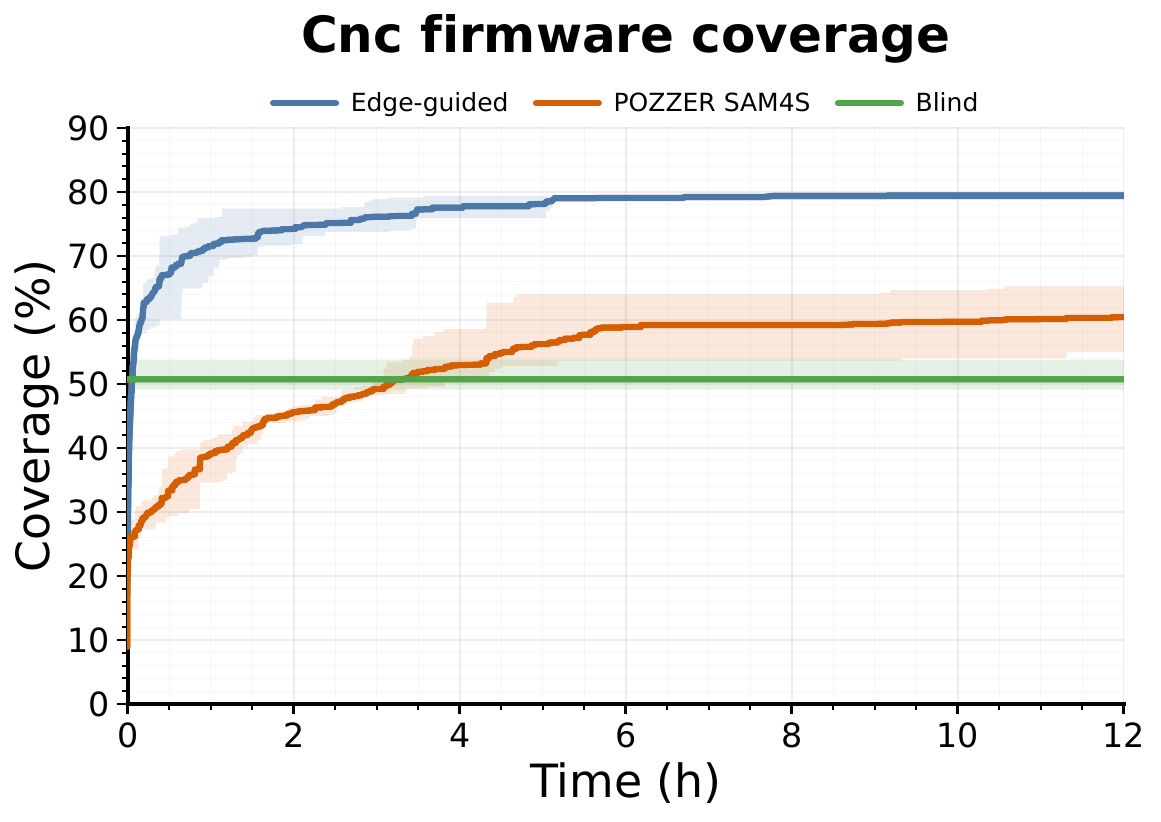}
    \end{subfigure}
    \begin{subfigure}{0.23\textwidth}
        \centering
        \includegraphics[width=\linewidth]{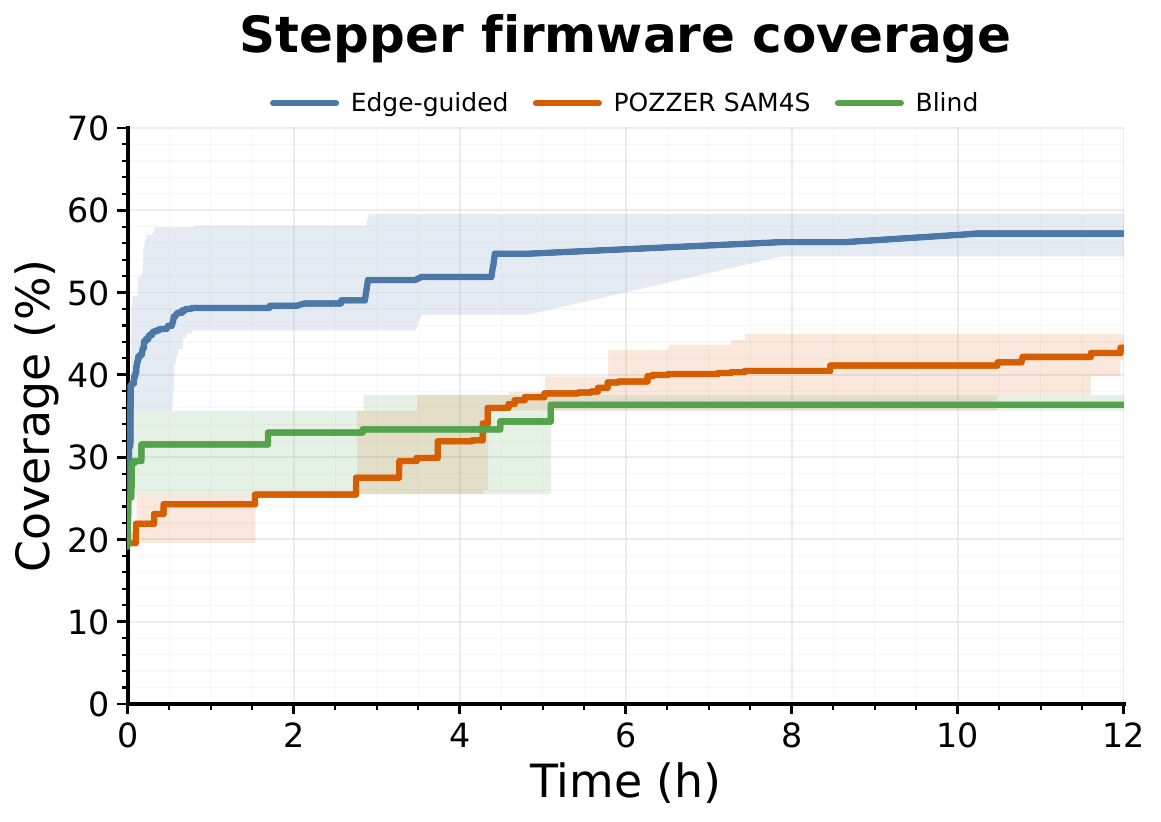}
    \end{subfigure}
    \begin{subfigure}{0.23\textwidth}
        \centering
        \includegraphics[width=\linewidth]{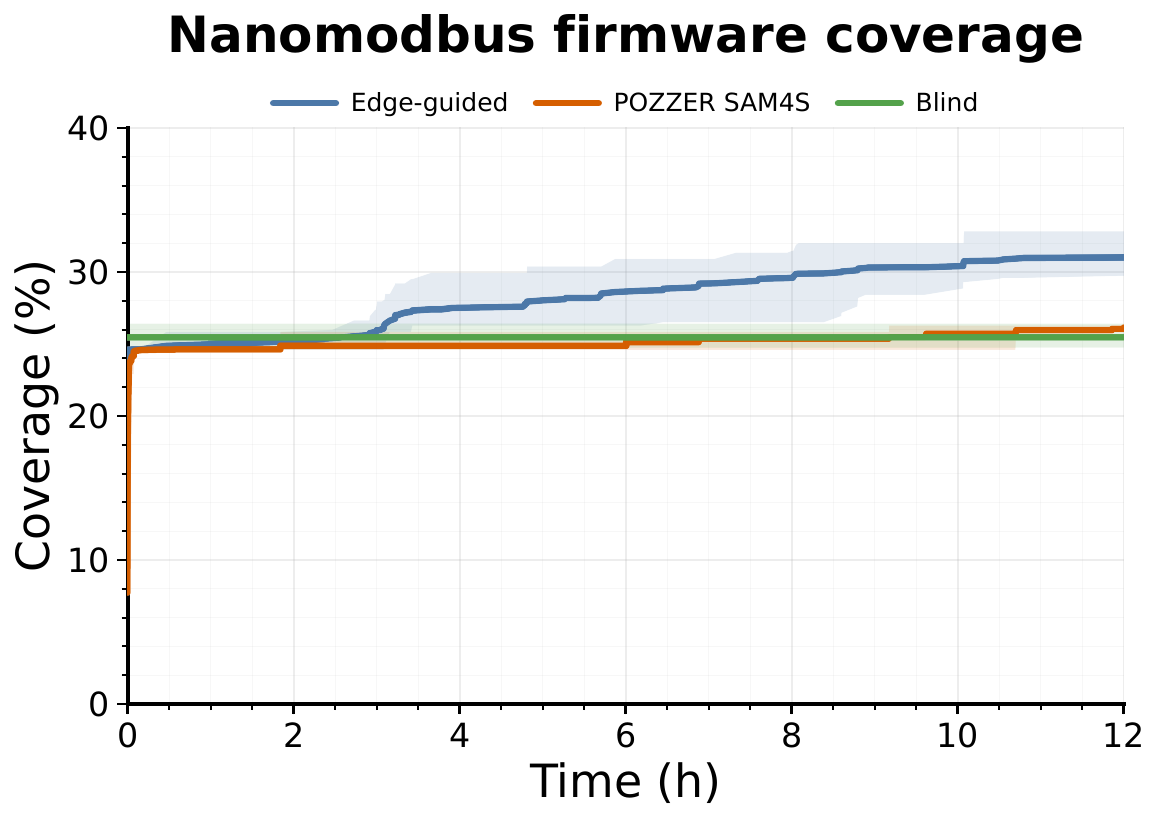}
    \end{subfigure}
    \caption{This figure presents coverage plots for different targets running on \emph{SAM4S}.}
    \label{fig:sam4s_plots}
\end{figure*}

\begin{figure*}[!t]
    \centering
    \begin{subfigure}{0.23\textwidth}
        \centering
        \includegraphics[width=\linewidth]{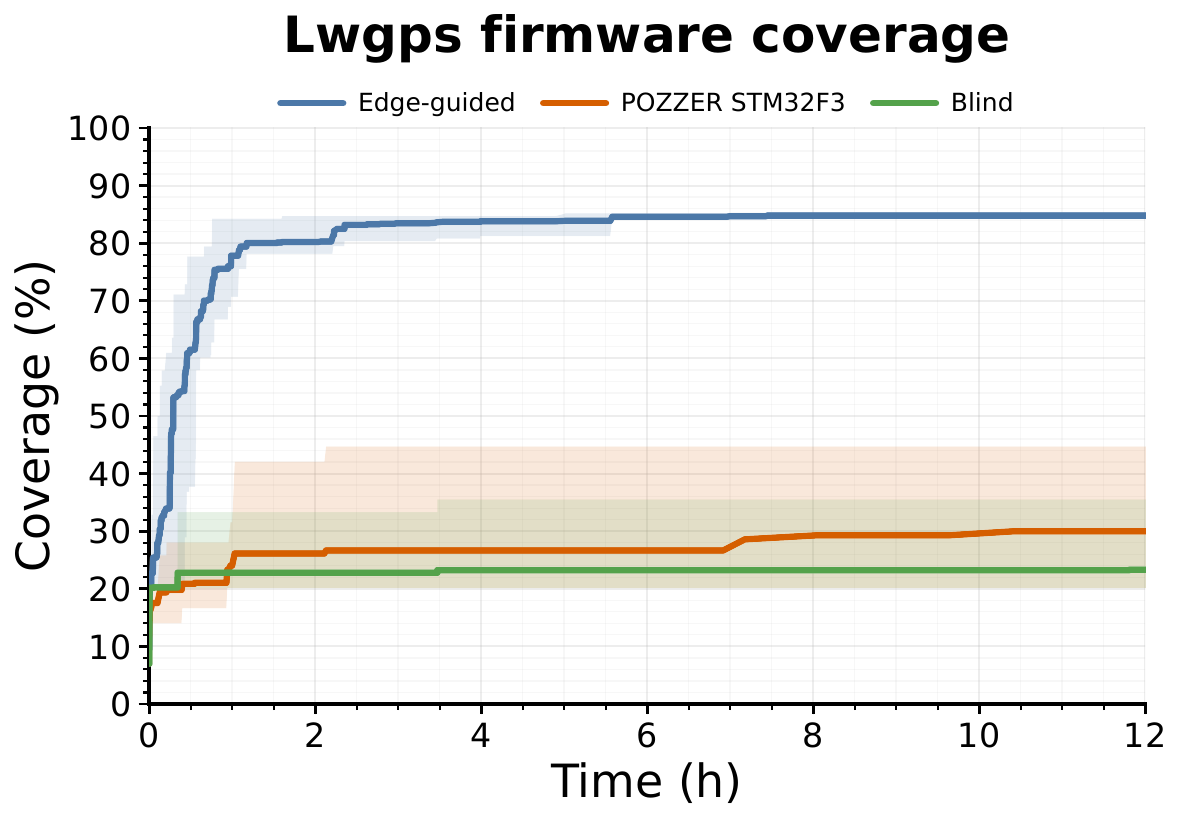}
    \end{subfigure}
    \begin{subfigure}{0.23\textwidth}
        \centering
        \includegraphics[width=\linewidth]{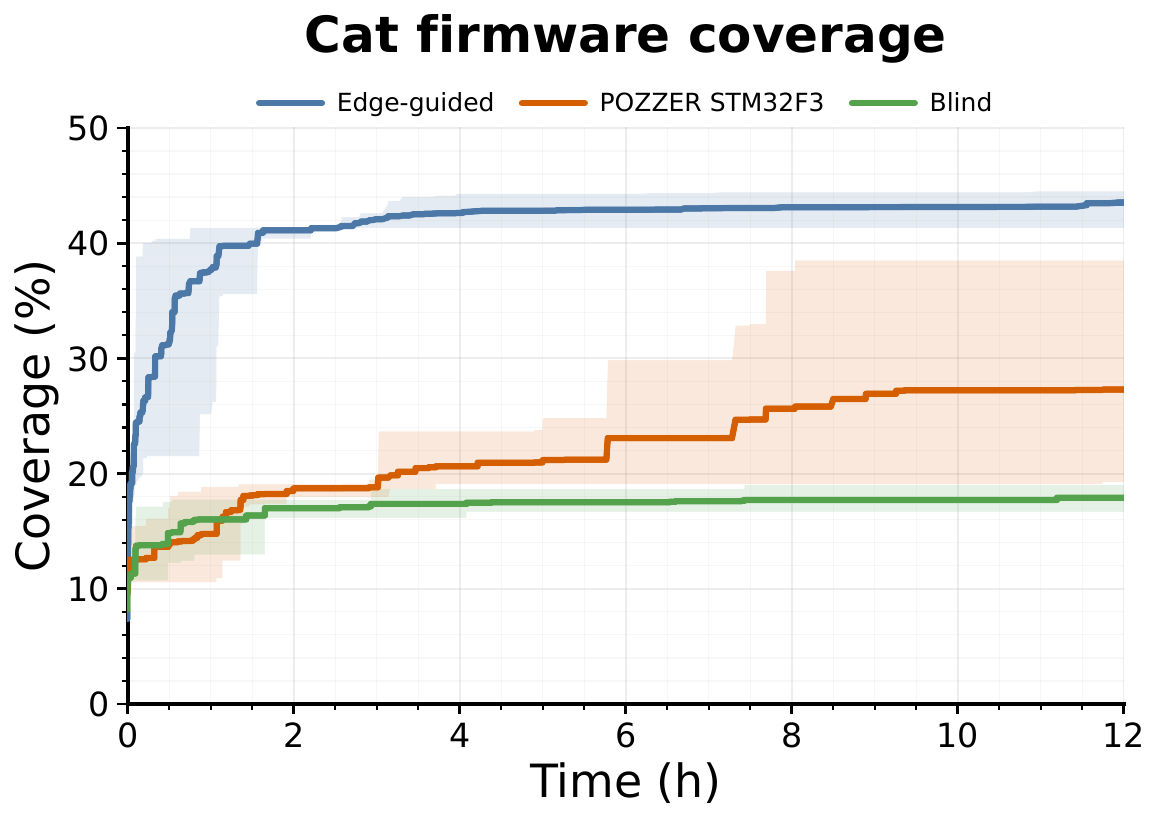}
    \end{subfigure}
    \begin{subfigure}{0.23\textwidth}
        \centering
        \includegraphics[width=\linewidth]{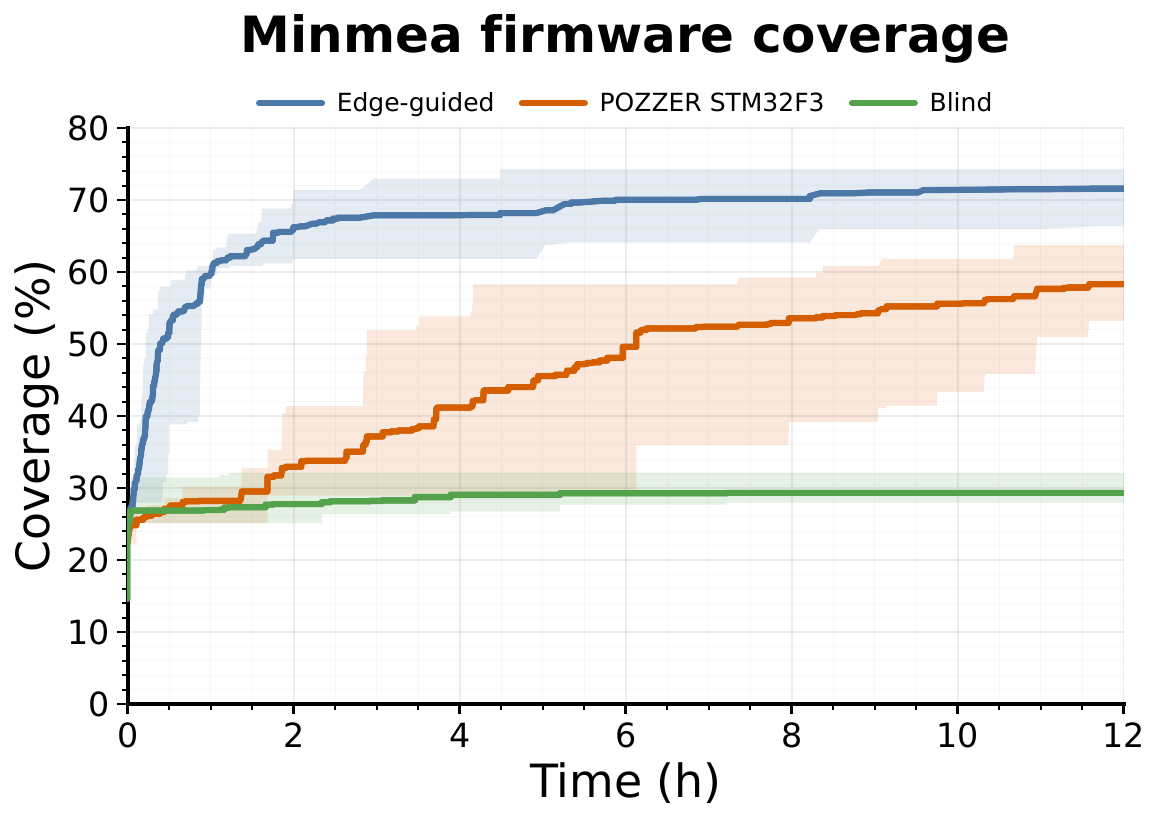}
    \end{subfigure}
    \begin{subfigure}{0.23\textwidth}
        \centering
        \includegraphics[width=\linewidth]{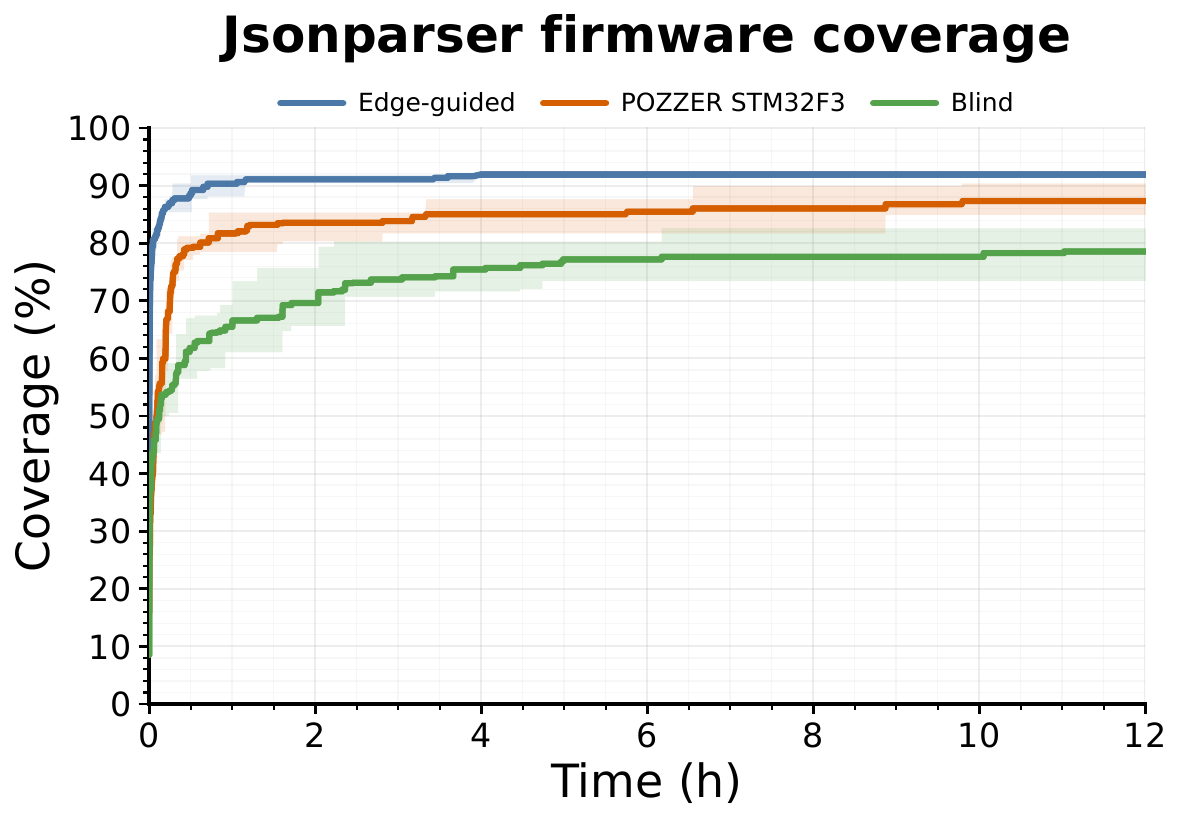}
    \end{subfigure}
    
    \begin{subfigure}{0.23\textwidth}
        \centering
        \includegraphics[width=\linewidth]{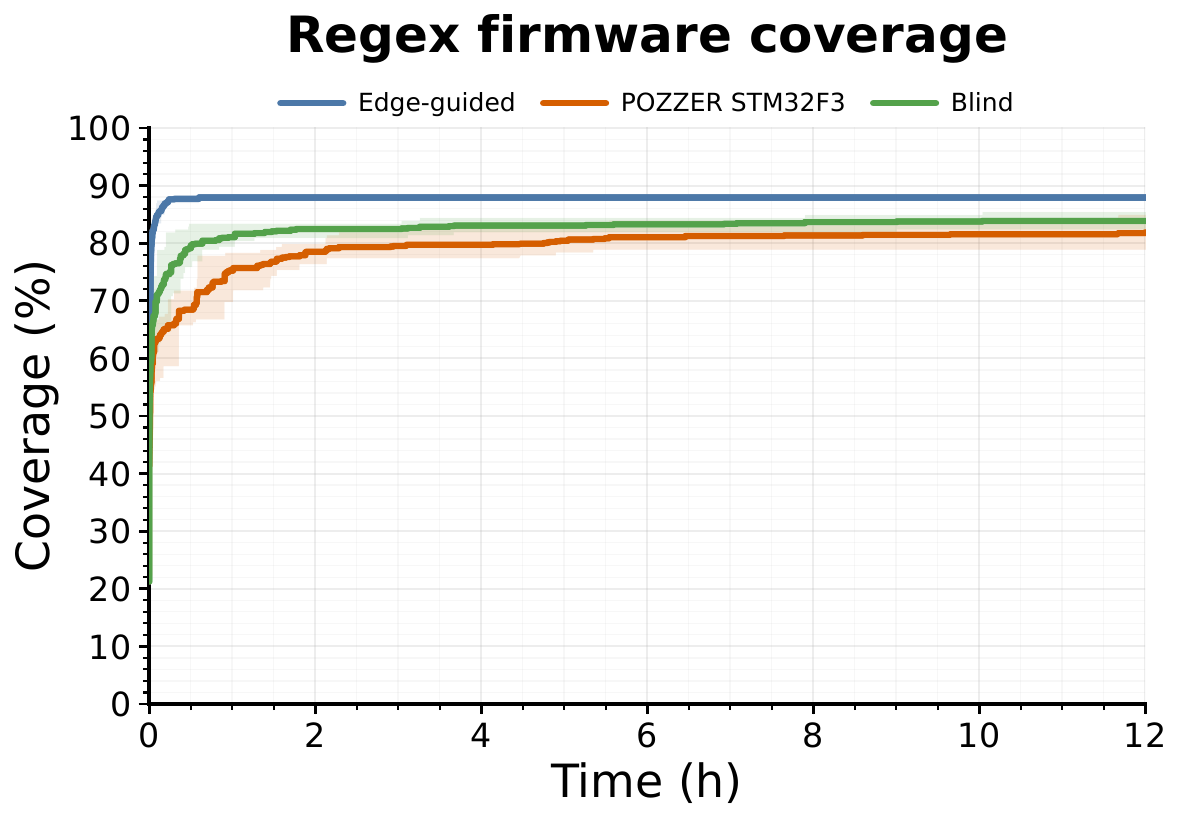}
    \end{subfigure}
    \begin{subfigure}{0.23\textwidth}
        \centering
        \includegraphics[width=\linewidth]{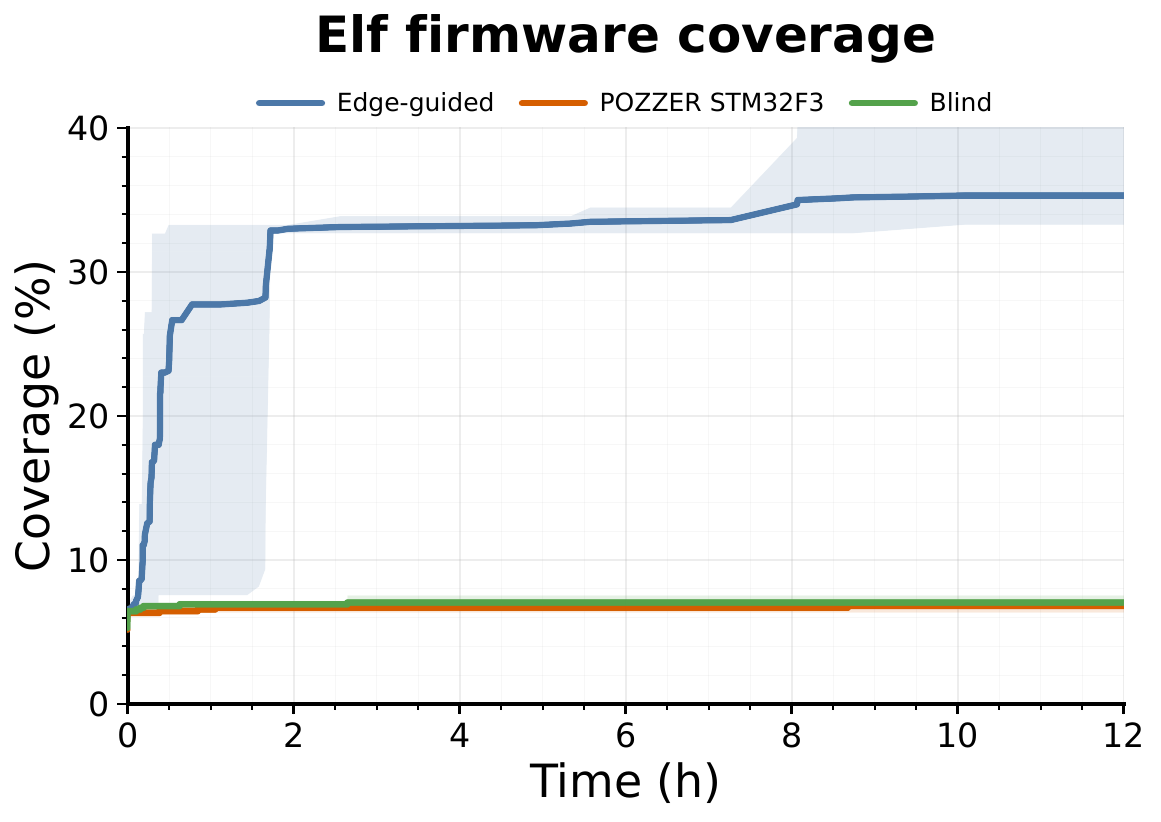}
    \end{subfigure}
    \begin{subfigure}{0.23\textwidth}
        \centering
        \includegraphics[width=\linewidth]{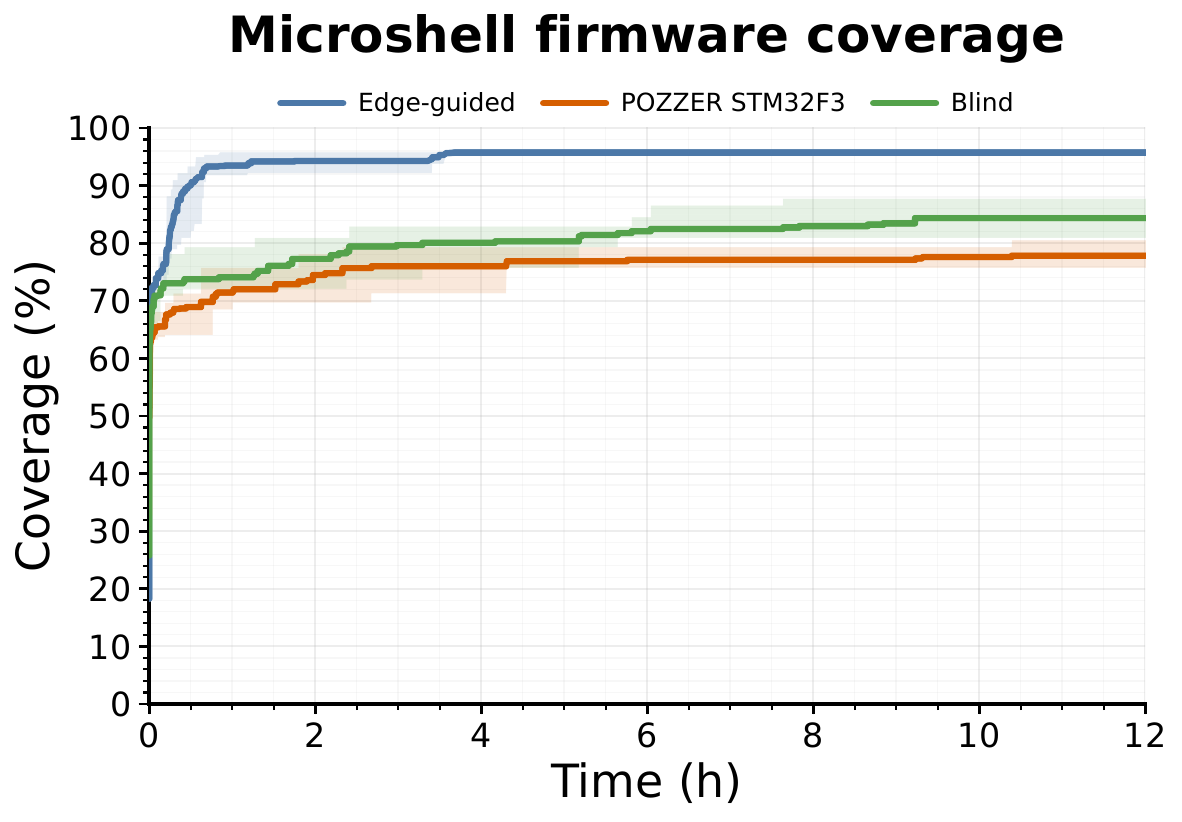}
    \end{subfigure}
    \begin{subfigure}{0.23\textwidth}
        \centering
        \includegraphics[width=\linewidth]{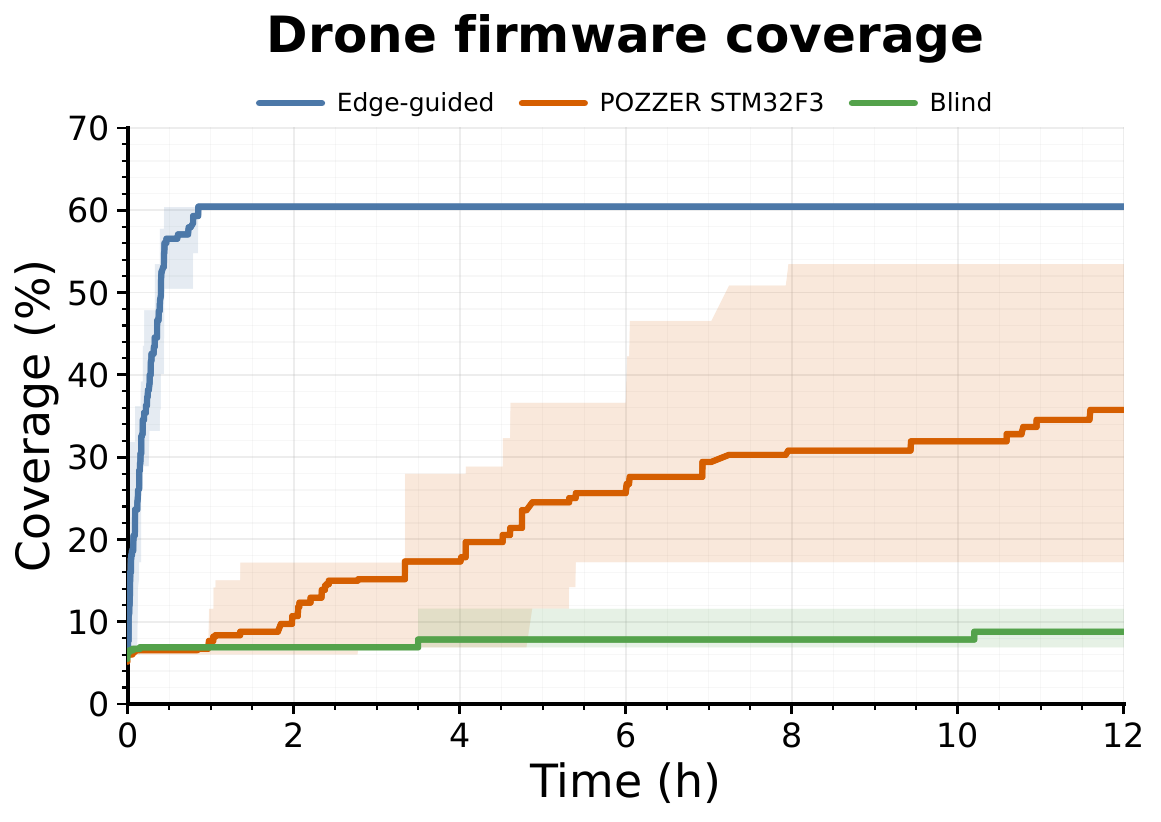}
    \end{subfigure}

    \begin{subfigure}{0.23\textwidth}
        \centering
        \includegraphics[width=\linewidth]{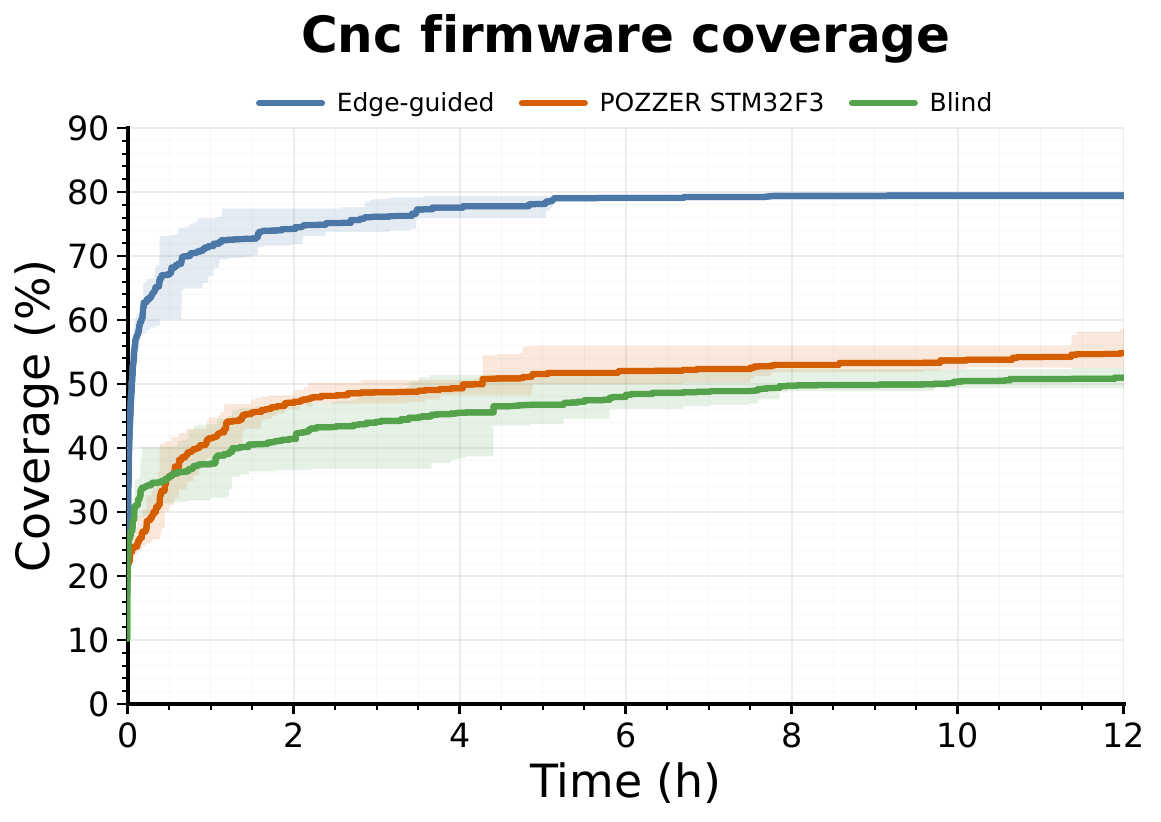}
    \end{subfigure}
    \begin{subfigure}{0.23\textwidth}
        \centering
        \includegraphics[width=\linewidth]{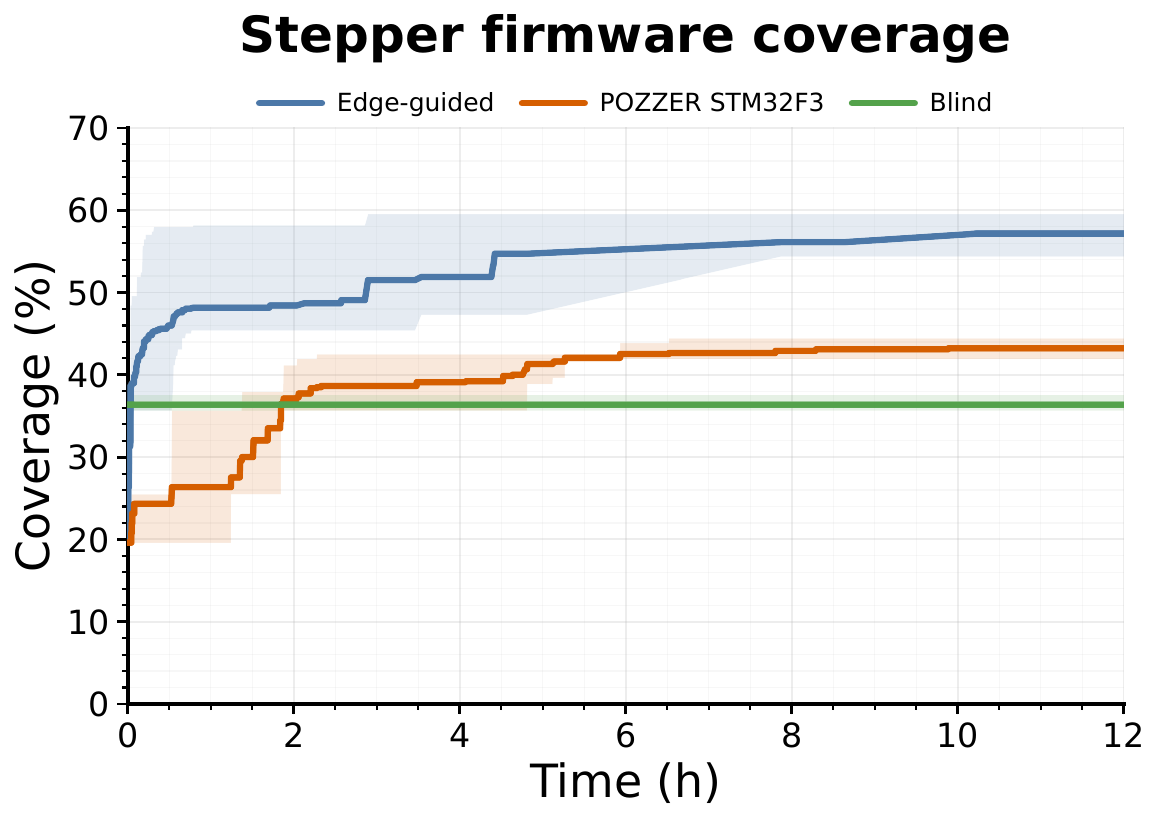}
    \end{subfigure}
    \begin{subfigure}{0.23\textwidth}
        \centering
        \includegraphics[width=\linewidth]{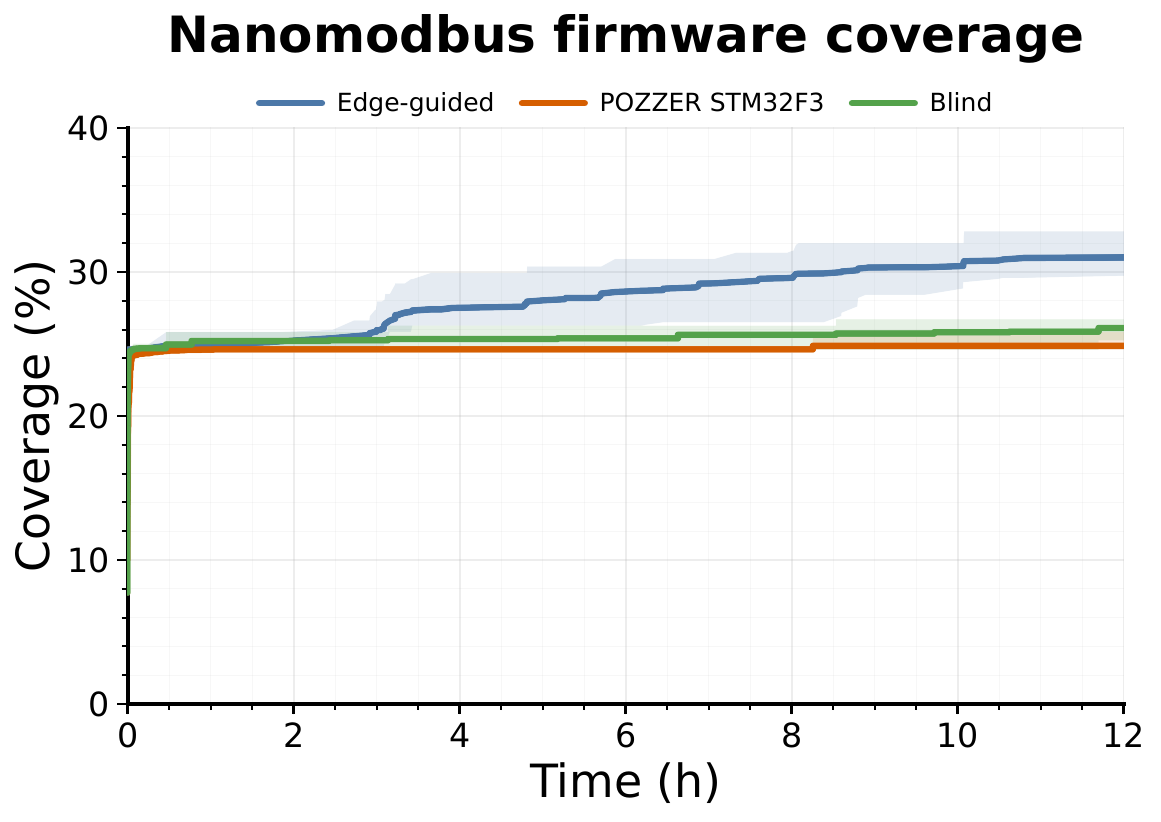}
    \end{subfigure}
    \caption{This figure presents coverage plots for different targets running on \emph{STM32F3}.}
    \label{fig:stm32f3_plots}
\end{figure*}

In this appendix, we show the remaining coverage plots from Table~\ref{table:cwresults}. In total, we had 15 targets in Table~\ref{table:cwresults} reporting results on two different hardware platforms. Therefore, here we present the remaining coverage plots for each platform separately. Figure~\ref{fig:sam4s_plots} shows the coverage plots for the target firmwares running on \emph{SAM4S}, while Figure~\ref{fig:stm32f3_plots} shows the coverage plots for the target firmware running on \emph{STM32F3}.

\end{document}